\documentclass{article}
\usepackage{epstopdf}
\usepackage{tikz}
\usepackage[mathscr]{eucal}
\usepackage{amsthm}
\usepackage{amsmath}
\usepackage{amssymb}
\usepackage{wasysym}
\usepackage{upgreek}
\usepackage{mathrsfs}
\usepackage{subcaption}
\usepackage{marginnote}
\usepackage{float}
\usepackage{graphicx}
\usepackage{color}
\usepackage{array,booktabs}
\usepackage[utf8]{inputenc}
\usepackage{newunicodechar}
\newunicodechar{é}{\'e}
\usepackage{cite}

\numberwithin{equation}{section}

\newcommand{\appropto}{\mathrel{\vcenter{
			\offinterlineskip\halign{\hfil$##$\cr
				\propto\cr\noalign{\kern2pt}\sim\cr\noalign{\kern-2pt}}}}}

\newcommand{\aSt}{\mathcal{W}}

\newcommand{\AStTL}[2]{\aSt_{#1,#2}}

\newcommand{\bAStTL}[2]{\overline{\mathcal{W}}_{#1,#2}}

\newcommand{\q}{\mathfrak{q}}

\begin{document}

\title{Geometrical four-point functions in the two-dimensional critical $Q$-state Potts model: The interchiral conformal bootstrap}

\author{Yifei He$^{1}$, Jesper Lykke Jacobsen$^{1,2,3}$, Hubert Saleur$^{1,4}$\\
[2.0mm]
${}^1$ \small Universit\'e Paris-Saclay, CNRS, CEA, Institut de Physique Th\'eorique, 91191, Gif-sur-Yvette, France \\
${}^2$ \small Laboratoire de Physique de l'Ecole Normale Sup\'erieure, ENS, Universit\'e PSL, CNRS,\\ 
      \small Sorbonne Universit\'e, Universit\'e de Paris, F-75005 Paris, France \\
${}^3$ \small Sorbonne Universit\'e, \'Ecole Normale Sup\'erieure, CNRS, Laboratoire de Physique (LPENS), 75005 Paris, France \\
${}^4$ \small Department of Physics, University of Southern California, Los Angeles, CA 90089, USA}

\date{}

\maketitle

\begin{abstract}
Based on the spectrum identified in our earlier work \cite{Jacobsen:2018pti}, we numerically solve the bootstrap to determine four-point correlation functions of the geometrical connectivities in the $Q$-state Potts model. Crucial in our approach is the existence of ``interchiral conformal blocks'', which arise from the degeneracy of fields with conformal weight $h_{r,1}$, with $r\in\mathbb{N}^*$, and are related to the underlying presence of the ``interchiral algebra'' introduced in \cite{Gainutdinov:2012nq}.  We also find evidence for the existence of ``renormalized'' recursions, replacing those that follow from the degeneracy of the field $\Phi_{12}^D$ in Liouville theory, and obtain the first few such recursions in closed form. This hints at the possibility of the full analytical determination of correlation functions in this model. 
\end{abstract}

\tableofcontents

\newpage

\section{Introduction and summary}

Recent years have witnessed the power of the modern bootstrap approach to conformal field theories (CFT) starting with the seminal work of \cite{Rattazzi:2008pe}. Since then many rigorous results on critical phenomena in dimension $d>2$ have been obtained; some significant examples can be found in \cite{ElShowk:2012ht,El-Showk:2014dwa,Kos:2016ysd}. While this approach and its ramifications rely on the unitarity of the CFT, an alternative method was also proposed for bootstrapping in the non-unitary case \cite{Gliozzi:2013ysa,Gliozzi:2014jsa} which has subsequently been applied to interesting geometrical models such as percolation and polymers \cite{LeClair:2018edq,Hikami:2017sbg}.

As exciting as these developments in $d>2$ CFT may be, important questions in $d=2$ CFT still remain to be answered. Prime among those is the issue of geometrical critical phenomena, where the definition of correlation functions involves non-local aspects. One typical example is the $Q$-state Potts model which, in the $Q\to1$ limit, describes the percolation problem. In such models, one focuses on the so-called geometrical correlations describing the connectivities in terms of the non-local, extended degrees of freedom, such as the Fortuin-Kasteleyn (FK) clusters \cite{FORTUIN1972536} in the case of the Potts model. The determination of such correlations is a difficult problem, because their very definition renders the underlying CFT non-unitary.

In the past decade, the understanding of the geometrical three-point functions was gradually achieved \cite{Delfino:2010xm,Picco:2013nga,Ikhlef:2015eua} in a development that revealed interesting connections to a so-called imaginary (or time-like) variant of Liouville theory with central charge $c < 1$. The next natural, yet highly non-trivial step, is to extend this development to geometrical four-point functions \cite{Delfino:2011sc}. We stress here that we are exclusively interested in the bulk geometry, which presents fundamental difficulties not present in the boundary case \cite{GoriViti:2018}.

An interesting strategy towards the determination of the geometrical four-point functions was proposed in a recent work \cite{Picco:2016ilr} using the conformal bootstrap philosophy. The idea can be stated simply: To obtain the amplitudes of the primary fields entering a given correlation function, one solves the crossing equation numerically with a proposed spectrum for the conformal weights of the participating primaries. This has led to a simple conjecture for the Potts spectrum in \cite{Picco:2016ilr} with apparent agreement with Monte-Carlo simulations \cite{Picco:2016ilr,Picco:2019dkm}.\footnote{See \cite{Javerzat:2019ujh,Javerzat:2019ohi} for related studies on the torus. See also \cite{Dotsenko:2019dcu} for a recent study of the four-spin correlations using the Coulomb Gas approach.}

It was however shown in \cite{Jacobsen:2018pti} that, unfortunately, the simple spectrum of \cite{Picco:2016ilr} does not correctly describe the geometrical correlations in the Potts model, although to the precision of Monte-Carlo simulations it appears as a rather convincing approximation. Moreover, \cite{Jacobsen:2018pti} made a more involved proposal for the spectrum, based on the representation theory of the affine Temperley-Lieb algebra, and verified its correctness through analytical checks in a number of solvable cases, by analytically arguing that the extra states in the corrected spectrum are actually necessary to avoid certain singularities which would otherwise be present, and finally by carrying out high-precision numerical verifications using a transfer matrix approach which is capable of targeting the amplitudes of the added parts of the spectrum.

To summarize, the spectrum of \cite{Jacobsen:2018pti} is now understood to provide the correct description of geometrical correlations in the Potts model.
Meanwhile, the correlation functions associated with  the simpler spectrum of \cite{Picco:2016ilr} were solved analytically in \cite{Migliaccio:2017dch} and understood later to provide a certain analytic continuation of correlation functions in type-D minimal models, or a non-diagonal generalization of the Liouville theory.

Although the spectrum used in \cite{Picco:2016ilr} is thus not correct for describing the geometrical correlation functions of interest, the other main idea of that work---namely, to study numerically the bootstrap equations---is certainly valid and worth further exploitation. The obvious suggestion is thus to revisit this idea, but in the context of the corrected Potts spectrum obtained in \cite{Jacobsen:2018pti}. This investigation is the focus of our work here.

\bigskip

\begin{figure}[t]
	\centering
	\includegraphics[width=\textwidth]{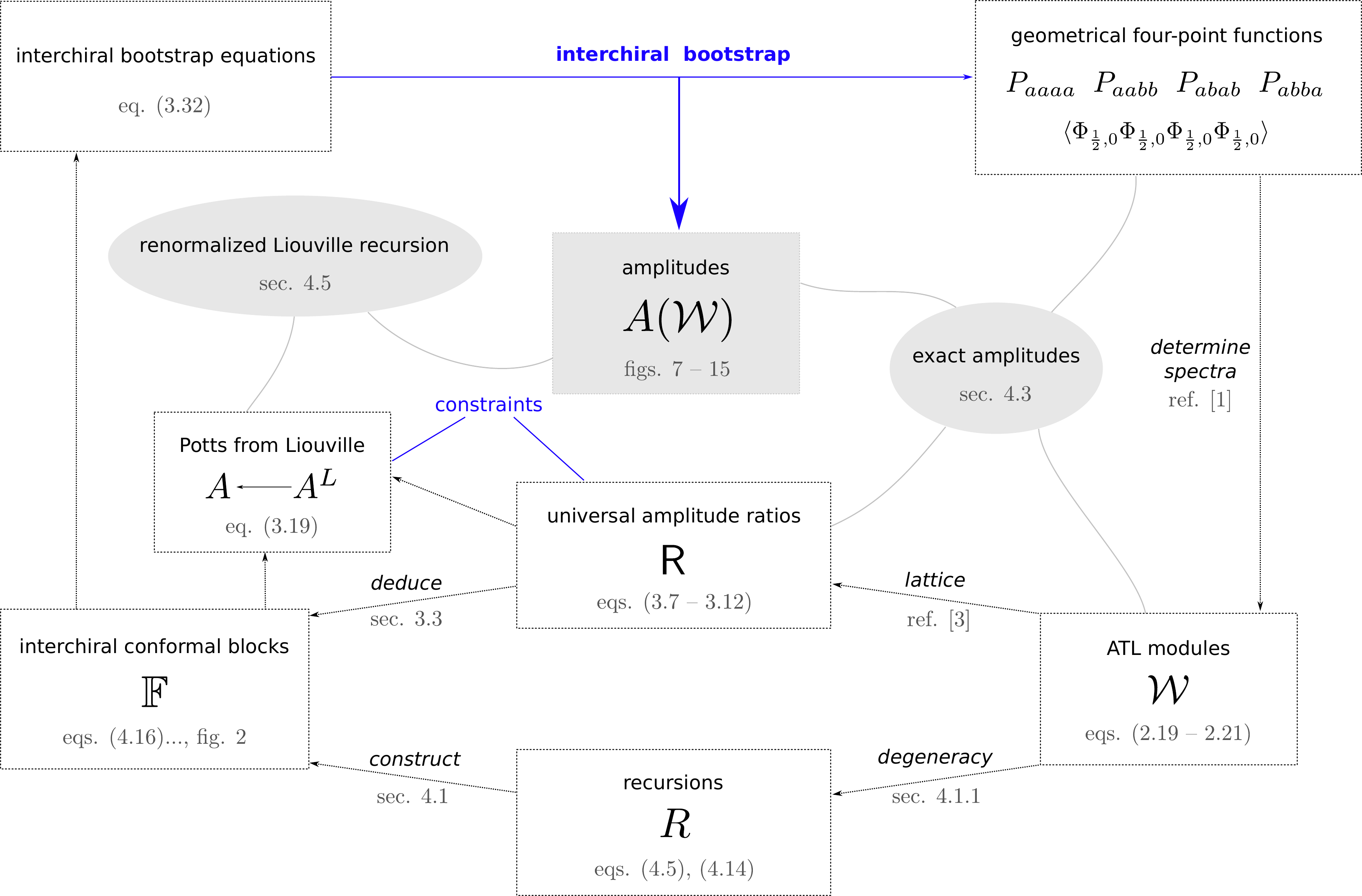}
	\caption{Chart of the contents of this paper.}
	\label{summary}
\end{figure}

To guide the readers through the bulk of this paper, we draw in fig.~\ref{summary} a chart which highlights the logical relations between the parts of this work, while locating the ``landmarks" of our findings.

\smallskip

We consider geometrical four-point functions which involve one or two FK clusters, i.e., the probabilities of the four points belonging to one or two distinct clusters. They are denoted as
\begin{equation}
P_{aaaa},\;\;P_{aabb},\;\;P_{abab},\;\;P_{abba}, \nonumber
\end{equation}
which can be seen as variants of CFT four-point functions of the spin operator $\Phi_{\frac{1}{2},0}$: 
\begin{equation}
\langle\Phi_{\frac{1}{2},0}\Phi_{\frac{1}{2},0}\Phi_{\frac{1}{2},0}\Phi_{\frac{1}{2},0}\rangle. \nonumber
\end{equation}
In \cite{Jacobsen:2018pti}, the $s$-channel spectra of the probabilities were obtained using a combination of algebraic and numerical methods. They are encoded by the affine Temperley-Lieb (ATL) modules
\begin{equation}
\mathcal{W}_{j,\mathrm{z}^2} \nonumber
\end{equation}
whose continuum limit gives rise to the conformal fields. The $t$- and $u$-channel spectra follow from geometric considerations. See eqs.~\eqref{Pspectra}--\eqref{ATLCFT}. 

Further studies on the lattice model, carried out in \cite{He:2020mhb}, reveal interesting features regarding how these fields contribute to the geometrical four-point functions: There exist universal ratios which relate either the amplitudes of the fields in different Potts probabilities, or the Potts amplitudes themselves, with the amplitudes appearing in the non-diagonal Liouville theory of \cite{Picco:2016ilr}. We have already established in \cite{He:2020mhb}, that the CFT four-point functions in the Liouville theory have a geometric interpretation in terms of clusters very similar to that of the Potts model, however with different weights assigned to the topologically non-trivial clusters. The amplitude ratios are universal in the sense that they depend only on the ATL module to which a given field belongs, and on the parameter $Q$. We briefly review these results in section \ref{geo4pt} and \ref{aratios} and redefine the universal amplitude ratios
\begin{equation}
\mathsf{R}_{\beta},\;\; \mathsf{R}_{\alpha},\;\;\mathsf{R}_{\bar{\alpha}}\nonumber
\end{equation}
for further convenience. See eqs.~\eqref{Rbeta}--\eqref{ratioforbootstrap} for definitions and their explicit expressions up to a certain level, as obtained from the lattice computations in \cite{He:2020mhb}.

\smallskip

The existence of the universal amplitude ratios strongly hints at the objects we call the ``interchiral conformal blocks"
\begin{equation}
\mathbb{F}_{j,\mathrm{z}^2}, \nonumber
\end{equation}
which organize the fields in the spectra according to the ATL modules they belong to (this ``organization'' corresponds, in the continuum limit, to the action of an interchiral algebra \cite{Gainutdinov:2012nq}, whence the name). Our crucial observation is the following: since the universal amplitude ratios depend only on the ATL modules, it is the same interchiral conformal blocks that enter various Potts probabilities as well as the non-diagonal Liouville theory of \cite{Picco:2016ilr}. Only the global amplitudes associated with entire ATL modules are modified by the change in the cluster weights implied \cite{He:2020mhb} by the passage from the true Potts correlators to non-diagonal Liouville correlators. By contrast, the relations among the fields within the same ATL module remain the same, i.e., the structure of the interchiral blocks $\mathbb{F}_{j,\mathrm{z}^2}$ is rigid. We discuss this in details in section \ref{pp}. There, using the interchiral block expansion of the four-point function in the non-diagonal Liouville theory and comparing with the Potts probabilities, we see that the bootstrap problem originally considered in \cite{Picco:2016ilr} has non-unique solutions. Furthermore, using the amplitude ratios $\mathsf{R}$, one can in fact extract some of the Potts amplitudes $A$ from the known amplitudes $A^L$ of the non-diagonal Liouville theory as obtained in \cite{Migliaccio:2017dch}:
\begin{equation}
A\leftarrow A^L. \nonumber
\end{equation}
The results are given in eq.~\eqref{PAusingL}.

\smallskip

With the existence of the interchiral conformal blocks established, the determination of the geometrical four-point functions reduces to solving for the global amplitudes
\begin{equation}
A_{aaaa}(\mathcal{W}_{j,\mathrm{z}^2}),\;\;A_{aabb}(\mathcal{W}_{j,\mathrm{z}^2}),\;\;A_{abab}(\mathcal{W}_{j,\mathrm{z}^2}),\;\; A_{abba}(\mathcal{W}_{j,\mathrm{z}^2})\nonumber
\end{equation}
of the entire ATL modules. The bootstrap idea proposed in \cite{Picco:2016ilr} then comes into play. We proceed by fully exploiting this idea using the interchiral block expansions of all four probabilities as related through crossing, and writing down the interchiral bootstrap equations \eqref{superP}. This is a linear system of the global amplitudes $A(\mathcal{W})$ in \eqref{ATLamp} whose relations are further constrained through the amplitude ratios $\mathsf{R}_{\alpha}$, $\mathsf{R}_{\bar{\alpha}}$. In addition, the Potts amplitudes $A$ extracted from the Liouville amplitudes $A^L$ further constrain the bootstrap problem. The precise ingredients of the bootstrap we carry out are indicated in blue on the chart (see fig.~\ref{summary}).

To implement the bootstrap, we need to construct the interchiral blocks $\mathbb{F}_{j,\mathrm{z}^2}$. For this, we first observe the degeneracy in the Potts spectra: the ATL module $\overline{\mathcal{W}}_{0,\q^2}$ consists of Kac modules in the CFT, i.e., they are degenerate representations of the Virasoro algebra. This includes in particular the field $\Phi^D_{2,1}$. Degeneracy of this field (as well as $\Phi^D_{1,2}$) are known to appear in the diagonal and non-diagonal Liouville theories \cite{Zamolodchikov:1995aa,Teschner:1995yf,Estienne:2015sua,Migliaccio:2017dch} which lead to recursions in the amplitudes when the Kac indices $(r,s)$ are shifted by $2$ units, eventually providing the full analytic solutions to those theories. Here, in the Potts model, with the sole degeneracy of $\Phi^D_{2,1}$ (but {\bf not} $\Phi^D_{1,2}$), by focusing on four-point functions of the spin operator, we obtain instead recursions
\begin{equation}
R^D,\;\;R^N\nonumber
\end{equation}
of amplitudes where the Kac index $r$ is shifted by $1$ unit. The explicit expressions are given in eqs.~\eqref{Rshift1D} and \eqref{Rshift1N} and the $D,N$ here label the diagonal and non-diagonal fields in the spectra. Such recursions exactly relate the amplitudes of the fields within the same ATL module, and we use them to re-sum the ordinary Virasoro conformal blocks $\mathcal{F}$ into the interchiral conformal blocks $\mathbb{F}$. This construction is given in section \ref{recursion} where the explicit interchiral blocks are given after eq.~\eqref{3superblocks} and illustrated with fig.~\ref{superblocksstructure}. We present the detailed derivation of the recursions from the degeneracy in section \ref{degeneracyrecursion}.

\medskip

The results of the numerical bootstrap are given in section \ref{bootresults} for $A(\mathcal{W}_{j,\mathrm{z}^2})$, with the ATL index $j\le 4$. We plot these amplitudes in the whole range of $0<Q<4$ in figs. \ref{aaaa0m1}--\ref{aabb41pole}. From these plots, we see clearly the analytic structures of the amplitudes, in particular their poles in $Q$ at certain rational values of the central charge. The following section \ref{singexact} is then devoted to analyzing these poles in details from a combination of perspectives: the requirement of smoothness as a function of $Q$ for the geometrical four-point functions, the amplitude ratios $\mathsf{R}$ which relate the amplitudes in the different geometries, and the corresponding difference in their respective spectrum. Through this analysis, we obtain certain exact amplitudes at special values of $Q$ which interpolate smoothly between the numerical bootstrap results as displayed in figs.~\ref{W4ires}--\ref{abab41analyticQ}. The bootstrap results are subsequently compared with lattice computations and the approximate description given by the non-diagonal Liouville theory of \cite{Picco:2016ilr} in section \ref{com}.

One interesting observation from the bootstrap on the Potts amplitudes is the following: While the degeneracy of the field $\Phi^D_{1,2}$ in the (non-diagonal) Liouville theory, and therefore the resulting recursion for shifting the Kac $s$-index, are absent in the case of the Potts model, there exists a ``renormalized'' version of the Liouville recursion, with the renormalization factors given by ratios of polynomials in $Q$. On the one hand, this is obtained from the extraction of the Potts amplitudes $A$ from the Liouville amplitudes $A^L$. On the other hand, we also obtain the remaining renormalized Liouville recursion up to level $j=4$ from the accurate numerical bootstrap results. They are given in eqs.~\eqref{renoLrecu} and \eqref{4020ratio}. It is natural to speculate that by fully understanding these renormalized Liouville recursions, combined with our interchiral block constructions, it would be possible to solve the Potts geometrical four-point functions analytically, which we will leave for future work.

\bigskip

We have in the above provided a complete guide and summary of the results contained in this paper. For the readers' convenience, we also give background reviews and supplementary materials. In section \ref{bootstrapPotts}, we give a review of the critical Potts model and the conformal bootstrap with emphasis on the application of the bootstrap approach to the Potts geometrical four-point functions. While the numerical aspects are important in the determination of the Potts amplitudes we present in section \ref{bootresults}, we leave the technical details to appendix \ref{bootstrapapp}. In addition, we recall in appendix \ref{ratiosappend} the original amplitude ratios obtained in \cite{He:2020mhb} and in appendix \ref{AL} the relevant analytic results of the non-diagonal Liouville theory, as they are used in various places in the paper.

\section{The conformal bootstrap approach to the Potts model}\label{bootstrapPotts}

We recall in this section the ingredients necessary to set up the conformal bootstrap for the geometrical correlation functions in the Potts model.

\subsection{The Potts model}
The $Q$-state Potts model \cite{potts_1952} is defined on a lattice where at each site resides a spin variable taking $Q$ possible values $\sigma_i=1,...,Q$ and the nearest neighbors have interaction energy $-K\delta_{\sigma_i,\sigma_j}$. The partition function is given by \
\begin{equation}\label{Z_spin}
Z=\sum_{\{\sigma\}}\prod_{\{ij\}}e^{K\delta_{\sigma_i,\sigma_j}} \,,
\end{equation} 
where $\{ij\}$ indicate the edges on the lattice and the sum is over all spin configurations $\{\sigma\}$. It is easy to recognize that the familiar Ising model corresponds to the case of $Q=2$. 

While the original definition \eqref{Z_spin} is restricted to integer values of $Q$, a more general definition is given by the Fortuin-Kasteleyn (FK) clusters \cite{FORTUIN1972536} where, by setting $v=e^K-1$, the partition function \eqref{Z_spin} becomes
\begin{equation}\label{Z_FK}
Z=\sum_{\mathcal{D}} v^{|\mathcal{D}|}Q^{\kappa(\mathcal{D})} \,.
\end{equation}
In this formulation, the partition function in given by configurations of bonds formed between neighboring lattice sites when they share the same spin value, with a probability $v/(1+v)$. The sum in \eqref{Z_FK} is over all diagrams $\mathcal{D}$, where $|\mathcal{D}|$ is the number of bonds and $\kappa(\mathcal{D})$ denotes the number of connected components---the so-called FK clusters---within a diagram. We henceforth focus on the two-dimensional square lattice.
At the critical value \cite{Baxter_1973},
\begin{equation}\label{cri}
v_{\rm c}=\sqrt{Q} \,,
\end{equation}
for $0\le Q\le 4$, the system goes through a second-order phase transition and is described by a conformal field theory (CFT) \cite{potts_1952,Baxter_1973}. Notice that in the FK-cluster description, the number of states $Q$ from the original definition enters the partition funciton \eqref{Z_FK} as a parameter and therefore the model is analytically continued to real values of $Q$.

Another equivalent formulation of the Potts model is through loops \cite{Baxter_1976}. Taking the midpoint of each edge to form another lattice, the loops are formed by connecting the nearest-neighboring sites such that they bounce on the FK clusters and internal cycles. As a result, two distinct Potts clusters are separated by an even number of loops.\footnote{For more details of the loop formulations, see section 2.1 of \cite{He:2020mhb}.} The partition function in this case becomes
\begin{equation}\label{Z_loop}
Z = Q^{|V|/2} \sum_\mathcal{D} \left( \frac{v}{n} \right)^{|\mathcal{D}|} n^{\ell(\mathcal{D})} \,.
\end{equation}
The $\ell(\mathcal{D})$ is the number of loops in a certain configuration and the loop fugacity is
\begin{equation}
n = \sqrt{Q} = \mathfrak{q} + \mathfrak{q}^{-1} \,, \label{loop-fugacity}
\end{equation}
where $\q$ is a quantum-group related parameter. At criticality \eqref{cri}, the partition function \eqref{Z_loop} only depends on the number of loops. In particular, all contractible and non-contractible loops get weight $n$.

\bigskip

On the lattice, one naturally considers the correlation functions
\begin{equation}
G_{a_1,a_2,\ldots,a_N} = \left \langle {\cal O}_{a_1}(\sigma_{i_1}) {\cal O}_{a_2}(\sigma_{i_2}) \cdots {\cal O}_{a_N}(\sigma_{i_N}) \right \rangle,
\label{order-param-corr}
\end{equation}
where the spin operator (the order parameter) is defined by
\begin{equation}
{\cal O}_a(\sigma_i) \equiv Q \delta_{\sigma_i,a} - 1 \,. \label{spin-op}
\end{equation}
More generic correlations are defined as a probability:
\begin{equation}\label{Pdef}
P_{\cal P} = \frac{1}{Z} \sum v^{|\mathcal{D}|} Q^{\kappa(\mathcal{D})} {\cal I}_{\cal P}(i_1,i_2,\ldots,i_N),
\end{equation}
usually labelled by $N$ ordered symbols in $\mathcal{P}$. The indicator function $\cal{I}_\mathcal{P}$ is defined such that two sites $i_j,i_k$ belong to the same FK cluster if and only if the corresponding ordered symbols in $\mathcal{P}$ are the same. These probabilities, which are well-defined for arbitrary real value of $Q$, are the geometrical correlations that we will study. In particular, we focus on the four-point geometrical correlations involving one or two clusters: $P_{aaaa}$, $P_{aabb}$, $P_{abba}$ and $P_{abab}$. On the lattice, they can be formally related to the lattice spin correlation functions by studying the combinatorics \cite{Delfino:2011sc}:
\begin{subequations} \label{sumrules}
	\begin{eqnarray}
	G_{aaaa} &=&(Q-1)(Q^2-3Q+3) P_{aaaa}+(Q-1)^2(P_{aabb}+P_{abba}+P_{abab}) \,, \label{Gaaaa}\\
	G_{aabb} &=&(2Q-3) P_{aaaa}+(Q-1)^2P_{aabb}+P_{abba}+P_{abab} \,, \\
	G_{abba} &=&(2Q-3)P_{aaaa}+P_{aabb}+(Q-1)^2P_{abba}+P_{abab} \,, \\
	G_{abab} &=&(2Q-3)P_{aaaa}+P_{aabb}+P_{abba}+(Q-1)^2P_{abab} \,.
	\end{eqnarray}
\end{subequations}
Notice that the left-hand side is only well-defined for integer values of $Q$.

\bigskip

In the continuum limit at criticality, the Potts model can be parameterized as \cite{CGreview}
\begin{equation}
\sqrt{Q}=2\cos \left( {\pi\over x+1} \right), \quad x \in [1,\infty],
\end{equation}
where the parameter $x$ is related to the central charge of the CFT by
\begin{equation}
c=1-{6\over x(x+1)} \,,
\end{equation}
and the quantum group related parameter $\q=e^{\frac{i\pi}{x+1}}$. One can adopt a Kac parameterization for the conformal dimensions of the primary fields as:
\begin{equation}\label{cw}
h_{r,s}={[(x+1)r-xs]^2-1\over 4x(x+1)}
\end{equation}
with 
$h_{-r,-s}=h_{r,s}$, although in this case the $x$ is not restricted to be an integer as in the minimal models and furthermore, the Kac indices $(r,s)$ can be fractions. In particular, the order parameter---i.e., the spin operator---is known to be given by the Kac indices $(r,s)=(\frac{1}{2},0)$ \cite{denNijs:1983zz,Nienhuis:1984wm}.
For convenience, we will also use another parameterization $\upbeta$ for the central charge:
\begin{equation}\label{upbetadef}
\upbeta^2=\frac{x}{x+1}, \quad \frac{1}{2}\le\upbeta^2\le 1,
\end{equation}
which is closely related to $c<1$  Liouville theory \cite{Zamolodchikov:2005fy}.

Throughout the paper, we will refer to diagonal and non-diagonal primaries with the Kac indices $(r,s)$ whose left and right conformal dimensions are given by
\begin{equation}\label{DN}
	(h,\bar{h})=\begin{cases}
	(h_{r,s},h_{r,s}),&\text{diagonal},\\
	(h_{r,s},h_{r,-s}),&\text{non-diagonal}.
	\end{cases}
\end{equation}
We shall often label such a primary by the superscript $D$ for diagonal, or $N$ for non-diagonal.
Its total conformal dimension is 
\begin{equation}\label{totalh}
h+\bar{h}
\end{equation} and the conformal spin is 
\begin{equation}\label{spin}
\bar{h}-h=\begin{cases}
0,&\text{diagonal},\\
rs,&\text{non-diagonal}.
\end{cases}
\end{equation}

\medskip

The four lattice sites in the continuum limit become $(i_1,i_2,i_3,i_4)\to(z_1,z_2,z_3,z_4)$ and the $s,t,u$-channels are defined as
\begin{equation}\label{channels}
\mbox{$s$-channel}:\;z_1 \to z_2\,,\;\;
\mbox{$t$-channel}:\;z_1 \to z_4\,,\;\;
\mbox{$u$-channel}:\;z_1 \to z_3\,.
\end{equation}
From the relations with the lattice spin correlations \eqref{sumrules}, it is natural to consider the four-point geometrical correlations $P_{aaaa}$, $P_{aabb}$, $P_{abba}$ and $P_{abab}$ as four-point functions of the spin operator $\Phi_{\frac{1}{2},0}$ 
of the type:\footnote{The spin operator $\Phi_{1/2,0}$, as determined by the representation theory of the symmetric group $S_Q$, has a non-zero $N$-point function as long as each point belongs to the same FK cluster as at least one other point. The four-point function thus provides non-zero contributions to precisely the probabilities $P_{aaaa}$, $P_{aabb}$, $P_{abba}$ and $P_{abab}$, with an overall multiplicity that can be computed from the representation theory \cite{Vasseur:2012tz,Vasseur:2013baa,Couvreur:2017inl}. Whenever we wish to single out one of these contributions, we must therefore specify the corresponding labels. \label{fn2}}
\begin{equation}\label{4spin}
\langle\Phi_{\frac{1}{2},0}(z_1,\bar{z}_1)\Phi_{\frac{1}{2},0}(z_2,\bar{z}_2)\Phi_{\frac{1}{2},0}(z_3,\bar{z}_3)\Phi_{\frac{1}{2},0}(z_4,\bar{z}_4)\rangle
\end{equation}
and they are therefore given by the conformal blocks of the spectra in the fusion channels. Accounting for the geometry,
the spectra of each channel in the geometrical correlations are related through crossing:
\begin{align}\label{Pspectra}
\setlength{\arraycolsep}{6mm}
\renewcommand{\arraystretch}{1.2}
\begin{array}{c|c|c|c}
\mbox{probability} & \mbox{$s$-channel} & \mbox{$t$-channel} & \mbox{$u$-channel}\\
\hline
P_{aaaa} & \mathcal{S}_1 & \mathcal{S}_1 & \mathcal{S}_1 \\
P_{aabb} & \mathcal{S}_2 & \mathcal{S}_3 & \mathcal{S}_3 \\
P_{abab} & \mathcal{S}_3 & \mathcal{S}_3 & \mathcal{S}_2 \\
P_{abba} & \mathcal{S}_3 & \mathcal{S}_2 & \mathcal{S}_3 \\
\end{array}
\end{align}
In particular, under $s\leftrightarrow t$ which we will focus on below, the spectra of $P_{aaaa}$ and $P_{abab}$ are symmetric while $P_{aabb}$ and $P_{abba}$ get interchanged.

\medskip

The spectra \eqref{Pspectra} were determined in \cite{Jacobsen:2018pti} by focusing on the $s$-channel, using a combination of algebraic and numerical methods. They are given in terms of the affine Temperley-Lieb (ATL) modules $\mathcal{W}_{j,\mathrm{z}^2}$:
\begin{align}\label{Pottsspectrum}
\setlength{\arraycolsep}{6mm}
\renewcommand{\arraystretch}{1.2}
\begin{array}{c|ll}
\mbox{spectrum} & \mbox{ATL modules} & \mbox{Parities}
\\
\hline
\mathcal{S}_1 &  \AStTL{0}{-1}\cup\AStTL{j}{e^{2i\pi p/M}} & j \in 2 \mathbb{N}^*, \ jp/M\hbox{ even}    \\
\mathcal{S}_2 & \AStTL{0}{-1}\cup\bAStTL{0}{\q^2}\cup\AStTL{j}{e^{2i\pi p/M}}& j \in 2 \mathbb{N}^*, \ jp/M\hbox{ even}     \\
\mathcal{S}_3 & \AStTL{j}{e^{2i\pi p/M}} & j \in 2 \mathbb{N}^*, \ jp/M\hbox{ integer}    \\
\hline
\end{array}
\end{align}
where $p,M$ are coprime integers with in particular $p=0$ allowed. The ATL modules each contains a tower of primary fields and their descendants with the following Kac indices:
\begin{subequations}\label{ATLCFT}
\begin{eqnarray}
\mathcal{W}_{j,e^{2i\pi p/M}}:&\;\;(r,s)=(\mathbb{Z}+\frac{p}{M},j)^N,&\;\;\text{non-diagonal}\\
\overline{\mathcal{W}}_{0,\q^2}:&\;\;(r,s)=(\mathbb{N}^*,1)^D,&\;\;\text{diagonal}
\end{eqnarray}
\end{subequations}
where we recall that $N,D$ stand for non-diagonal and diagonal respectively, with the conformal dimensions given by \eqref{DN}. In particular, the Virasoro modules in $\overline{\mathcal{W}}_{0,\q^2}$ are given by Kac modules, where the null descendant of the primaries with $(r,s)\in\mathbb{N}^*$ at level $rs$ is removed.

Notice that $jp/M$ indicates the conformal spin $rs$ of the leading primary in a module $\mathcal{W}_{j,e^{2i\pi p/M}}$ and in the following we will often need to refer to the modules with even and odd spins separately. We will thus use the notation
\begin{subequations}
\begin{eqnarray}
\mathcal{W}^+_{j,e^{2i\pi p/M}},&&jp/M\;\text{even},\label{Wplus}\\
\mathcal{W}^-_{j,e^{2i\pi p/M}},&&jp/M\;\text{odd},	
\end{eqnarray}
\end{subequations}
for the rest of the paper.

\subsection{The conformal bootstrap approach}
Consider a generic four-point function of identical operators:
\begin{equation}\label{4ptgeneric}
\langle\Phi(z_1,\bar{z}_1)\Phi(z_2,\bar{z}_2)\Phi(z_3,\bar{z}_3)\Phi(z_4,\bar{z}_4)\rangle.
\end{equation}
After mapping the four points $(z_1,z_2,z_3,z_4)$ to $(z,0,\infty,1)$ through a global conformal transformation, the $s,t,u$ channels \eqref{channels} become
\begin{equation}
\mbox{$s$-channel}:\;z \to 0\,, \quad
\mbox{$t$-channel}:\;z \to 1\,, \quad
\mbox{$u$-channel}:\;z \to \infty\,,
\end{equation}
and the four-point function \eqref{4ptgeneric} can be written in terms of the conformal block expansions:
\begin{equation}\label{GCB}
G(z,\bar{z})=\sum_{(h,\bar{h})\in\mathcal{S}^{(\mathsf{c})}}A^{(\mathsf{c})}(h,\bar{h})\mathcal{F}^{(\mathsf{c})}_{h}(z)\bar{\mathcal{F}}^{(\mathsf{c})}_{\bar{h}}(\bar{z}),  \quad \mbox{with } \mathsf{c}=s,t,u.
\end{equation}
The constant coefficient $A^{(\mathsf{c})}(h,\bar{h})$ here---which we will henceforth refer to as the amplitude for field $(h,\bar{h})$---arises from the structure constant in the fusion
\begin{equation}\label{fusion}
\Phi\times\Phi\overset{(\mathsf{c})}{\longrightarrow}(h,\bar{h})
\end{equation}
as%
\footnote{As explained in footnote~\ref{fn2}, the correlation function decomposes into probabilities $P_{aaaa}$, $P_{aabb}$, $P_{abba}$ and $P_{abab}$, depending on the chosen geometry. The amplitudes similarly depend on the geometry, but in the general reasoning presented here we shall keep that dependence implicit and only specify it when needed.}
\begin{equation}\label{AC}
A^{(\mathsf{c})}(h,\bar{h})=C(\Phi,\Phi,(h,\bar{h}))C((h,\bar{h}),\Phi,\Phi),
\end{equation}
where we have chosen the normalization of the two-point functions of identical primaries to be $1$ besides position-dependent factor. Note that our discussions below are independent of this normalization which we take merely for convenience and notation simplicity. The structure constant is symmetric under permutation of the three fields and in the following we will also use the notation
\begin{equation}
C_{(r_i,s_i)(r_j,s_j)(r_k,s_k)} \,,
\end{equation}
where the indices $(r,s)=(r,s)^{D,N}$ represent diagonal or non-diagonal fields, as specified by the superscript. Notice that the fusion with the identity operator $\Phi^D_{(1,1)}$ gives:
\begin{equation}
\Phi^D_{1,1}\times(h,\bar{h})\to(h,\bar{h}) \,,
\end{equation}
suggesting the structure constants
\begin{equation}\label{id3pt}
C_{(1,1)^D(r,s)(r,s)}=1 \,,
\end{equation}
since this reduces to the normalized two-point function of $(r,s)$.

The essence of the conformal bootstrap approach lies in the crossing equation for the four-point functions, which states the equivalence of the conformal-block expansions in different fusion channels \cite{Belavin:1984vu}. In the case of the four-point function \eqref{4ptgeneric}, this is to say that the conformal-block expansions \eqref{GCB} in different channels $\mathsf{c}=s,t,u$ give the same four-point functions, as a result of the associativity of the fusion algebra \eqref{fusion}. Such equivalence puts strong constraints on the spectra $\mathcal{S}^{(\mathsf{c})}$ and the amplitudes $A^{(\mathsf{c})}(h,\bar{h})$, and in certain cases can uniquely define the theory of interest. Combining with positivity constraints from unitarity\footnote{In unitary CFTs, the amplitudes \eqref{AC} are positive, as the squares of the structure constants or the matrix constructed from pairwise products of the structure constants are positive-definite.} and powerful numerical implementations, the conformal bootstrap approach has recently led to many rigorous results in $d>2$ unitary theories \cite{ElShowk:2012ht,El-Showk:2014dwa}. 

In the non-unitary case, one cannot resort to such positivity constraints. However, with a proposed spectrum and a clever algorithm, it is possible to solve a target theory numerically using the crossing equation alone. This is the idea recently proposed in \cite{Picco:2016ilr}, in the context of non-unitary geometric models such as the Potts model. It is this type of bootstrap approach that we are applying here to study the Potts geometrical correlations using the spectra \eqref{Pottsspectrum}.

\medskip

Consider in our case the conformal block expansion of geometrical correlations 
\eqref{4spin}. The crossing-symmetric probabilities $P(z,\bar{z}) = P_{aaaa},P_{abab}$ have the same spectrum in the $s$- and $t$-channel and are given by:
\begin{equation}\label{crsymP}
P(z,\bar{z})=\sum_{(h,\bar{h})\in\mathcal{S}}A(h,\bar{h})\mathcal{F}^{(s)}_{h}(z)\mathcal{F}^{(s)}_{\bar{h}}(\bar{z})=\sum_{(h,\bar{h})\in\mathcal{S}}A(h,\bar{h})\mathcal{F}^{(t)}_{h}(z)\mathcal{F}^{(t)}_{\bar{h}}(\bar{z}),
\end{equation}
where $\mathcal{S}=\mathcal{S}^{(s)}=\mathcal{S}^{(t)}$. For $P_{aaaa}$ one has $\mathcal{S}=\mathcal{S}_1$ and $A(h,\bar{h})=A_{aaaa}(h,\bar{h})$ while for $P_{abab}$, $\mathcal{S}=\mathcal{S}_3$ and $A(h,\bar{h})=A_{abab}(h,\bar{h})$.
Rewritten as
\begin{equation}\label{booteq1}
\sum_{(h,\bar{h})\in\mathcal{S}}A(h,\bar{h})\big(\mathcal{F}^{(s)}_{h}(z)\mathcal{F}^{(s)}_{\bar{h}}(\bar{z})-\mathcal{F}^{(t)}_{h}(z)\mathcal{F}^{(t)}_{\bar{h}}(\bar{z})\big)=0,
\end{equation}
this is a linear system for the amplitudes $A(h,\bar{h})$. Using the method proposed in \cite{Picco:2016ilr}, one can numerically solve this linear system
by sampling the points $z_i$. Doing this multiple times provides statistics on the amplitudes which were used in \cite{Picco:2016ilr} as a measurement of crossing symmetry.

Here we are going to take this approach one step further since we have the spectra \eqref{Pottsspectrum} for all four probabilities. While the crossing-symmetric probabilities, $P_{aaaa}$ and $P_{abab}$, can be expanded using \eqref{crsymP}, the other two probabilities, $P_{abba}$ and $P_{aabb}$, get interchanged under $s\leftrightarrow t$ and thus have the following conformal block expansions:
\begin{eqnarray}\label{booteq2}
&&P_{aabb}=\sum_{(h,\bar{h})\in\mathcal{S}_2}A_{aabb}(h,\bar{h})\mathcal{F}^{(s)}_{h}(z)\mathcal{F}^{(s)}_{\bar{h}}(\bar{z})=\sum_{(h,\bar{h})\in\mathcal{S}_3}A_{abba}(h,\bar{h})\mathcal{F}^{(t)}_{h}(z)\mathcal{F}^{(t)}_{\bar{h}}(\bar{z}),\\
&&P_{abba}=\sum_{(h,\bar{h})\in\mathcal{S}_3}A_{abba}(h,\bar{h})\mathcal{F}^{(s)}_{h}(z)\mathcal{F}^{(s)}_{\bar{h}}(\bar{z})=\sum_{(h,\bar{h})\in\mathcal{S}_2}A_{aabb}(h,\bar{h})\mathcal{F}^{(t)}_{h}(z)\mathcal{F}^{(t)}_{\bar{h}}(\bar{z}).\label{booteq3}
\end{eqnarray}
As discussed in \cite{Jacobsen:2018pti}, the fields $(r,s)$ with even and odd spins have the following amplitude relations:
\begin{subequations}\label{evenodd}
	\begin{eqnarray}
	A_{abab}&=&A_{abba},\;\;\;rs\;\text{even},\label{evenA}\\
	A_{abab}&=&-A_{abba},\;\;rs\;\text{odd},
	\end{eqnarray}
\end{subequations}
and therefore the symmetric and anti-symmetric combinations, $P_{abab}+P_{abba}$ and $P_{abab}-P_{abba}$, only involve fields with even and odd conformal spin, respectively.

Eq.~\eqref{booteq1} for $P_{aaaa}$, $P_{abab}$ and eqs.~\eqref{booteq2}--\eqref{booteq3} for $P_{aabb}$, $P_{abba}$ together define our problem of solving the Potts geometrical correlations.

\subsubsection{Conformal blocks}\label{CB}
One main ingredient in the conformal bootstrap approach to the four-point functions is the computation of conformal blocks. For practical implementations, we use the Zamolodchikov recursive formula \cite{Zamolodchikovrecursion} to compute the Virasoro conformal blocks of the primary fields appearing in \eqref{Pottsspectrum}. In particular, in the case of the four-spin correlations \eqref{4spin} with external dimensions $h_{\frac{1}{2},0}$, the $s$-channel conformal block for an internal field with dimension $h$ is given by:
\begin{equation}\label{CBZamo}
\mathcal{F}_{h}^{(s)}(z)=(16q)^{h-\frac{c-1}{24}}(z(1-z))^{-\frac{c-1}{24}-\frac{1}{8\upbeta^2}}\theta_3(q)^{-\frac{c-1}{6}-\frac{1}{\upbeta^2}}H_{h}(q),
\end{equation}
and in the $t$-channel we have:
\begin{equation}\label{st}
\mathcal{F}^{(t)}_{h}(z)=\mathcal{F}^{(s)}_{h}(1-z).
\end{equation}
In the above expressions, the elliptic nome $q$ and the Jacobi theta function $\theta_3(q)$ are given by:
\begin{equation}
q(z)=e^{i\pi\tau}, \quad \tau=i \, \frac{K(1-z)}{K(z)}, \quad \theta_3(q)=\sum_{n=-\infty}^{\infty}q^{n^2},
\end{equation}
where $K(z)$ is the complete elliptical integral of the first kind. The $H_{h}(q)$ is given by the recursive relation
\begin{equation}\label{CBH}
H_{h}(q)=1+\sum_{m,n=1}^{\infty}\frac{(16q)^{mn}}{h-h_{m,n}}\mathrm{R}_{m,n}H_{h_{m,-n}}(q),
\end{equation}
where the $\mathrm{R}_{m,n}$ in the case of four-spin conformal blocks are given explicitly by \cite{Picco:2016ilr}:
\begin{equation}\label{Rblock}
\mathrm{R}_{m,n}=\begin{cases}
0,\;&\hbox{$n$ odd,}\\
-2^{1-4mn}\lambda_{mn}\prod_{m'=1-m}^{m}\prod_{n'=1-n}^{n}\lambda_{m',n'}^{(-1)^{n'+1}},\; &\hbox{$n$ even,}
\end{cases}
\end{equation}
with $\lambda_{m',n'}=-\frac{m'}{2\upbeta}+\frac{n'\upbeta}{2}$
and the products exclude $(m',n')=(0,0)$.

Notice that in the spectrum \eqref{Pottsspectrum}, there are primary fields with degenerate indices, i.e., $r,s\in\mathbb{N}^*$, appearing in the modules $\mathcal{W}_{j,1}$. This poses problems for the computation of the conformal blocks for these primaries which have poles for $r,s\in\mathbb{N}^*$, as is obvious in \eqref{CBH}. Of course, the appearance of such pole terms in the Zamolodchikov recursive formula is associated with the fact that in minimal models the degenerate representations have the null descendants decoupling from the spectrum, while in the case of the Potts model, we do not expect such decoupling to occur. This means that the theory is {\sl generically logarithmic} \cite{Gainutdinov_2013}.  
The fields with $r,s\in\mathbb{N}^*$ in $\mathcal{W}_{j,1}$ are expected to have logarithmic partners, and the presence of Jordan cells for $L_0,\bar{L}_0$ should lead to finite confomal blocks regularizing the naive divergences in Zamolodchikov's formula, with, in particular, a $\ln(z\bar{z})$ dependency.\footnote{We thank S.~Ribault for various discussions on the topic of conformal blocks.} This will be studied in detail in a forthcoming paper. For the time being, 
we content ourselves with a ``naive" regularization procedure for numerical implementations. As can be seen from \eqref{CBZamo} and \eqref{CBH}, the residue of the pole at $h=h_{m,n}$ is given by
\begin{equation}\label{reg1}
\mathrm{R}_{m,n}\mathcal{F}^{(s)}_{h_{m,-n}} \,,
\end{equation}
where $\mathcal{F}^{(s)}_{h_{m,-n}}$ is the conformal block of the descendant with conformal dimension $h=h_{m,-n}=h_{m,n}+mn$. We therefore subtract this pole term from the (left and right) block of $\mathcal{F}^{(s)}_{h_{m,n}}$ and include the term
\begin{equation}\label{reg2}
\mathcal{F}^{(s)}_{h_{m,-n}}(z)\mathcal{F}^{(s)}_{h_{m,-n}}(\bar{z})
\end{equation}
in the four-point function with a free coefficient.\footnote{This is similar to the regularization procedure in \cite{Picco:2016ilr}, however we do not assume a specific $z$-dependence.} This takes into account certain contributions of the descendants in the four-point functions. However, it is worth stressing that this serves as an approximation in the numerical bootstrap, since in general the coefficient involved should have logarithmic dependence in $z$ and the modification here could change the higher-level structure of the blocks. While this may introduce instabilities into the numerics, we will discuss below extra constraints to impose on the numerical bootstrap in order to stabilize the solution.

Note that in the module $\overline{\mathcal{W}}_{0,\q^2}$ of \eqref{Pottsspectrum}, the fields also have degenerate indices $(r,1)$. However, for the case of four-spin conformal blocks, the residue is zero for these fields due to \eqref{Rblock}, which exactly removes the null descendants and generate the Kac modules appearing in $\overline{\mathcal{W}}_{0,\q^2}$. The conformal blocks in this case are thus exact.

\section{From minimal models to the Potts model}\label{MMtoPotts}

In \cite{Picco:2016ilr}, the authors conjectured a simple spectrum for some of the geometrical correlations in the Potts model which, using the bootstrap approach, was checked to satisfy the crossing equation \eqref{booteq1}. While it provides a numerical description of the Potts probabilities that appeared to be in accord with Monte-Carlo simulations \cite{Picco:2019dkm}, the proposed spectrum was finally shown in \cite{Jacobsen:2018pti} to be only a subset of the true Potts spectrum \eqref{Pottsspectrum}. Later it was understood \cite{Migliaccio:2017dch} that the spectrum of \cite{Picco:2016ilr} was in fact valid for a generalization of type-D minimal models, when the $\upbeta^2$ in \eqref{upbetadef} was taken to irrational values.

In \cite{He:2020mhb}, we studied the CFT four-point functions given by the spectrum of \cite{Picco:2016ilr} from the lattice point of view and revealed its connection with the Potts probabilities: the four-point functions of the operators of interest involve the same types of diagrammatic expansion in terms of clusters/loops as the Potts probabilities we consider here, however with different weights assigned to the topologically non-trivial loops. We referred to the geometrical correlations thus obtained as the ``pseudo-probabilities" (see eq.~\eqref{pseudoprobdef} below for a precise definition). In the work \cite{He:2020mhb} we have also studied the Potts probabilities in a lattice regularization---i.e., on semi-infinite cylinders of finite circumference $\mathsf{L}$---and observed, to arbitrarily high numerical precision, several striking facts regarding the contribution of the fields to the Potts probabilities and to the pseudo-probabilities. Crucially, these facts were observed to be independent of $\mathsf{L}$, and can hence be presumed to carry over to the continuum limit as well.
In this section, we briefly summarize these results and explain how they can be used to extract information about the Potts model from minimal models (i.e., the generalization to generic central charges) and also as input for the bootstrap of the Potts model itself.

\subsection{A geometric picture of the correlation functions}\label{geo4pt}

In \cite{Picco:2016ilr}, the authors found a crossing-symmetric spectrum $\mathcal{S}_{r,s}=\mathcal{S}_{\mathbb{Z}+\frac{1}{2},2\mathbb{Z}}$ at generic values of $0< Q< 4$ \footnote{In \cite{Picco:2016ilr}, the spectrum was found to be crossing symmetric for complex values of $Q$. Here we focus on real $0\le Q\le 4$ corresponding to the second-order phase transition in the lattice model for which the continuum limit is known to be conformally invariant.} for the $s\leftrightarrow t$ crossing-symmetric four-point function conjectured to describe the Potts probabilities\footnote{The factor $\frac{1}{Q-2}$ was fixed later in \cite{Picco:2019dkm} and the claim in \cite{Picco:2016ilr} that \eqref{PRS4pt} was exact was modified to an approximation in \cite{Picco:2019dkm}.
We show here the approximate nature of the proportionality by the symbol $\appropto$.}
\begin{equation}\label{PRS4pt}
\langle V^D_{\frac{1}{2},0}V^N_{\frac{1}{2},0}V^D_{\frac{1}{2},0}V^N_{\frac{1}{2},0}\rangle\appropto P_{aaaa}+\frac{2}{Q-2} P_{abab} \,,
\end{equation} 
which is approximately true for generic $Q$ and becomes exact for $Q=0,3,4$. The fields $V_{\frac{1}{2},0}^D$ and $V_{\frac{1}{2},0}^N$ have conformal dimensions $(h_{\frac{1}{2},0},h_{\frac{1}{2},0})$, i.e., same as the spin operator, and have their origin in the diagonal and non-diagonal sectors, respectively, of the type-D minimal models (here and below $D$ and $N$ stand for diagonal and non-diagonal). Initially proposed as the spectrum for the Potts probabilities, it is now understood \cite{Migliaccio:2017dch} that this spectrum arises from a certain limit of minimal models when the $\upbeta^2$ in \eqref{upbetadef} is taken to irrational values, although numerically it gives a reasonable approximation of some of the Potts probabilities \cite{Picco:2019dkm}. The structure constants appearing in the four-point function were later obtained analy\-tically in \cite{Migliaccio:2017dch} and the corresponding CFT at generic central charges is in fact a non-diagonal generalization of Liouville theory \cite{Estienne:2015sua}.\footnote{The exact relation between this non-diagonal Liouville theory and the well-known diagonal Liouville theory is however unclear. See \cite{Ribault:2019qrz} for a recent study on this.} From now on, we will refer to the analytically-known amplitudes (the square of the structure constants) in this four-point function \eqref{PRS4pt} as $A^{L}$, where $L$ stands for Liouville. See appendix \ref{AL} for explicit expressions of $A^L$ that are relevant in this paper.

Regarding this intriguing relation between the Potts model and minimal models, we studied in \cite{He:2020mhb} the cluster interpretation of the minimal-model four-point functions and its irrational limit and thus provided a geometric picture of \eqref{PRS4pt}. We have seen there that the four-point function in question is, in fact, given by the cluster expansion on the lattice of the type
\begin{equation}\label{PRS4ptCFT}
\langle V^D_{\frac{1}{2},0}V^N_{\frac{1}{2},0}V^D_{\frac{1}{2},0}V^N_{\frac{1}{2},0}\rangle\propto P_{aaaa}+\tilde{P}_{abab} \,,
\end{equation}
where we have defined the pseudo-probability
\begin{eqnarray}\label{pseudoprobdef}
\tilde{P}_{abab}=
{1\over Z_{\rm Potts}}\sum_{\mathcal{D}\in \mathcal{D}_{abab}} W_{\rm Potts} (\mathcal{D}) M(k(\mathcal{D})),
\end{eqnarray}
with the sum over all diagrams of the type $\mathcal{D}_{abab}$, i.e., points $1,3$ and $2,4$ belonging to two distinct FK clusters. The multiplicity $M(k(\mathcal{D}))$ is defined as the weight of a diagram $\mathcal{D}$ with respect to the Potts weight when the two marked clusters---i.e., those labelled $a$ and $b$---are separated by $k$ (necessarily even) non-contractible loops.\footnote{Non-contractible on the four-time punctured sphere at the marked points.} This difference in weighing a certain diagram is ultimately due to the different weights assigned to the non-contractible loops in the Potts model and minimal models.\footnote{In the case of the Potts probabilities, the non-contractible loops each gets the weight $\sqrt{Q}$ as in \eqref{loop-fugacity} while in the pseudo-probability \eqref{pseudoprobdef}, one sums over the algebra of the type-D Dynkin diagram for the non-contractible loop weight. See \cite{He:2020mhb} for more details.}  In contradistinction to \eqref{pseudoprobdef}, the true Potts probability is
\begin{eqnarray}\label{Pottsprob}
P_{abab}=
{1\over Z_{\rm Potts}}\sum_{\mathcal{D}\in \mathcal{D}_{abab}} W_{\rm Potts} (\mathcal{D}).
\end{eqnarray}
Note that the two quantities \eqref{pseudoprobdef} and \eqref{Pottsprob} are expanded by the same set of diagrams $\mathcal{D}\in \mathcal{D}_{abab}$, with the difference in the weight summarized into the multiplicity $M(k(\mathcal{D}))$. The explicit expression of $M$ is given by \cite{He:2020mhb}
\begin{equation}
M(k=2l)={2\over Q^l}\sum_{m=-l}^l \left(\begin{array}{c}
2l
\\l+m
\end{array}\right){1\over \q^{2m}+\q^{-2m}},\label{Dmulti}
\end{equation}
and can be written in terms of ratios of polynomials in $Q$
\begin{subequations}
	\begin{eqnarray}
	M(k=2)&=&{2\over Q-2} \,, \\
	M(k=4)&=&{2(3Q-10)\over (Q-2)(Q^2-4Q+2)} \,, \\
	\vdots \nonumber
	\end{eqnarray}
\end{subequations}
Similar expressions hold for the geometries $\mathcal{D}_{abba},\mathcal{D}_{aabb}$ giving rise to the relations between probabilities $P_{abba},P_{aabb}$ and pseudo-probabilities $\tilde{P}_{abba},\tilde{P}_{aabb}$.

\subsection{Universal amplitude ratios}\label{aratios}
Through numerical studies on the lattice, we have further investigated in \cite{He:2020mhb} the relation between the geometry of the lattice models and the contribution of the spectrum to the geometrical correlation functions. We found facts about how the fields in \eqref{Pottsspectrum} contribute through their amplitudes to various Potts probabilities, and to their counterparts---the pseudo-probabilities---where the geometric content is modified. These facts state the existence of universal amplitude ratios of eigenvalues of the lattice transfer matrix and, amazingly, such ratios do not depend on the lattice size. It is therefore natural to assume that the same ratios hold in the continuum limit in the corresponding CFT and is translated into ratios of amplitudes of the fields.%
\footnote{It is crucial for this translation between lattice quantities and the continuum limit that the affine Temperley-Lieb modules $\mathcal{W}_{j,\mathrm{z}^2}$---the centerpiece of our algebraic understanding of the lattice model---have well-defined continuum limits, and in particular their labels $j$ and $\mathrm{z}^2$ can be cleanly interpreted in both contexts \cite{Jacobsen:2018pti}.}
Here we restate these facts directly in the CFT language and they are of the following two types:
\begin{enumerate}
\item When the same field contributes to both a Potts probability and to the corresponding pseudo-probability, the ratio of the two corresponding amplitudes depends only on the ATL module that the field belongs to, and on $Q$.
\item When the same field contributes to two different Potts probabilities, the ratio of the corresponding amplitudes depends only on the ATL module that the field  belongs to, and on $Q$.
\end{enumerate}

The facts of the first type give the ratios between the amplitudes $A$ in the true Potts probabilities ($P_{abab}$, $P_{abba}$ and $P_{aabb}$) and $\tilde{A}$ in the pseudo-probabilities ($\tilde{P}_{abab}$, $\tilde{P}_{abba}$ and $\tilde{P}_{aabb}$). Note that the probability $P_{aaaa}$ does not involve any non-contractible loops\footnote{This is because any loop surrounding the four points can be contracted at ``infinity" on the sphere. See \cite{He:2020mhb} for more details.} and therefore there is no corresponding pseudo-probability. We now define the following ratios for a certain ATL module related to the $\beta$ given in appendix \ref{ratiosappend}:
\begin{equation}\label{Rbeta}
\mathsf{R}_{\beta}(\mathcal{W}_{j,\mathrm{z}^2})\equiv\frac{\tilde{A}_{abab}}{A_{abab}}(\mathcal{W}_{j,\mathrm{z}^2})=\frac{\sum_{k=2\;\text{even}}^{j}\beta_{j,\mathrm{z}^2}^{(k)}M(k)}{\sum_{k=2\;\text{even}}^{j}\beta_{j,\mathrm{z}^2}^{(k)}}.
\end{equation}
Using the explicit expressions of $\beta^{(k)}_{j,\mathrm{z}^2}$ given in appendix \ref{ratiosappend}, we have the following:
\begin{subequations}\label{ratioforextract}
	\begin{eqnarray}
\mathsf{R}_{\beta}(\mathcal{W}_{2,-1})&=&\frac{2}{Q-2},\\
\mathsf{R}_{\beta}(\mathcal{W}_{4,-1})&=&-\frac{4}{(Q-1)(Q-2)(Q^2-4Q+2)}.\label{Rbeta4m1}
	\end{eqnarray}
\end{subequations}
Notice that for $\mathcal{W}_{4,-1}$, the denominator in the last expression of \eqref{Rbeta} actually vanishes at $Q=1$ and $Q=4$, indicating that the module decouples from $P_{abab}$ at these values of $Q$. This is partially taken care of by the factor of $Q-1$ in the denominator of \eqref{Rbeta4m1}, while at $Q=4$ the module disappears from $\tilde{P}_{abab}$ as well, since $M(k)=1$ at $Q=4$.
One can similarly define $\mathsf{R}_{\gamma}$ as 
\begin{equation}
\mathsf{R}_{\gamma}(\mathcal{W}_{j,\mathrm{z}^2})\equiv\frac{\tilde{A}_{aabb}}{A_{aabb}}(\mathcal{W}_{j,\mathrm{z}^2}) \,,
\end{equation} 
which is related to the $\gamma$ in appendix \ref{ratiosappend}. We shall however not use its explicit expression in this paper.

In addition, we also have the ratios relating amplitudes in different Potts probabilities from the second type of facts:
\begin{equation}\label{Ralpha}
\mathsf{R}_{\bar{\alpha}}(\mathcal{W}^+_{j,\mathrm{z}^2})\equiv\frac{\bar{\alpha}_{j,\mathrm{z}^2}}{2}=\frac{A_{abab}}{A_{aaaa}}(\mathcal{W}^+_{j,\mathrm{z}^2})=\frac{A_{abba}}{A_{aaaa}}(\mathcal{W}^+_{j,\mathrm{z}^2}),
\end{equation}
and
\begin{equation}\label{Ralphabar}
\mathsf{R}_{\alpha}(\mathcal{W}^+_{j,\mathrm{z}^2})\equiv\alpha_{j,\mathrm{z}^2}=\frac{A_{aabb}}{A_{aaaa}}(\mathcal{W}^+_{j,\mathrm{z}^2}),
\end{equation}
where we have used the definitions of $\alpha$ and $\bar{\alpha}$ in \eqref{betadef} and that $A_{abab}(\mathcal{W}^+_{j,\mathrm{z}^2})=A_{abba}(\mathcal{W}^+_{j,\mathrm{z}^2})$ due to \eqref{evenA} and \eqref{Wplus}.
Note that $\mathsf{R}_{\bar{\alpha}}$ and $\mathsf{R}_{\alpha}$ are not defined for $\mathcal{W}^-_{j,\mathrm{z}^2}$, i.e., $\frac{jp}{M}$ odd, since $A_{aaaa}(\mathcal{W}^-_{j,\mathrm{z}^2})=0$.
Using the expressions of $\alpha$ and $\bar{\alpha}$ in appendix \ref{ratiosappend}, we have the following expressions for $\mathsf{R}_{\alpha}$ and $\mathsf{R}_{\bar{\alpha}}$:
\begin{subequations}\label{ratioforbootstrap}
	\begin{eqnarray}
	\mathsf{R}_{\alpha}(\mathcal{W}_{0,-1})&=&-1 \,, \label{0m1alpha}\\
	\mathsf{R}_{\alpha}(\mathcal{W}_{2,1})&=&{1\over 1-Q} \,, \label{Ralpha21}\\
	\mathsf{R}_{\bar{\alpha}}(\mathcal{W}_{2,1})&=&\frac{2-Q}{2} \,,\label{Rbar21}\\
	\mathsf{R}_{\alpha}(\mathcal{W}_{4,-1})&=&{2-Q\over 2}\,,\label{Ralpha4m1}\\
	\mathsf{R}_{\bar{\alpha}}(\mathcal{W}_{4,-1})&=&{(Q-1)(Q-4)\over 4} \,,\label{Ralphabar4m1}\\
	\mathsf{R}_{\alpha}(\mathcal{W}_{4,1})&=&-\frac{Q^5-7 Q^4+15 Q^3-10 Q^2+4 Q-2}{2 (Q^2-3Q+1)} \,, \label{Ralpha41}\\ 
	\mathsf{R}_{\bar{\alpha}}(\mathcal{W}_{4,1}) &=& -\frac{ (Q^2 - 4 Q+2) (Q^2 - 3 Q -2)}{4}\,.\label{Ralphabar41}
	\end{eqnarray}
\end{subequations}

As discussed in \cite{He:2020mhb}, the ratios \eqref{ratioforextract} and \eqref{ratioforbootstrap} were obtained as a numerical lattice observation whose first-principle derivation is still unknown. These thus comprise all the ratios of the two types for ATL modules up to $j=4$.\footnote{As mentioned in appendix D of \cite{He:2020mhb}, it is numerically impossible to obtain the complete set of ratios for $j=6$ using the current lattice-computation approach, although we have provided in that reference some partial results.} In the following, we will first use the ratios \eqref{ratioforextract} to analytically extract certain Potts amplitudes from the well-known Liouville amplitudes $A^L$. We will then use this together with the ratios \eqref{ratioforbootstrap} to bootstrap the Potts probabilities.

\subsection{Probabilities and pseudo-probabilities}\label{pp}

The key observation now from the results we summarized above is that, while the geometric feature is changed from the Potts probability \eqref{Pottsprob} to the pseudo-probability \eqref{pseudoprobdef} through the multiplicity $M(k)$, only the global amplitudes $A(\mathcal{W}_{j,\mathrm{z}^2})$ associated with entire ATL modules are modified. The relations between the amplitudes of fields belonging to the same ATL module remain the same and this relation permeates into the continuum---as manifested in the existence of the universal amplitude ratios. This allows us to define the ``interchiral conformal blocks" $\mathbb{F}_{j,\mathrm{z}^2}$, which group the Virasoro conformal blocks according to the ATL modules they belong to. From \eqref{Rbeta}, \eqref{Ralpha} and \eqref{Ralphabar} we see that it is the same interchiral conformal blocks $\mathbb{F}_{j,\mathrm{z}^2}$ that enter various probabilities and pseudo-probabilities. The existence of these blocks is ultimately due to the degeneracy of the field $\Phi^D_{2,1}$ in $\overline{\mathcal{W}}_{0,\q^2}$ and can be constructed explicitly as an infinite sum  of products of left and right conformal blocks, which we shall discuss in details in section \ref{recursion}. As suggested in \cite{Gainutdinov:2012nq}, the underlying algebra can be considered as an extension of the product of left and right Virasoro algebras via fusion with $\Phi^D_{2,1}$, leading to the object dubbed the interchiral algebra in that reference. This algebra, in turn, can be obtained as the continuum limit of the affine Temperley-Lieb algebra.

\medskip

Consider now the combination
\begin{equation}\label{PRSPP}
P_{aaaa}+\tilde{P}_{abab}.
\end{equation}
The corresponding CFT correlation function \eqref{PRS4ptCFT} is well-known to be given by the non-diagonal Liouville theory of \cite{Picco:2016ilr}. Note that its spectrum $\mathcal{S}_{\mathbb{Z}+\frac{1}{2},2\mathbb{Z}}$ belongs to the ATL modules $\mathcal{W}_{j,-1}$. We can then expand it in terms of the interchiral conformal blocks as
\begin{equation}\label{PRSPPcft}
P_{aaaa}+\tilde{P}_{abab}=\sum_{j=0\;\text{even}}^\infty A^L(\mathcal{W}_{j,-1}) \mathbb{F}_{j,-1}.
\end{equation}
Meanwhile, from the lattice study in \cite{He:2020mhb}, we have seen that the modules $\mathcal{W}_{0,-1}$ and $\mathcal{W}_{j,\mathrm{z}^2}$ from the Potts spectrum \eqref{Pottsspectrum} all appear, where the amplitudes for $\mathcal{W}_{j,\mathrm{z}^2}$ are modified from their corresponding values in the Potts model through the ratios $\mathsf{R}_{\beta}(\mathcal{W}_{j,\mathrm{z}^2})$ defined in \eqref{Rbeta}. The combination \eqref{PRSPP} can thus be written as 
\begin{equation}\label{PRSPPlattice}
\begin{aligned}
P_{aaaa}+\tilde{P}_{abab}=&A_{aaaa}(\mathcal{W}_{0,-1})\mathbb{F}_{0,-1}
+\sum_{\frac{j}{2}\;\text{odd}}\tilde{A}_{abab}(\mathcal{W}^-_{j,-1})\mathbb{F}_{j,-1}\\
&+\sum_{\frac{j}{2}\;\text{even}}\big(A_{aaaa}(\mathcal{W}^+_{j,-1})+\tilde{A}_{abab}(\mathcal{W}^+_{j,-1})\big)\mathbb{F}_{j,-1}\\
&+\sum_{{jp\over M}\;\text{even}}\big(A_{aaaa}(\mathcal{W}^+_{j,\mathrm{z}^2})+\tilde{A}_{abab}(\mathcal{W}^+_{j,\mathrm{z}^2})\big)\mathbb{F}_{j,\mathrm{z}^2}+\sum_{{jp\over M}\;\text{odd}}\tilde{A}_{abab}(\mathcal{W}^-_{j,\mathrm{z}^2})\mathbb{F}_{j,\mathrm{z}^2}\\
=&A_{aaaa}(\mathcal{W}_{0,-1})\mathbb{F}_{0,-1}
+\sum_{\frac{j}{2}\;\text{odd}}\mathsf{R}_{\beta}  A_{abab}(\mathcal{W}^-_{j,-1})\mathbb{F}_{j,-1}\\
&+\sum_{\frac{j}{2}\;\text{even}}A_{aaaa}(\mathcal{W}^+_{j,-1})\big(1+\mathsf{R}_{\beta}\mathsf{R}_{\bar{\alpha}}\big)\mathbb{F}_{j,-1}\\
&+\sum_{{jp\over M}\;\text{even}}A_{aaaa}(\mathcal{W}^+_{j,\mathrm{z}^2})\big(1+\mathsf{R}_{\beta}\mathsf{R}_{\bar{\alpha}}\big)\mathbb{F}_{j,\mathrm{z}^2}+\sum_{{jp\over M}\;\text{odd}}\mathsf{R}_{\beta} A_{abab}(\mathcal{W}^-_{j,\mathrm{z}^2})\mathbb{F}_{j,\mathrm{z}^2}.
\end{aligned}
\end{equation}
In the case the the true Potts probabilities, one has instead the combination:
\begin{equation}\label{Pottsabab}
\begin{aligned}
P_{aaaa}+P_{abab}=&A_{aaaa}(\mathcal{W}_{0,-1})\mathbb{F}_{0,-1}
+\sum_{\frac{j}{2}\;\text{odd}}A_{abab}(\mathcal{W}^-_{j,-1})\mathbb{F}_{j,-1}\\
&+\sum_{\frac{j}{2}\;\text{even}}\big(A_{aaaa}(\mathcal{W}^+_{j,-1})+A_{abab}(\mathcal{W}^+_{j,-1})\big)\mathbb{F}_{j,-1}\\
&+\sum_{{jp\over M}\;\text{even}}\big(A_{aaaa}(\mathcal{W}^+_{j,\mathrm{z}^2})+A_{abab}(\mathcal{W}^+_{j,\mathrm{z}^2})\big)\mathbb{F}_{j,\mathrm{z}^2}+\sum_{{jp\over M}\;\text{odd}}A_{abab}(\mathcal{W}^-_{j,\mathrm{z}^2})\mathbb{F}_{j,\mathrm{z}^2}\\
=&A_{aaaa}(\mathcal{W}_{0,-1})\mathbb{F}_{0,-1}
+\sum_{\frac{j}{2}\;\text{odd}} A_{abab}(\mathcal{W}^-_{j,-1})\mathbb{F}_{j,-1}\\
&+\sum_{\frac{j}{2}\;\text{even}}A_{aaaa}(\mathcal{W}^+_{j,-1})\big(1+\mathsf{R}_{\bar{\alpha}}\big)\mathbb{F}_{j,-1}\\
&+\sum_{{jp\over M}\;\text{even}}A_{aaaa}(\mathcal{W}^+_{j,\mathrm{z}^2})\big(1+\mathsf{R}_{\bar{\alpha}}\big)\mathbb{F}_{j,\mathrm{z}^2}+\sum_{{jp\over M}\;\text{odd}} A_{abab}(\mathcal{W}^-_{j,\mathrm{z}^2})\mathbb{F}_{j,\mathrm{z}^2}.
\end{aligned}
\end{equation}
Note that $\mathbb{F}$ in \eqref{PRSPPlattice} and \eqref{Pottsabab} represent $s$- or $t$-channel blocks since in both cases we have combinations that are crossing-symmetric under $s\leftrightarrow t$.

\smallskip

One remark on the bootstrap problem follows immediately. By comparing eqs.~\eqref{PRSPPlattice}--\eqref{Pottsabab}, it is obvious that the crossing-symmetric spectrum proposed in \cite{Picco:2016ilr}, which gives a complete description of eq.~\eqref{PRSPPlattice}, is also a solution to the conformal block expansion of the true Potts probabilities \eqref{Pottsabab} using the full spectrum \eqref{Pottsspectrum}, with of course the amplitudes for the whole ATL modules given by $\tilde{A}_{abab}$ instead of $A_{abab}$ as in \eqref{PRSPPlattice}. This means that within the full spectrum \eqref{Pottsspectrum} of the Potts model, the states which do not appear in the spectrum of \cite{Picco:2016ilr} have $A_{aaaa}+\tilde{A}_{abab}=0$, i.e.,
\begin{subequations} \label{decoupling-ratios}
\begin{eqnarray}
1+\mathsf{R}_{\beta}(\mathcal{W}^+_{j,\mathrm{z}^2})\mathsf{R}_{\bar{\alpha}}(\mathcal{W}^+_{j,\mathrm{z}^2})&=&0,\\
\mathsf{R}_{\beta}(\mathcal{W}^-_{j,\mathrm{z}^2})&=&0,
\end{eqnarray}
\end{subequations}
for $\frac{p}{M}\neq\frac{1}{2}$, as we have checked explicitly in \cite{He:2020mhb} for all the ATL modules up to $j=4$. This suggests that the solution to the original bootstrap problem considered in \cite{Picco:2016ilr}, i.e., the bootstrap of the probabilities $P_{aaaa}+P_{abab}$, is not unique: The spectrum and amplitudes in \cite{Picco:2016ilr} is one solution, while the true Potts spectrum with its amplitudes provides another one, and possibly there exists (infinitely many?) further solutions. Geometrically this can be understood as the freedom of assigning weights to cluster/loop configurations, a mechanism which we have seen explicitly at play above, where it involves two different ways of assigning weights to the non-contractible loops. This complexity of the solution to the bootstrap problem is perhaps rooted in the irrationality of the theory in general, and the simple spectrum given by \cite{Picco:2016ilr}---being itself a generalization of minimal models---stands out as special, or ``minimal" in a(nother) sense, as it sees a large number of fields decouple from the spectrum, cf.~\eqref{decoupling-ratios}.

\smallskip

Now focus on eqs.~\eqref{PRSPPlattice}--\eqref{Pottsabab}. It is fascinating to see that we can, in fact, extract the true Potts amplitudes from the known Liouville amplitudes $A^L$ for the modules that contribute to both combinations:
\begin{subequations}
\begin{eqnarray}
A_{aaaa}(\mathcal{W}_{0,-1})&=&A^L(\mathcal{W}_{0,-1}),\\
\mathsf{R}_{\beta}A_{abab}(\mathcal{W}^-_{j,-1})&=&A^L(\mathcal{W}^-_{j,-1}),\;\;\frac{j}{2}\;\text{odd},\\
A_{aaaa}(\mathcal{W}^+_{j,-1})(1+\mathsf{R}_{\beta}\mathsf{R}_{\bar{\alpha}})&=&A^L(\mathcal{W}^+_{j,-1}),\;\;\frac{j}{2}\;\text{even}.
\end{eqnarray}
\end{subequations}
Using eqs.~\eqref{ratioforextract} and \eqref{ratioforbootstrap}, we have the following expressions of the Potts amplitudes in terms of the Liouville amplitudes:
\begin{subequations}\label{PAusingL}
\begin{eqnarray}
A_{aaaa}(\mathcal{W}_{0,-1})&=&A^L(\mathcal{W}_{0,-1}),\label{Aaaaa0m1ana}\\
A_{abab}(\mathcal{W}_{2,-1})&=&{Q-2\over 2}A^L(\mathcal{W}_{2,-1}),\label{abab2m1ana}\\
A_{aaaa}(\mathcal{W}_{4,-1})&=&{(Q-2)(Q^2-4Q+2)\over Q(Q-3)^2}A^L(\mathcal{W}_{4,-1}).\label{A4m1ana}
\end{eqnarray}
\end{subequations}
Writing this out explicitly, the Potts probabilities are given by the $A^L$ and the interchiral blocks as
\begin{subequations}
\begin{eqnarray}
P_{aaaa}&=&A^L(\mathcal{W}_{0,-1}) \mathbb{F}_{0,-1}+{(Q-2)(Q^2-4Q+2)\over Q(Q-3)^2}A^L(\mathcal{W}_{4,-1})\mathbb{F}_{4,-1}+\ldots,\label{Paaaaanalytic}\\
P_{abab}&=&{Q-2\over 2}A^L(\mathcal{W}_{2,-1}) \mathbb{F}_{2,-1}+
{(Q-1)(Q-4)(Q-2)(Q^2-4Q+2)\over 4Q(Q-3)^2}A^L(\mathcal{W}_{4,-1})\mathbb{F}_{4,-1}+\ldots,\label{Pababanalytic}\;\;\;\;\;\;\;\;\;\;\;
\end{eqnarray}
\end{subequations}
where we have used \eqref{Ralphabar4m1} in writing the second term of \eqref{Pababanalytic}, and the left-out terms ($\ldots$) are to be determined by the bootstrap computations.

\bigskip

Let us now turn to another combination of probabilities
\begin{equation}
P_{aaaa}+\tilde{P}_{aabb}.
\end{equation}
In the $s$-channel, as studied in \cite{He:2020mhb}, we have
\begin{equation}\label{aabbtilde}
\begin{aligned}
P_{aaaa}+\tilde{P}_{aabb}=&\big(A_{aaaa}(\mathcal{W}_{0,-1})+\tilde{A}
_{aabb}(\mathcal{W}_{0,-1})\big)\mathbb{F}^{(s)}_{0,-1}\\
&+\sum_{{jp\over M}\;\text{even}}\big(A_{aaaa}(\mathcal{W}^+_{j,\mathrm{z}^2})+\tilde{A}_{aabb}(\mathcal{W}^+_{j,\mathrm{z}^2})\big)\mathbb{F}^{(s)}_{j,\mathrm{z}^2} +\sum_{\mathsf{a}}\tilde{A}_{aabb}(\mathcal{W}_{0,\q^{2\mathsf{a}}})\mathbb{F}^{(s)}_{0,\q^{2\mathsf{a}}}\\
=&\big(A_{aaaa}(\mathcal{W}_{0,-1})+\mathsf{R}_{\gamma}A
_{aabb}(\mathcal{W}_{0,-1})\big)\mathbb{F}^{(s)}_{0,-1}\\
&+\sum_{{jp\over M}\;\text{even}}\big(A_{aaaa}(\mathcal{W}^+_{j,\mathrm{z}^2})+\mathsf{R}_{\gamma}A
_{aabb}(\mathcal{W}^+_{j,\mathrm{z}^2})\big)\mathbb{F}^{(s)}_{j,\mathrm{z}^2}+\sum_{\mathsf{a}}\tilde{A}_{aabb}(\mathcal{W}_{0,\q^{2\mathsf{a}}})\mathbb{F}^{(s)}_{0,\q^{2\mathsf{a}}} \,,
\end{aligned}
\end{equation}
where the last term involves diagonal Verma modules with the conformal dimensions
\begin{equation}\label{aabbspec}
(h_{e+\frac{\mathsf{a}}{x+1},0},h_{e+\frac{\mathsf{a}}{x+1},0}),\;\;e\in\mathbb{Z}.
\end{equation}
It was argued in \cite{He:2020mhb}, by comparison with the CFT results \cite{Ribault:2018jdv}, that the $s$-channel here involves purely diagonal fields so the first two terms disappear from \eqref{aabbtilde} with the amplitude ratios $\mathsf{R}_{\gamma},\mathsf{R}_{\alpha}$ as we have checked explicitly in \cite{He:2020mhb}. In the limit of irrational $\upbeta^2$, the sum in the last term becomes an integral over a compact set,
\begin{equation}
\int_{0}^{\pi} {\rm d}\theta \,,
\end{equation}
where we have introduced the variable $\theta=\frac{\mathsf{a}\pi}{x+1}$, and \eqref{aabbspec} becomes
\begin{equation}
(h_{r,0},h_{r,0}),\;\;r\in\mathbb{R},
\end{equation}
i.e., a continuous diagonal spectrum.
Geometrically speaking this indicates that the non-contractable loop weights are integrated over the fugacities 
\begin{equation}
n_\mathrm{z}=\mathrm{z}+\mathrm{z}^{-1}, \quad \mbox{with } \mathrm{z}=e^{i\theta}.
\end{equation}

In contrast with this, we have in the Potts model:
\begin{equation}\label{aabb}
\begin{aligned}
P_{aaaa}+P_{aabb}=&\big(A_{aaaa}(\mathcal{W}_{0,-1})+A
_{aabb}(\mathcal{W}_{0,-1})\big)\mathbb{F}^{(s)}_{0,-1}\\
&+\sum_{{jp\over M}\;\text{even}}\big(A_{aaaa}(\mathcal{W}_{j,\mathrm{z}^2})+A
_{aabb}(\mathcal{W}_{j,\mathrm{z}^2})\big)\mathbb{F}^{(s)}_{j,\mathrm{z}^2}+A_{aabb}(\overline{\mathcal{W}}_{0,\q^{2}})\mathbb{F}^{(s)}_{0,\q^{2}}\\
=&A_{aaaa}(\mathcal{W}_{0,-1})\big(1+\mathsf{R}_{\alpha}(\mathcal{W}_{0,-1})\big)\mathbb{F}^{(s)}_{0,-1}\\
&+\sum_{{jp\over M}\;\text{even}}A_{aaaa}(\mathcal{W}_{j,\mathrm{z}^2})\big(1+\mathsf{R}_{\alpha}(\mathcal{W}_{j,\mathrm{z}^2})\big)\mathbb{F}^{(s)}_{j,\mathrm{z}^2}+A_{aabb}(\overline{\mathcal{W}}_{0,\q^{2}})\mathbb{F}^{(s)}_{0,\q^{2}} \,.
\end{aligned}
\end{equation}
First notice that for the first two terms, we do not expect in general a cancellation and thus, the $s$-channel spectrum here in the true Potts probabilities involves non-diagonal modules $\mathcal{W}_{j,\mathrm{z}^2}$. Furthermore, the last term, as argued in \cite{Jacobsen:2018pti} already, comes from the requirement that in the Potts model, all loops---contractible or non-contractible---carry the weight $\sqrt{Q}$. Therefore one fixes \cite{Jacobsen:2018pti}:
\begin{equation}
\mathrm{z}^2=\q^{\pm2},
\end{equation}
and obtains the module $\overline{\mathcal{W}}_{0,\q^2}$. As mentioned before, this includes the Kac modules of diagonal primaries with conformal dimensions
\begin{equation}
(h_{r,1},h_{r,1}),\;\;r\in\mathbb{N}^*,
\end{equation} 
a discrete spectrum.
By comparing \eqref{aabbtilde} with \eqref{aabb}, we see that the diagonal spectrum in the $s$-channel of the geometry $\mathcal{D}_{aabb}$, where the two FK clusters are separated by a large number of non-contractible loops, encodes important geometric information on the non-contractible loop weight in the lattice model.

\subsection{The interchiral bootstrap equations}\label{supereq}

We will consider the bootstrap problem of the following probabilities as related by crossing:
\begin{subequations}\label{Ps}
\begin{eqnarray}
P^{(s)}_{aaaa}&=&P^{(t)}_{aaaa},\\
P^{(s)}_{abab}&=&P^{(t)}_{abab},\\
P^{(s)}_{aabb}&=&P^{(t)}_{abba},\\
P^{(s)}_{abba}&=&P^{(t)}_{aabb}.
\end{eqnarray}
\end{subequations}
Eqs.~\eqref{Ps} are simply a shorthand rewriting of eqs.~\eqref{booteq1}, \eqref{booteq2} and \eqref{booteq3} which should be interpreted as follows: the subscripts indicate the spectrum for the interchiral block expansions and thus the corresponding amplitudes, while the superscripts indicate which channel of the blocks to use. The basic idea is now to write the Potts probabilities in terms of the interchiral block expansions with the amplitudes $A(\mathcal{W})$ associated with the whole ATL modules. Eqs.~\eqref{Ps} are then a coupled linear system for these amplitudes and will be used for the ``interchiral conformal bootstrap".

The amplitudes involved here are:
\begin{equation}\label{ATLamp}
\begin{aligned}
&A_{aaaa}(\mathcal{W}_{0,-1}),\;\;
A_{aaaa}(\mathcal{W}^{+}_{j,\mathrm{z}^2}),\;\;\\
&A_{abab}(\mathcal{W}^{+}_{j,\mathrm{z}^2}),\;\;
A_{abab}(\mathcal{W}^{-}_{j,\mathrm{z}^2}),\;\;\\
&A_{abba}(\mathcal{W}^{+}_{j,\mathrm{z}^2}),\;\;
A_{abba}(\mathcal{W}^{-}_{j,\mathrm{z}^2}),\;\;\\
&A_{aabb}(\mathcal{W}_{0,-1}),\;\;
A_{aabb}(\mathcal{W}^{+}_{j,\mathrm{z}^2}),\;\;A_{aabb}(\overline{\mathcal{W}}_{0,\q^{2}}),
\end{aligned}
\end{equation}
and we can then write eq. \eqref{Ps} as the {\bf interchiral bootstrap equations}:
\begin{subequations}\label{superP}
			\begin{eqnarray}
			A_{aaaa}\mathbb{F}^{(s)}_{0,-1}+\sum_{\{\mathcal{W}^+_{j,\mathrm{z}^2}\}} \!\!\! A_{aaaa}\mathbb{F}^{(s)}_{j,\mathrm{z}^2}&=&A_{aaaa}\mathbb{F}^{(t)}_{0,-1}+\sum_{\{\mathcal{W}^+_{j,\mathrm{z}^2}\}} \!\!\! A_{aaaa}\mathbb{F}^{(t)}_{j,\mathrm{z}^2} \,, \label{superaaaa}\\
			\sum_{\{\mathcal{W}^+_{j,\mathrm{z}^2}\}} \!\!\! A_{abab}\mathbb{F}^{(s)}_{j,\mathrm{z}^2}+\sum_{\{\mathcal{W}^-_{j,\mathrm{z}^2}\}} \!\!\! A_{abab}\mathbb{F}^{(s)}_{j,\mathrm{z}^2} 
			&=&\sum_{\{\mathcal{W}^+_{j,\mathrm{z}^2}\}} \!\!\! A_{abab}\mathbb{F}^{(t)}_{j,\mathrm{z}^2}+\sum_{\{\mathcal{W}^-_{j,\mathrm{z}^2}\}} \!\!\! A_{abab}\mathbb{F}^{(t)}_{j,\mathrm{z}^2}\label{superabab} \,, \\
			A_{aabb}\mathbb{F}^{(s)}_{0,-1}+ A_{aabb}\mathbb{F}^{(s)}_{0,\q^2}+\sum_{\{\mathcal{W}^+_{j,\mathrm{z}^2}\}} \!\!\! A_{aabb}\mathbb{F}^{(s)}_{j,\mathrm{z}^2}&=&\sum_{\{\mathcal{W}^+_{j,\mathrm{z}^2}\}} \!\!\! A_{abba}\mathbb{F}^{(t)}_{j,\mathrm{z}^2}+\sum_{\{\mathcal{W}^-_{j,\mathrm{z}^2}\}} \!\!\! A_{abba}\mathbb{F}^{(t)}_{j,\mathrm{z}^2} \,, \\
			\sum_{\{\mathcal{W}^+_{j,\mathrm{z}^2}\}} \!\!\! A_{abba}\mathbb{F}^{(s)}_{j,\mathrm{z}^2}+\sum_{\{\mathcal{W}^-_{j,\mathrm{z}^2}\}} \!\!\! A_{abba}\mathbb{F}^{(s)}_{j,\mathrm{z}^2}
			&=&A_{aabb}\mathbb{F}^{(t)}_{0,-1}+ A_{aabb}\mathbb{F}^{(t)}_{0,\q^2}+\sum_{\{\mathcal{W}^+_{j,\mathrm{z}^2}\}} \!\!\! A_{aabb}\mathbb{F}^{(t)}_{j,\mathrm{z}^2} \,,\;\;\;\;\;\;\;\;\;
			\end{eqnarray}
\end{subequations}
where we have omitted the arguments for the amplitudes for notation simplicity. Notice that we can further impose the constraints $\mathsf{R}_{\alpha}$ and $\mathsf{R}_{\bar{\alpha}}$ from \eqref{ratioforbootstrap}. Recall also that $\mathcal{W}^+_{j,\mathrm{z}^2}$ have the same amplitudes $A_{abab}$ and $A_{abba}$, while $\mathcal{W}^-_{j,\mathrm{z}^2}$ have opposite amplitudes due to \eqref{evenodd}. This gives us:
\begin{subequations}\label{constraints}
\begin{eqnarray}
A_{aaaa}\mathsf{R}_{\alpha}&=&A_{aabb},\;\;\text{for}\;\;\mathcal{W}_{0,-1},\mathcal{W}^+_{j,\mathrm{z}^2}\\
A_{aaaa}\mathsf{R}_{\bar{\alpha}}&=&A_{abab},\;\;\text{for}\;\;\mathcal{W}^+_{j,\mathrm{z}^2}\label{Rabab}\\
A_{abab}&=&A_{abba},\;\;\text{for}\;\;\mathcal{W}^+_{j,\mathrm{z}^2}\label{Aeven}\\
A_{abab}&=&-A_{abba},\;\;\text{for}\;\;\mathcal{W}^-_{j,\mathrm{z}^2}.\label{Aodd}
\end{eqnarray}
\end{subequations}
In addition, we have obtained analytically \eqref{PAusingL} which, with proper normalizations, can be imposed as extra constraints into the bootstrap.

\section{The interchiral conformal bootstrap}\label{ATLsuper}

With the setup given by eqs.~\eqref{superP} and \eqref{constraints}, we are now ready to bootstrap the amplitudes \eqref{ATLamp}. In this section, we start by constructing the interchiral conformal blocks $\mathbb{F}_{j,\mathrm{z}^2}$ explicitly and show how they arise from the degeneracy of the field $\Phi^D_{2,1}$. We then present and study the bootstrap results on the amplitudes. The numerical details of the bootstrap will be discussed in appendix \ref{bootstrapapp}.
 
\subsection{Recursions and the interchiral conformal blocks}\label{recursion}
In the conformal bootstrap approach to the diagonal Liouville theory \cite{Zamolodchikov:1995aa,Teschner:1995yf}, the degeneracy of the diagonal fields $\Phi_{r,s}=\Phi^D_{1,2}\text{ and }\Phi^D_{2,1}$ are used to obtain the recursions when the structure constants, and thus the amplitudes, are related through shifting the Kac indices by 2 units: $s\pm 1$ or $r\pm 1$, which eventually leads to a full solution of the theory. The key idea is to consider the four-point functions involving these degenerate fields which, in the conformal block expansions, truncate to only two terms. One can then write the relations of the structure constants using the fusing matrix---the linear transformation between the conformal blocks in the two channels as constructed from the solutions of BPZ equations. This technique was further generalized to the non-diagonal case in \cite{Estienne:2015sua,Migliaccio:2017dch} which gives the analytic amplitudes $A^L$ of the non-diagonal Liouville theory in \cite{Picco:2016ilr}.

In the case of the Potts model, the degeneracy of $\Phi^{D}_{1,2}$ is absent and therefore this technique does not apply directly. However, we see in the spectrum \eqref{Pottsspectrum} that the field $\Phi^{D}_{2,1}\in\overline{\mathcal{W}}_{0,\q^2}$ is degenerate and one expects the recursions in the (diagonal and non-diagonal) Liouville theory for shifting the first Kac index, $r\pm 1$, to hold in this case. In fact, for the Potts geometrical correlations $P_{aaaa},P_{abab},P_{aabb},P_{abba}$ which can be considered (up to the remarks in footnote~\ref{fn2}) as four-point functions of the spin operator \eqref{4spin}:
\begin{equation}\label{4ptspin}
\langle\Phi_{\frac{1}{2},0}\Phi_{\frac{1}{2},0}\Phi_{\frac{1}{2},0}\Phi_{\frac{1}{2},0}\rangle,
\end{equation}
the degeneracy of $\Phi^{D}_{2,1}$ indicates a recursion of the amplitudes with $r$ shifted by $1$ which we will give explicitly below, deferring the derivation to section \ref{degeneracyrecursion}. Such recursion exactly relates the amplitudes of the primaries belonging to the same ATL module and organizes the corresponding Virasoro conformal blocks into the interchiral conformal blocks.

\medskip

Recall from the spectrum \eqref{Pottsspectrum} that the non-diagonal primary fields in the modules $\mathcal{W}_{j,e^{2i\pi p/M}}$ have the conformal dimensions
\begin{equation} (h_{\frac{p}{M}+e,j},h_{\frac{p}{M}+e,-j}),\;\;e\in\mathbb{Z}.
\end{equation}
The module $\mathcal{W}_{j,e^{2i\pi p/M}}$ therefore contains fields related by $e\to e+1$, with the leading amplitude $A(h_{\frac{p}{M},j},h_{\frac{p}{M},-j})$. We will therefore take the overall amplitude associated with the interchiral blocks to be:
\begin{equation}\label{superA}
A(\mathcal{W}_{j,e^{2i\pi p/M}})=
\begin{cases}
A(h_{0,j},h_{0,-j}), & \mbox{for } p=0,\\
2A(h_{\frac{p}{M},j},h_{\frac{p}{M},-j}), & \text{otherwise},
\end{cases}
\end{equation}
where the factor of $2$ in the second line accounts for the identification of the amplitudes:
\begin{equation}\label{AAA}
A(h_{r,s},h_{r,-s})=A(h_{r,-s},h_{r,s})=A(h_{-r,s},h_{r,s}),
\end{equation}
since the two non-diagonal fields $(r,s)$ and $(r,-s)$ have the same total conformal dimension \eqref{totalh} and spin \eqref{spin} (up to a sign).
The amplitudes of the other fields within the module are related by the recursion
\begin{equation}\label{Rshift1N}
\begin{aligned}
R^N_{e,j}=&\frac{A(h_{e+1,j},h_{e+1,-j})}{A(h_{e,j},h_{e,-j})}\\
=&\frac{\Gamma(-j-\frac{e}{\upbeta^2})		 
	\Gamma(1+j-\frac{1+e}{\upbeta^2})
	\Gamma(\frac{1-j}{2}+\frac{e}{2\upbeta^2})
\Gamma(\frac{1+j}{2}+\frac{e}{2\upbeta^2})
	\Gamma(\frac{1-j}{2}+\frac{1+e}{2\upbeta^2})
	\Gamma(\frac{1+j}{2}+\frac{1+e}{2\upbeta^2})
	}
{\Gamma(j+\frac{1+e}{\upbeta^2})\Gamma(1-j+\frac{e}{\upbeta^2})
\Gamma(\frac{1+j}{2}-\frac{1+e}{2\upbeta^2})
\Gamma(\frac{1-j}{2}-\frac{1+e}{2\upbeta^2})
	\Gamma(\frac{1+j}{2}-\frac{e}{2\upbeta^2})
\Gamma(\frac{1-j}{2}-\frac{e}{2\upbeta^2})
	} \,,
\end{aligned}
\end{equation}
which has the properties
\begin{subequations}
	\begin{eqnarray}
	\frac{1}{R^N_{-e-1,-j}}=R^N_{e,j},\label{property1}\\
	\frac{1}{R^N_{-e-1,j}}=R^N_{e,j},\label{property2}
	\end{eqnarray}
\end{subequations}
where \eqref{property1} is explicit while \eqref{property2} is expected due to \eqref{AAA} and can be easily checked using that $j\in\mathbb{Z}$. Notice that in the special case of $e=0$, the expression \eqref{Rshift1N} includes the divergent factor $\frac{\Gamma(-j)}{\Gamma(1-j)}$. In this case one can use the property \eqref{property2} to obtain instead:
\begin{equation}\label{RshiftNdeg}
R^N_{0,j}=\frac{\Gamma(j)		 
	\Gamma(1-j-\frac{1}{\upbeta^2})\Gamma(\frac{1-j}{2}+\frac{1}{2\upbeta^2})\Gamma(\frac{1+j}{2}+\frac{1}{2\upbeta^2})
}
{\Gamma(1+j)\Gamma(-j+\frac{1}{\upbeta^2})
\Gamma(\frac{1-j}{2}-\frac{1}{2\upbeta^2})\Gamma(\frac{1+j}{2}-\frac{1}{2\upbeta^2})}
\end{equation}
which is divergence-free.

Note that due to the definition \eqref{superA} and the identification \eqref{AAA}, we have also
\begin{equation}\label{pm1pm}
A(\mathcal{W}_{j,e^{2i\pi p/M}})=2A(h_{\frac{p}{M},j},h_{\frac{p}{M},-j})=2A(h_{-\frac{p}{M},j},h_{-\frac{p}{M},-j})=\frac{1}{R^N_{-\frac{p}{M},j}}A(\mathcal{W}_{j,e^{2i\pi(1-p/M)}}),
\end{equation}
i.e., the recursion \eqref{Rshift1N} also relates the global amplitudes for the modules
\begin{equation}\label{pm1}
\mathcal{W}_{j,e^{2i\pi p/M}}\leftrightarrow\mathcal{W}_{j,e^{2i\pi (1-p/M)}},
\end{equation}
which corresponds to\footnote{Recall that $\mathrm{z}$ is related with the phase acquired by the non-contractible lines as they wind around the axis of the cylinder\cite{Gainutdinov:2014foa}. Switching  $\mathrm{z}$ and $\mathrm{z}^{-1}$ amounts to switching clockwise and counterclockwise.}
\begin{equation}
\mathrm{z}\leftrightarrow \mathrm{z}^{-1}.
\end{equation}
 This reduces the independent non-diagonal amplitudes to $A(\mathcal{W}_{j,e^{2i\pi p/M}})$ with
\begin{equation}
0\le\frac{p}{M}\le\frac{1}{2}\;.
\end{equation}

The module $\overline{\mathcal{W}}_{0,\q^2}$ involves diagonal primaries with conformal dimensions
\begin{equation}
(h_{1+e,1},h_{1+e,1}),\;\;e\in\mathbb{N}.
\end{equation}
Again the corresponding Kac modules are related by $e\to e+1$ with the leading $(h_{1,1},h_{1,1})$ and therefore we take
\begin{equation}\label{superAdeg}
A(\overline{\mathcal{W}}_{0,\q^2})=A(h_{1,1},h_{1,1}).
\end{equation}
The amplitudes of diagonal fields have the recursion:\footnote{In the spectrum \eqref{Pottsspectrum} we have only $s=1$, but here we give the most general recursion as the result of the degenerate $\Phi^D_{2,1}$. Note that compared to \eqref{Rshift1N} we have changed the notation $j\to s$ since the module $\overline{\mathcal{W}}_{0,\q^2}$ corresponds to $j=0$ but the conformal weights can be written with $s=1$ using \eqref{cw}. See eq.~(5.21) in \cite{Jacobsen:2018pti}.}
\begin{equation}\label{Rshift1D}
\begin{aligned}
R^{D}_{e,s}=&\frac{A(h_{e+1,s},h_{e+1,s})}{A(h_{e,s},h_{e,s})}\\
=&\frac{\Gamma(s-\frac{e}{\upbeta^2})\Gamma(1+s-\frac{1+e}{\upbeta^2})\Gamma(\frac{1-s}{2}+\frac{1+e}{2\upbeta^2})^2\Gamma(\frac{1-s}{2}+\frac{e}{2\upbeta^2})^2}{\Gamma(-s+\frac{1+e}{\upbeta^2})\Gamma(1-s+\frac{e}{\upbeta^2})\Gamma(\frac{1+s}{2}-\frac{e}{2\upbeta^2})^2\Gamma(\frac{1+s}{2}-\frac{1+e}{2\upbeta^2})^2},
\end{aligned}
\end{equation}
which has the explicit property
\begin{equation}
\frac{1}{R^D_{-e-1,-s}}=R^D_{e,s}
\end{equation}
as expected since $h_{r,s}=h_{-r,-s}$.

\bigskip

The interchiral conformal blocks involved in the bootstrap equations \eqref{superP} are the following:
\begin{equation}\label{3superblocks}
\mathbb{F}^{(\mathsf{c})}_{j,-1}, \quad \mathbb{F}^{(\mathsf{c})}_{j,\mathrm{z}^2=e^{2i\pi p/M}}, \quad \mathbb{F}^{(\mathsf{c})}_{0,\q^2},
\end{equation}
where here and below the channels are denoted with $\mathsf{c}=s,t$.
Writing explicitly, we define:
\begin{equation}\label{superjm1}
\begin{aligned}
\mathbb{F}^{(\mathsf{c})}_{j,-1}=&\frac{1}{2}\sum_{e \in \mathbb{N}}\mathcal{R}_{e+\frac{1}{2},j}\bigg(\mathcal{F}^{(\mathsf{c})}_{h_{e+\frac{1}{2},j}}(z)\mathcal{F}^{(\mathsf{c})}_{h_{e+\frac{1}{2},-j}}(\bar{z})+\mathcal{F}^{(\mathsf{c})}_{h_{e+\frac{1}{2},-j}}(z)\mathcal{F}^{(\mathsf{c})}_{h_{e+\frac{1}{2},j}}(\bar{z})\bigg),\\
=&\sum_{e \in \mathbb{N}}\mathcal{R}_{e+\frac{1}{2},j}\text{Re}\left[\mathcal{F}^{(\mathsf{c})}_{h_{e+\frac{1}{2},j}}(z)\mathcal{F}^{(\mathsf{c})}_{h_{e+\frac{1}{2},-j}}(\bar{z})\right]
\end{aligned}
\end{equation}
where
\begin{equation}\label{Rjm1}
\mathcal{R}_{e+\frac{1}{2},j}\equiv\frac{A(h_{e+\frac{1}{2},j},h_{e+\frac{1}{2},-j})}{A(h_{\frac{1}{2},j},h_{\frac{1}{2},-j})}=\begin{cases}
1,&\mbox{for } e=0,\\
\prod^{e-1}_{i=0}R^N_{i+\frac{1}{2},j},& \mbox{for } e\neq 0,
\end{cases}
\end{equation}
and we have used \eqref{AAA} to write the block of $(h_{-e-\frac{1}{2},j},h_{e+\frac{1}{2},j})$ as
\begin{equation}
\mathcal{F}^{(\mathsf{c})}_{h_{e+\frac{1}{2},-j}}(z)\mathcal{F}^{(\mathsf{c})}_{h_{e+\frac{1}{2},j}}(\bar{z})
\end{equation}
to restrict the summation to $e\ge 0$.

\begin{figure}[t!]
   \centering
	\includegraphics[width=0.8\textwidth]{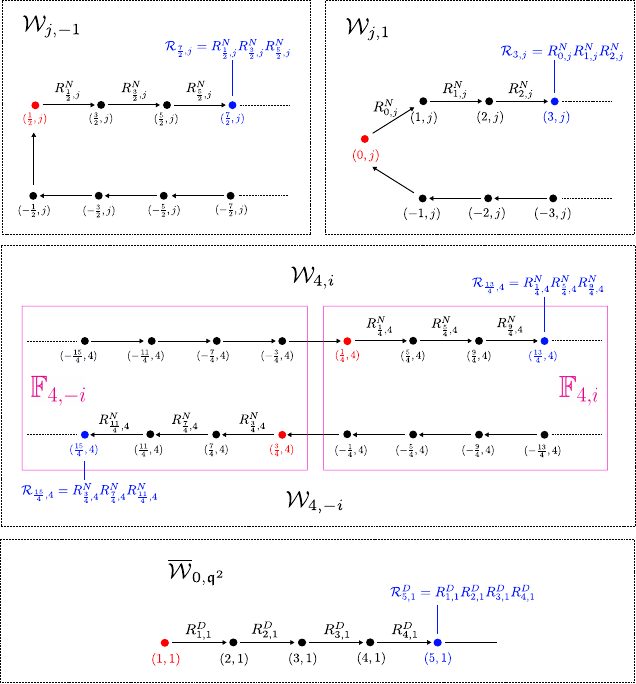}
	\caption{Examples of the construction of interchiral conformal blocks $\mathbb{F}_{j,\mathrm{z}^2}$ and $\mathbb{F}_{0,\mathfrak{q}^2}$. The dots indicate the fields $(r,s)=(e+\frac{p}{M},j)^N,(e+1,1)^D$ whose amplitudes satisfy the recursion $R^{D,N}$ in eqs.~\eqref{Rshift1D} and \eqref{Rshift1N} under $e\to e+1$ as illustrated by the arrows. The leading primary in each ATL module whose amplitude is taken as $A(\mathcal{W})$ in \eqref{superA} and \eqref{superAdeg} is labeled {\color{red}red}. We give explicit examples of the coefficients $\mathcal{R}$ for sub-leading fields entering the blocks indicated with {\color{blue}blue}. The fields in modules $\mathcal{W}_{j,e^{2i\pi p/M}}$ and $\mathcal{W}_{j,e^{2i\pi(1-p/M)}}$ are regrouped into the blocks $\mathbb{F}_{j,e^{2i\pi p/M}}$ and $\mathbb{F}_{j,e^{-2i\pi p/M}}$ as illustrated in the {\color{magenta}magenta} boxes.}
	\label{superblocksstructure}
\end{figure}

For $\mathbb{F}^{(\mathsf{c})}_{j,\mathrm{z}^2=e^{2i\pi p/M}}$ with $\frac{p}{M}\neq\frac{1}{2}$ we have
\begin{equation}
\begin{aligned}
\mathbb{F}^{(\mathsf{c})}_{j,e^{2i\pi p/M}}=&\frac{1}{2}\sum_{e \in \mathbb{N}}\mathcal{R}_{e+\frac{p}{M},j}\bigg(\mathcal{F}^{(\mathsf{c})}_{h_{e+\frac{p}{M},j}}(z)\mathcal{F}^{(\mathsf{c})}_{h_{e+\frac{p}{M},-j}}(\bar{z})+\mathcal{F}^{(\mathsf{c})}_{h_{e+\frac{p}{M},-j}}(z)\mathcal{F}^{(\mathsf{c})}_{h_{e+\frac{p}{M},j}}(\bar{z})\bigg),\\
=&\sum_{e \in \mathbb{N}}\mathcal{R}_{e+\frac{p}{M},j}\text{Re}\left[\mathcal{F}^{(\mathsf{c})}_{h_{e+\frac{p}{M},j}}(z)\mathcal{F}^{(\mathsf{c})}_{h_{e+\frac{p}{M},-j}}(\bar{z})\right]
\end{aligned}
\end{equation}
with
\begin{equation}
\mathcal{R}_{e+\frac{p}{M},j}\equiv\frac{A(h_{e+\frac{p}{M},j},h_{e+\frac{p}{M},-j})}{A(h_{\frac{p}{M},j},h_{\frac{p}{M},-j})}=
\begin{cases}
1,\;&\mbox{for } e=0,\\
\prod^{e-1}_{i=0}R^N_{i+\frac{p}{M},j},\;\;\;& \mbox{for } e\neq 0.
\end{cases}
\end{equation}
Notice that we have used \eqref{AAA} to group the blocks in the module $\mathcal{W}_{j,e^{2i\pi (1-p/M)}}$ with $e<0$ with the blocks in the module $\mathcal{W}_{j,e^{2i\pi p/M}}$ with $e\ge 0$ to construct $\mathbb{F}_{j,e^{2i\pi p/M}}$, and vice versa. See figure 
\ref{superblocksstructure} for an illustration of this for the case of $\mathbb{F}_{4,i}$ and $\mathbb{F}_{4,-i}$. Also keep in mind that due to \eqref{pm1pm}, the blocks $\mathbb{F}_{j,e^{2i\pi p/M}}$ and $\mathbb{F}_{j,e^{2i\pi (1-p/M)}}$ can be further grouped into
\begin{equation}
\mathbb{F}_{j,e^{2i\pi p/M}}+R^N_{-\frac{p}{M},j}\mathbb{F}_{j,e^{2i\pi (1-p/M)}}.
\end{equation}
In the special case of $p=0$, we have
\begin{equation}
\begin{aligned}
\mathbb{F}^{(\mathsf{c})}_{j,1}=&\frac{1}{2}\sum_{e\ge 0}\mathcal{R}_{e,j}\bigg(\mathcal{F}^{(\mathsf{c})}_{h_{e,j}}(z)\mathcal{F}^{(\mathsf{c})}_{h_{e,-j}}(\bar{z})+\mathcal{F}^{(\mathsf{c})}_{h_{e,-j}}(z)\mathcal{F}^{(\mathsf{c})}_{h_{e,j}}(\bar{z})\bigg),\\
=&\sum_{e\ge 0}\mathcal{R}_{e,j}\text{Re}\left[\mathcal{F}^{(\mathsf{c})}_{h_{e,j}}(z)\mathcal{F}^{(\mathsf{c})}_{h_{e,-j}}(\bar{z})\right]
\end{aligned}
\end{equation}
with
\begin{equation}
\mathcal{R}_{e,j}\equiv
\begin{cases}
1,\;\;&\mbox{for } e=0,\\
\frac{2A(h_{e,j},h_{e,-j})}{A(h_{0,j},h_{0,-j})}=2\prod^{e-1}_{i=0}R^N_{i,j},\;\;& \mbox{for } e\neq 0,\\
\end{cases}
\end{equation}
where the difference in the definition of $\mathcal{R}_{e\neq 0,j}$ and $\mathcal{R}_{0,j}$ takes into account the special choice of \eqref{superA} for $p=0$.

Finally, the block $\mathbb{F}_{0,\q^2}$ is given by:
\begin{equation}\label{superB0q}
\mathbb{F}^{(\mathsf{c})}_{0,\q^2}=\sum_{e\in\mathbb{N}^*}\mathcal{R}_{e,1}\mathcal{F}^{(\mathsf{c})}_{h_{e,1}}(z)\mathcal{F}^{(\mathsf{c})}_{h_{e,1}}(\bar{z}),
\end{equation}
where
\begin{equation}\label{Rd}
\mathcal{R}_{e,1}\equiv\frac{A(h_{e,1},h_{e,1})}{A(h_{1,1},h_{1,1})}=\begin{cases}
1,\;\;& \mbox{for } e=1,\\
\prod^{e-1}_{i=1}R^D_{i,1},\;\;& \mbox{for } e\neq 1.
\end{cases}
\end{equation}

In figure \ref{superblocksstructure}, we give explicit examples of the construction of various interchiral blocks.

\subsubsection{Recursions from degeneracy}\label{degeneracyrecursion}
In this subsection, we derive the recursions \eqref{Rshift1N} and \eqref{Rshift1D} using the degeneracy of $\Phi^D_{2,1}$. The key is to study the four-point functions involving the degenerate field $\Phi^D_{2,1}$, as done in the conformal bootstrap approach to the diagonal and non-diagonal Liouville theory in \cite{Zamolodchikov:1995aa,Teschner:1995yf,Estienne:2015sua,Migliaccio:2017dch}. We first briefly summarize the general formalism and then explore its consequences on the geometrical correlation of the type \eqref{4ptspin}.

\smallskip

\begin{figure}[t]
	\centering
	\includegraphics[width=0.7\textwidth]{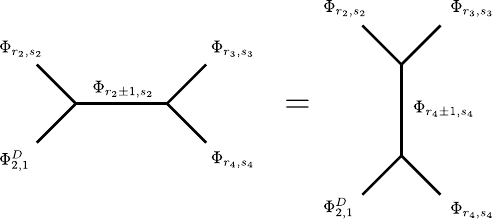}
	\caption{The $s$- and $t$-channel of the four-point function $\langle\Phi^D_{2,1}\Phi_{r_2,s_2}\Phi_{r_3,s_3}\Phi_{r_4,s_4}\rangle$.}
	\label{deg0}
\end{figure}

Consider a generic four-point function with the
degenerate field $\Phi^D_{2,1}$:
\begin{equation}\label{deg4}
\langle\Phi^D_{2,1}\Phi_{r_2,s_2}\Phi_{r_3,s_3}\Phi_{r_4,s_4}\rangle \,,
\end{equation}
where $\Phi_{r_i,s_i}$ represent either diagonal or non-diagonal fields. Due to the degeneracy of $\Phi^D_{2,1}$, the fusion involves only two terms
\begin{subequations}\label{DNfusion}
	\begin{eqnarray}
	\Phi^D_{2,1}\times\Phi^D_{r,s}&\to&\Phi^D_{r+1,s}+\Phi^D_{r-1,s},\\	\Phi^D_{2,1}\times\Phi^N_{r,s}&\to&\Phi^N_{r+1,s}+\Phi^N_{r-1,s}.
	\end{eqnarray}
\end{subequations}
and the $s$- and $t$-channels of the four-point function \eqref{deg4} are illustrated in figure \ref{deg0}.
The conformal block expansions of \eqref{deg4} in the $s$- and $t$-channel therefore truncate to two terms:
\begin{equation}\label{deg4pt}
\begin{aligned}
\langle\Phi^D_{2,1}\Phi_{r_2,s_2}\Phi_{r_3,s_3}\Phi_{r_4,s_4}\rangle=\left[\begin{array}{cc}
\mathcal{F}_+&\mathcal{F}_-
\end{array}\right]^{(s)}\mathbf{a}^{(s)}\left[\begin{array}{cc}
\bar{\mathcal{F}}_+\\
\bar{\mathcal{F}}_-
\end{array}\right]^{(s)}
=\left[\begin{array}{cc}
\mathcal{F}_+&\mathcal{F}_-
\end{array}\right]^{(t)}\mathbf{a}^{(t)}\left[\begin{array}{cc}
\bar{\mathcal{F}}_+\\
\bar{\mathcal{F}}_-
\end{array}\right]^{(t)},
\end{aligned}
\end{equation}
where $\pm$ represents $(r_i\pm 1,s_i)$ and we have omitted the dependence on the external fields. Note that depending on the external fields, the $s$- and $t$-channels can independently involve either diagonal or non-diagonal fields. We will in the following study four-point functions of various types and use the labels:
\begin{equation}\label{DNlabel}
DD,\;\;NN,\;\;DN,\;\;ND
\end{equation}
where the first letter indicates the type of field in the $s$-channel, and the second letter similarly specifies the type for the $t$-channel. (According to the fusion \eqref{DNfusion}, these are just the labels for the fields $\Phi_{r_2,s_2}$ and $\Phi_{r_4,s_4}$.)
The amplitude matrix $\mathbf{a}^{(\mathsf{c})}$ in \eqref{deg4pt} is given by
\begin{equation}\label{amatrix}
\mathbf{a}^{(\mathsf{c})}=
\begin{cases}
\left[\begin{array}{cc}
a^{(\mathsf{c})}_+ & 0 \\
0 & a^{(\mathsf{c})}_-
\end{array}\right],\;\;&\text{diagonal},\\
&

\\
\left[\begin{array}{cc}
0 & a^{(\mathsf{c})}_+\\
a^{(\mathsf{c})}_- & 0
\end{array}\right],\;\;&\text{non-diagonal},
\end{cases} \quad \mbox{for } \mathsf{c}=s,t.
\end{equation}
The amplitudes $a^{(\mathsf{c})}_{\pm}$ come from the structure constants:
\begin{equation}\label{aC}
a^{(\mathsf{c})}_{\pm}=\begin{cases}
C_{(2,1)^D(r_2,s_2)(r_2\pm 1,s_2)}C_{(r_2\pm 1,s_2)(r_3,s_3)(r_4,s_4)}, & \mbox{for } \mathsf{c}=s,\\
C_{(2,1)^D(r_4,s_4)(r_4\pm 1,s_4)}C_{(r_4\pm 1,s_4)(r_3,s_3)(r_2,s_2)}, & \mbox{for } \mathsf{c}=t,
\end{cases}
\end{equation}
where $(r,s)$ represent either diagonal or non-diagonal fields obeying the fusion \eqref{DNfusion}.
The $s$- and $t$-channel conformal blocks are related through the fusing matrix:
\begin{equation}\label{fusingF}
\left[\begin{array}{c}
\mathcal{F}_+\\
\mathcal{F}_-
\end{array}
\right]^{(s)}
=\left[\begin{array}{cc}
F_{++} & F_{+-} \\
F_{-+} & F_{--} 
\end{array}\right]
\left[\begin{array}{c}
\mathcal{F}_+\\
\mathcal{F}_-
\end{array}
\right]^{(t)},
\end{equation}
and similarly
\begin{equation}\label{fusingFbar}
\left[\begin{array}{c}
\bar{\mathcal{F}}_+\\
\bar{\mathcal{F}}_-
\end{array}
\right]^{(s)}
=\left[\begin{array}{cc}
\bar{F}_{++} & \bar{F}_{+-} \\
\bar{F}_{-+} & \bar{F}_{--} 
\end{array}\right]
\left[\begin{array}{c}
\bar{\mathcal{F}}_+\\
\bar{\mathcal{F}}_-
\end{array}
\right]^{(t)} \,,
\end{equation}
where $F_{\pm\pm}$ is given by
\begin{equation}
F_{\mathsf{s}\mathsf{t}}=\frac{\Gamma(1-\frac{2\mathsf{s}}{\upbeta}\lambda_{r_2,s_2})\Gamma(\frac{2\mathsf{t}}{\upbeta}\lambda_{r_4,s_4})}{\prod_{+,-}\Gamma(\frac{1}{2}\pm\frac{1}{\upbeta}\lambda_{r_3,s_3}-\frac{\mathsf{s}}{\upbeta}\lambda_{r_2,s_2}+\frac{\mathsf{t}}{\upbeta}\lambda_{r_4,s_4})}, \quad \mbox{with } \mathsf{s},\mathsf{t}=\pm
\end{equation}
and $\bar{F}$ is obtained by replacing $\lambda$ with $\bar{\lambda}$, defined as:\footnote{This corresponds to the Liouville momentum $P_{s,r}$ used in \cite{Migliaccio:2017dch}.}
\begin{subequations}
\begin{eqnarray}
\lambda_{r_i,s_i}&=&-\frac{r_i}{2\upbeta}+\frac{s_i\upbeta}{2},\\
\bar{\lambda}_{r_i,s_i}&=& \begin{cases}
\lambda_{r_i,s_i},&\text{diagonal},\\
\lambda_{-r_i,s_i},&\text{non-diagonal}.
\end{cases}
\end{eqnarray}
\end{subequations} 
Plugging \eqref{fusingF} and \eqref{fusingFbar} into \eqref{deg4pt}, we obtain
\begin{equation}\label{arelation}
\left[\begin{array}{cc}
F_{++} & F_{+-} \\
F_{-+} & F_{--} 
\end{array}\right]^T\mathbf{a}^{(s)}
\left[\begin{array}{cc}
\bar{F}_{++} & \bar{F}_{+-} \\
\bar{F}_{-+} & \bar{F}_{--} 
\end{array}\right]=\mathbf{a}^{(t)},
\end{equation}
which gives the relations among $a^{(s)}_{\pm}$ and $a^{(t)}_{\pm}$. Keep in mind that the explicit relations depend on the properties \eqref{DNlabel} and therefore the explicit form of $\mathbf{a}^{(\mathsf{c})}$ as in \eqref{amatrix}.

\medskip

In the conformal bootstrap approach to solve the diagonal \cite{Zamolodchikov:1995aa,Teschner:1995yf} and non-diagonal \cite{Estienne:2015sua,Migliaccio:2017dch} Liouville theory, the ratio
\begin{equation}
\rho=\frac{a^{(s)}_+}{a^{(s)}_-}
\end{equation}
has been exploited in various four-point functions of the type \eqref{deg4} to obtain the recursion for shifting the amplitudes with $r\pm1$ as we mentioned in the previous section. Here, we will focus on the other consequence of \eqref{arelation}, that is, the relation between $a^{(t)}_{\pm}$ and $a^{(s)}_{\pm}$. From \eqref{arelation}, this relation can be extracted for different types of the four-point function as labeled with \eqref{DNlabel}. Defining the ratios
\begin{equation}\label{chidef}
\chi_{\mathsf{s}\mathsf{t}}=\frac{a^{(t)}_{\mathsf{t}}}{a^{(s)}_{\mathsf{s}}}, \quad \mbox{with } \mathsf{s},\mathsf{t}=\pm,
\end{equation}
we will need the following explicit expressions from \eqref{arelation}:
\begin{equation}\label{chi}
\begin{aligned}
	&\chi_{-+}^{DN}=F_{-+}\bar{F}_{--}+\rho^{DN} F_{++}\bar{F}_{+-},\\
&\chi_{+-}^{ND}=F_{+-}\bar{F}_{--}+\frac{1}{\rho^{ND}}F_{--}\bar{F}_{+-},\\
&\chi_{-+}^{DD}=F_{-+}\bar{F}_{-+}+ \rho^{DD}F_{++}\bar{F}_{++},\\
&\chi_{+-}^{DD}=F_{+-}\bar{F}_{+-}+ \frac{1}{\rho^{DD}}F_{--}\bar{F}_{--},\\
&\chi_{--}^{DD}=F_{--}\bar{F}_{--}+\rho^{DD}F_{+-}\bar{F}_{+-},
\end{aligned}
\end{equation}
where
\begin{equation}\label{rho}
	\rho^{DN}=-\frac{F_{-+}\bar{F}_{-+}}{F_{++}\bar{F}_{++}}, \quad
	\rho^{ND}=-\frac{F_{-+}\bar{F}_{+-}}{F_{++}\bar{F}_{--}}, \quad
	\rho^{DD}=-\frac{F_{-+}\bar{F}_{--}}{F_{++}\bar{F}_{+-}}.
\end{equation}
The superscript should be interpreted as in \eqref{DNlabel}: for example, $\chi^{DN}_{-+}$ corresponds to the amplitude ratio of the non-diagonal $t$-channel field $(r_4+1,s_4)^N$ with the diagonal $s$-channel field $(r_2-1,s_2)^D$.

\bigskip

We are now ready to derive the recursions \eqref{Rshift1N} and \eqref{Rshift1D} for non-diagonal and diagonal fields in the four-point function \eqref{4ptspin}. 

\smallskip

For non-diagonal fields, consider the following four-point function of the type \eqref{4ptspin}:
\begin{equation}
\langle\Phi^N_{\frac{1}{2},0}\Phi^D_{\frac{1}{2},0}\Phi^N_{\frac{1}{2},0}\Phi^D_{\frac{1}{2},0}\rangle \,,
\end{equation}
where the fusion gives the non-diagonal fields in the Potts spectrum \eqref{Pottsspectrum}:
\begin{equation}
\Phi^N_{\frac{1}{2},0}\times\Phi^D_{\frac{1}{2},0}\to(h_{e,j},h_{e,-j})
\end{equation}
and the amplitudes arise from the structure constants:
\begin{equation}\label{ACspin}
A(h_{e,j},h_{e,-j})=C_{(\frac{1}{2},0)^N(\frac{1}{2},0)^D(e,j)^N}C_{(e,j)^N(\frac{1}{2},0)^D(\frac{1}{2},0)^N}.
\end{equation}
The desired recursion \eqref{Rshift1N} is then written as
\begin{equation}\label{Rshift1}
R^N_{e,j}=\frac{A(h_{e+1,j},h_{e+1,-j})}{A(h_{e,j},h_{e,-j})}=\frac{C_{(\frac{1}{2},0)^N(\frac{1}{2},0)^D(e+1,j)^N}C_{(e+1,j)^N(\frac{1}{2},0)^D(\frac{1}{2},0)^N}}{C_{(\frac{1}{2},0)^N(\frac{1}{2},0)^D(e,j)^N}C_{(e,j)^N(\frac{1}{2},0)^D(\frac{1}{2},0)^N}}.
\end{equation}

\begin{figure}[H]
	\centering
	\includegraphics[width=0.7\textwidth]{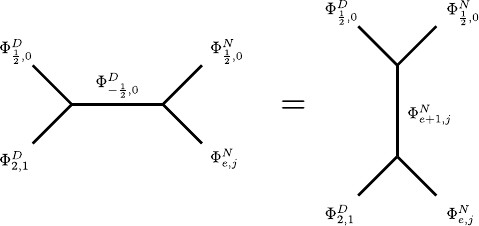}
	\caption{The $s$- and $t$-channels of the four-point function $\langle\Phi^D_{2,1}\Phi^D_{\frac{1}{2},0}\Phi^N_{\frac{1}{2},0}\Phi^N_{e,j}\rangle$. Notice that since $(\frac{1}{2},0)$ and $(-\frac{1}{2},0)$ both represent the same spin field, we obtain the structure constants $C_{(-\frac{1}{2},0)^D(\frac{1}{2},0)^N(e,j)^N}$ in the $s$-channel and $C_{(e+1,j)^N(\frac{1}{2},0)^D(\frac{1}{2},0)^N}$ in the $t$-channel which relates two fields with $e\to e+1$.}
	\label{deg1}
\end{figure}

Now consider the four-point function
\begin{equation}\label{G1}
G_1^{DN}=\langle\Phi^D_{2,1}\Phi^D_{\frac{1}{2},0}\Phi^N_{\frac{1}{2},0}\Phi^N_{e,j}\rangle \,,
\end{equation}
whose crossing equation and the relevant fusion channels are illustrated in figure \ref{deg1}.
The corresponding amplitudes in the two channels come from the structure constants and give the following ratio:
\begin{equation}
\chi^{DN}_{-+,1}=\frac{C_{(2,1)^D(e,j)^N(e+1,j)^N}C_{(e+1,j)^N(\frac{1}{2},0)^D(\frac{1}{2},0)^N}}{C_{(2,1)^D(\frac{1}{2},0)^D(-\frac{1}{2},0)^D}C_{(-\frac{1}{2},0)^D(\frac{1}{2},0)^N(e,j)^N}}.
\end{equation}
Keep in mind the identification
\begin{equation}\label{identification}
C_{(-\frac{1}{2},0)^D(\frac{1}{2},0)^N(e,j)^N}=C_{(\frac{1}{2},0)^D(\frac{1}{2},0)^N(e,j)^N} \,,
\end{equation}
since $(\frac{1}{2},0)$ and $(-\frac{1}{2},0)$ represent the same spin field.

\begin{figure}[H]
	\centering
	\includegraphics[width=0.7\textwidth]{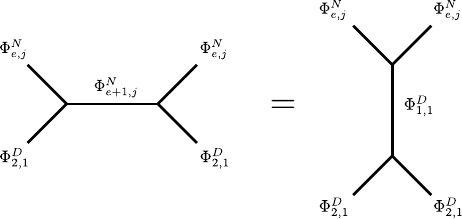}
	\caption{The $s$- and $t$-channels of the four-point function $\langle\Phi^D_{2,1}\Phi^N_{e,j}\Phi^N_{e,j}\Phi^D_{2,1}\rangle$. Notice the appearance of the identity field $\Phi^D_{1,1}$ in the $t$-channel and the corresponding amplitude $a^{(t)}_-=1$, due to the normalization of the two-point functions.}
	\label{deg2}
\end{figure}

We then turn to the four-point function:
\begin{equation}\label{G2}
G_2^{ND}=\langle\Phi^D_{2,1}\Phi^N_{e,j}\Phi^N_{e,j}\Phi^D_{2,1}\rangle
\end{equation}
as illustrated in figure \ref{deg2}. Notice that in this case we have
\begin{equation}\label{Cid}
a^{(t)}_{-}=C_{(2,1)^D(2,1)^D(1,1)^D}C_{(1,1)^D(e,j)^N(e,j)^N}=1,
\end{equation}
where $(1,1)^D$ represents the identity field and the structure constants in \eqref{Cid} are given by \eqref{id3pt}, due to the normalization of the two-point function.\footnote{We have, for convenience, chosen \eqref{id3pt} which means the constant in the two-point functions are normalized to 1. With a different normalization, the derivation here still holds, since all the normalization factors cancel in the final expression \eqref{Rasratios} below. Therefore, the recursions we obtain here are independent of the normalization.} Therefore, one has
\begin{equation}
\chi_{+-,2}^{ND}=\frac{1}{C_{(2,1)^D(e,j)^N(e+1,j)^N}C_{(e+1,j)^N(e,j)^N(2,1)^D}}.
\end{equation}

\begin{figure}[H]
	\centering
	\includegraphics[width=0.7\textwidth]{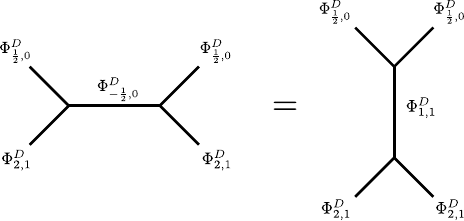}
	\caption{The $s$- and $t$-channels of the four-point function $\langle\Phi^D_{2,1}\Phi^D_{\frac{1}{2},0}\Phi^D_{\frac{1}{2},0}\Phi^D_{2,1}\rangle$.}
	\label{deg3}
\end{figure}

Finally, consider the four-point function
\begin{equation}
G_3^{DD}=\langle\Phi^D_{2,1}\Phi^D_{\frac{1}{2},0}\Phi^D_{\frac{1}{2},0}\Phi^D_{2,1}\rangle
\end{equation}
as illustrated in figure \ref{deg3}. Similar to the previous case, one has 
\begin{equation}
a^{(t)}_{-}=C_{(2,1)^D(2,1)^D(1,1)^D}C_{(1,1)^D(\frac{1}{2},0)^D(\frac{1}{2},0)^D}=1,
\end{equation}
and therefore
\begin{equation}
\chi_{--,3}^{DD}=\frac{1}{C_{(2,1)^D(\frac{1}{2},0)^D(-\frac{1}{2},0)^D}C_{(-\frac{1}{2},0)^D(\frac{1}{2},0)^D(2,1)^D}}.
\end{equation}
It is now easy to see that the recursion \eqref{Rshift1} can be expressed as
\begin{equation}\label{Rasratios}
R^N_{e,j}=\frac{\chi_{-+,1}^{DN}\chi_{+-,2}^{ND}\chi_{-+,1}^{DN}}{\chi_{--,3}^{DD}},
\end{equation}
where we have used the permutation symmetry of the three-point structure constants.
After plugging in the explicit expressions of \eqref{chi}--\eqref{rho}, eq.~\eqref{Rasratios} becomes 
\eqref{Rshift1N}.

\smallskip

In the diagonal case, the derivation of \eqref{Rshift1D} is completely analogous. We then consider the following four-point function of the type \eqref{4ptspin}:
\begin{equation}
\langle\Phi^D_{\frac{1}{2},0}\Phi^D_{\frac{1}{2},0}\Phi^D_{\frac{1}{2},0}\Phi^D_{\frac{1}{2},0}\rangle \,,
\end{equation}
where the diagonal fields arise from the fusion
\begin{equation}
\Phi^D_{\frac{1}{2},0}\times\Phi^D_{\frac{1}{2},0}\to(h_{e,s},h_{e,s})
\end{equation}
with the amplitudes
\begin{equation}\label{Adiagonal}
A(h_{e,s},h_{e,s})=C_{(\frac{1}{2},0)^D(\frac{1}{2},0)^D(e,s)^D}C_{(e,s)^D(\frac{1}{2},0)^D(\frac{1}{2},0)^D}.
\end{equation}
The recursion \eqref{Rshift1D} is then given by
\begin{equation}\label{RD}
R^D_{e,s}=\frac{A(h_{e+1,s},h_{e+1,s})}{A(h_{e,s},h_{e,s})}=\frac{C_{(\frac{1}{2},0)^D(\frac{1}{2},0)^D(e+1,s)^D}C_{(e+1,s)^D(\frac{1}{2},0)^D(\frac{1}{2},0)^D}}{C_{(\frac{1}{2},0)^D(\frac{1}{2},0)^D(e,s)^D}C_{(e,s)^D(\frac{1}{2},0)^D(\frac{1}{2},0)^D}}.
\end{equation}
Going through the same procedure as in the non-diagonal case, but replacing \eqref{G1}, \eqref{G2} with
\begin{subequations}
\begin{eqnarray}
&&G_1^{DD}=\langle\Phi^D_{2,1}\Phi^D_{\frac{1}{2},0}\Phi^D_{\frac{1}{2},0}\Phi^D_{e,s}\rangle,\\
&&G_2^{DD}=\langle\Phi^D_{2,1}\Phi^D_{e,s}\Phi^D_{e,s}\Phi^D_{2,1}\rangle,
\end{eqnarray}
\end{subequations}
one arrives at the expression for the recursion \eqref{RD} given by
\begin{equation}\label{RasratiosD}
R^D_{e,s}=\frac{\chi_{-+,1}^{DD}\chi_{+-,2}^{DD}\chi_{-+,1}^{DD}}{\chi_{--,3}^{DD}}.
\end{equation}
Plugging in \eqref{chi}--\eqref{rho}, we obtain \eqref{Rshift1D}.
\smallskip

Note that to obtain \eqref{Rasratios} and \eqref{RasratiosD}, it is important that we are studying a four-point function of the spin operator where the amplitudes are given by three-point structure constants as in \eqref{ACspin} and \eqref{Adiagonal}, since in this case the four-point function of figure \ref{deg1} involving $\Phi^D_{2,1}$ gives rise to both $C_{(\frac{1}{2},0)(\frac{1}{2},0)(e,j)}$ and $C_{(\frac{1}{2},0)(\frac{1}{2},0)(e+1,j)}$ in their $s$- and $t$-channels.


\subsection{Results}\label{bootresults}

In this section, we give the bootstrap results on the amplitudes \eqref{ATLamp} associated with the ATL modules up to $j=4$ and leave the numerical details to appendix \ref{bootstrapapp}. As discussed in section \ref{supereq}, this involves solving numerically the truncated interchiral bootstrap equations \eqref{superP}  combined with the constraints of the amplitude ratios \eqref{constraints} and the analytic results \eqref{PAusingL}. For this last constraint, we have in fact imposed the ratios of
\begin{equation}\label{analyticratios}
\frac{A_{abab}(\mathcal{W}_{2,-1})}{A_{aaaa}(\mathcal{W}_{0,-1})},\;\;\frac{A_{aaaa}(\mathcal{W}_{4,-1})}{A_{aaaa}(\mathcal{W}_{0,-1})}
\end{equation}  
as obtained from \eqref{PAusingL} without fixing the overall normalization. It is worth pointing out that \eqref{analyticratios} can in fact be (partially) bootstrapped as a consistency check. See figure \ref{Aanalattice} and the related discussions in appendix \ref{basiccheck}. Notice that in the Potts spectrum \eqref{Pottsspectrum}, the leading primary in the module $\overline{\mathcal{W}}_{0,\q^2}$ has the conformal dimension $(h_{1,1},\bar{h}_{1,1})=(0,0)$, corresponding to the identity field. This is in fact the field with the lowest conformal dimension, and since it only appears in the probability $P_{aabb}$, it is natural to use the normalization
\begin{equation}\label{Anorm}
A_{aabb}(\overline{\mathcal{W}}_{0,\q^2})=1
\end{equation}
for the bootstrap equations \eqref{superP}. 

From the discussions in \cite{Jacobsen:2018pti}, it is expected that some amplitudes should display singularities at rational values of $\upbeta^2$, the effect of which is to cancel the overall singularities and thus lead to smooth geometrical correlations. We will study this in more details in the next section, while here we simply point out the locations of the singularities.

\subsubsection{$A_{aaaa}$}
\begin{figure}[H]
	\centering
	\includegraphics[width=0.5\textwidth]{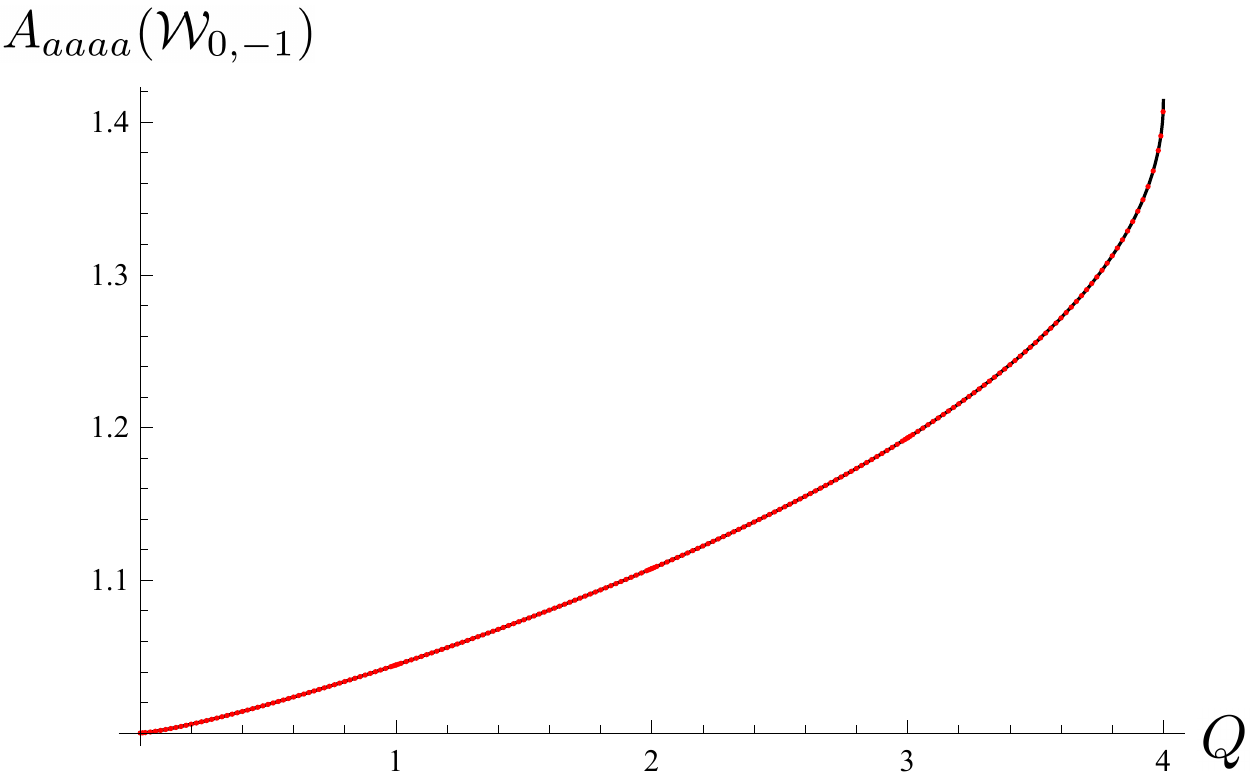}
	\caption{The amplitude $A_{aaaa}(\mathcal{W}_{0,-1})$. Red dots are the numerical bootstrap result and the black curve is the analytic expression \eqref{Aaaaa0m1ana}. They agree perfectly (the black curve being only visible behind the red points close to $Q=4$).}
	\label{aaaa0m1}
\end{figure}

Up to $j=4$, the following amplitudes appear in the interchiral block expansion of $P_{aaaa}$:
\begin{equation}
\begin{aligned}
A_{aaaa}(\mathcal{W}_{0,-1}),\;\;&A_{aaaa}(\mathcal{W}_{2,1}),\\
A_{aaaa}(\mathcal{W}_{4,-1}),\;\;&A_{aaaa}(\mathcal{W}_{4,1}).
\end{aligned}
\end{equation}
All the other amplitudes of the primaries can be obtained using the recursions which have been incorporated into the interchiral blocks for the numerical bootstrap.
With the normalization \eqref{Anorm}, we obtained the amplitude $A_{aaaa}(\mathcal{W}_{0,-1})$ given in figure \ref{aaaa0m1}, where we also plot the analytic amplitude \eqref{Aaaaa0m1ana}. The explicit expression of the latter is given in \eqref{A0m1} of appendix \ref{AL}, as obtained originally in \cite{Migliaccio:2017dch} and reproduced in \cite{Picco:2019dkm}, where it was also found to agree with Monte-Carlo simulations. It was pointed out in \cite{Picco:2019dkm} that this specific normalization for the amplitude $A(h_{\frac{1}{2},0},h_{\frac{1}{2},0})$ (i.e., our $A_{aaaa}(\mathcal{W}_{0,-1})$ here) underlies the three-point structure constants describing the probability $P_{aaa}$ of three points belonging to the same FK cluster \cite{Delfino:2010xm}. Here we can clearly see that the agreement with bootstrap result is perfect.

\begin{figure}[H]
	\begin{centering}
		\begin{subfigure}[h]{0.5\textwidth}
			\includegraphics[width=0.9\textwidth]{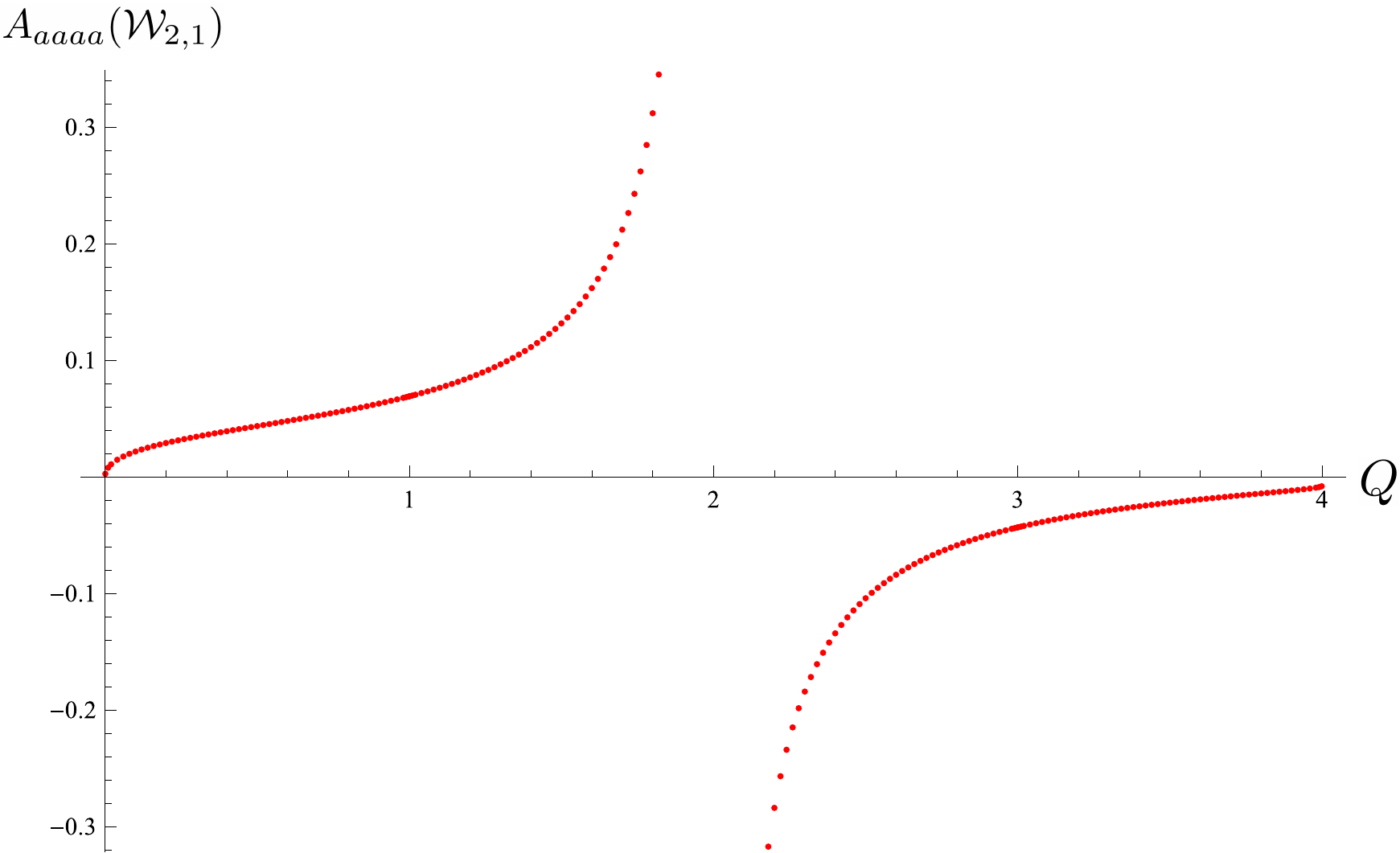}
			\caption{}
			\label{aaaa21}
		\end{subfigure}
		\begin{subfigure}[h]{0.5\textwidth}
			\includegraphics[width=0.9\textwidth]{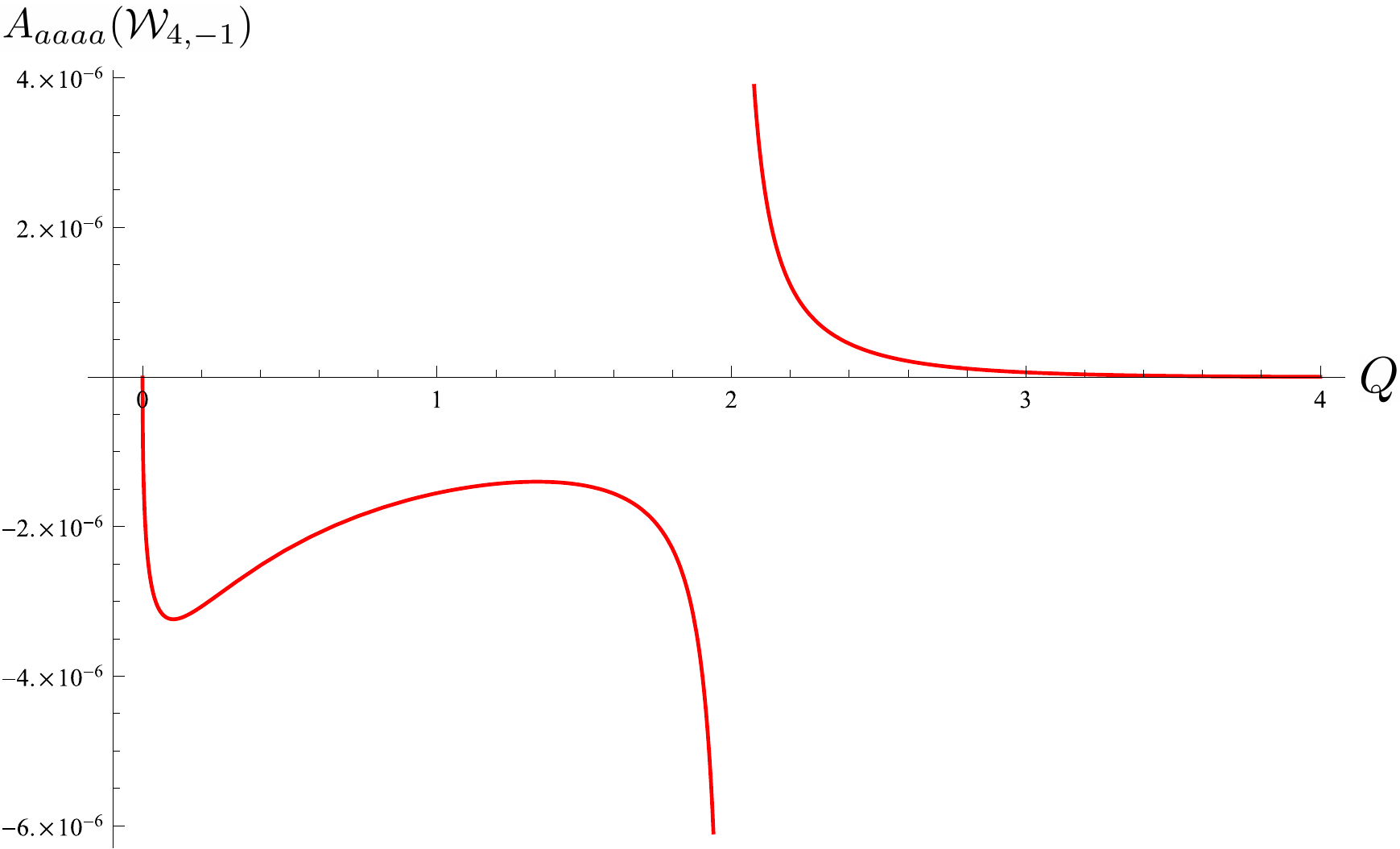}
			\caption{}
			\label{}
		\end{subfigure}
	\end{centering}
	\caption{The bootstrapped $A_{aaaa}(\mathcal{W}_{2,1})$ on the left and the analytic $A_{aaaa}(\mathcal{W}_{4,-1})$ on the right.}
	\label{aaaa214m1}
\end{figure}

In figure \ref{aaaa214m1}, we show on the left the amplitude $A_{aaaa}(\mathcal{W}_{2,1})$ and on the right $A_{aaaa}(\mathcal{W}_{4,-1})$ as given in \eqref{A4m1ana}. In both cases, the amplitudes have simple poles at $Q=2$ and no other poles in the range $0<Q<4$.

\begin{figure}[H]
	\begin{centering}
		\begin{subfigure}{0.5\textwidth}
			\includegraphics[width=0.9\textwidth]{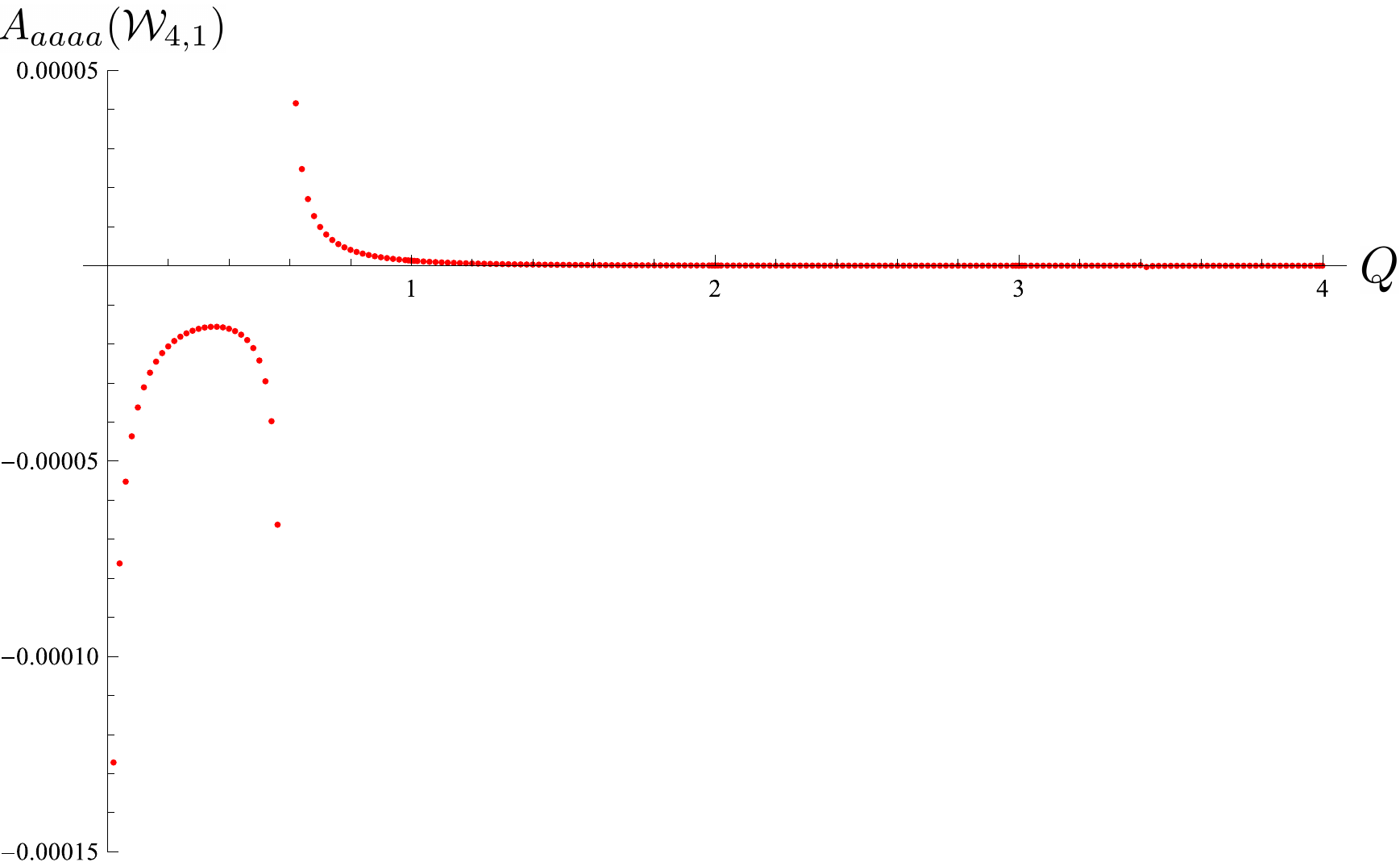}
		\end{subfigure}\\
		
		\begin{subfigure}{0.5\textwidth}
			\includegraphics[width=0.9\textwidth]{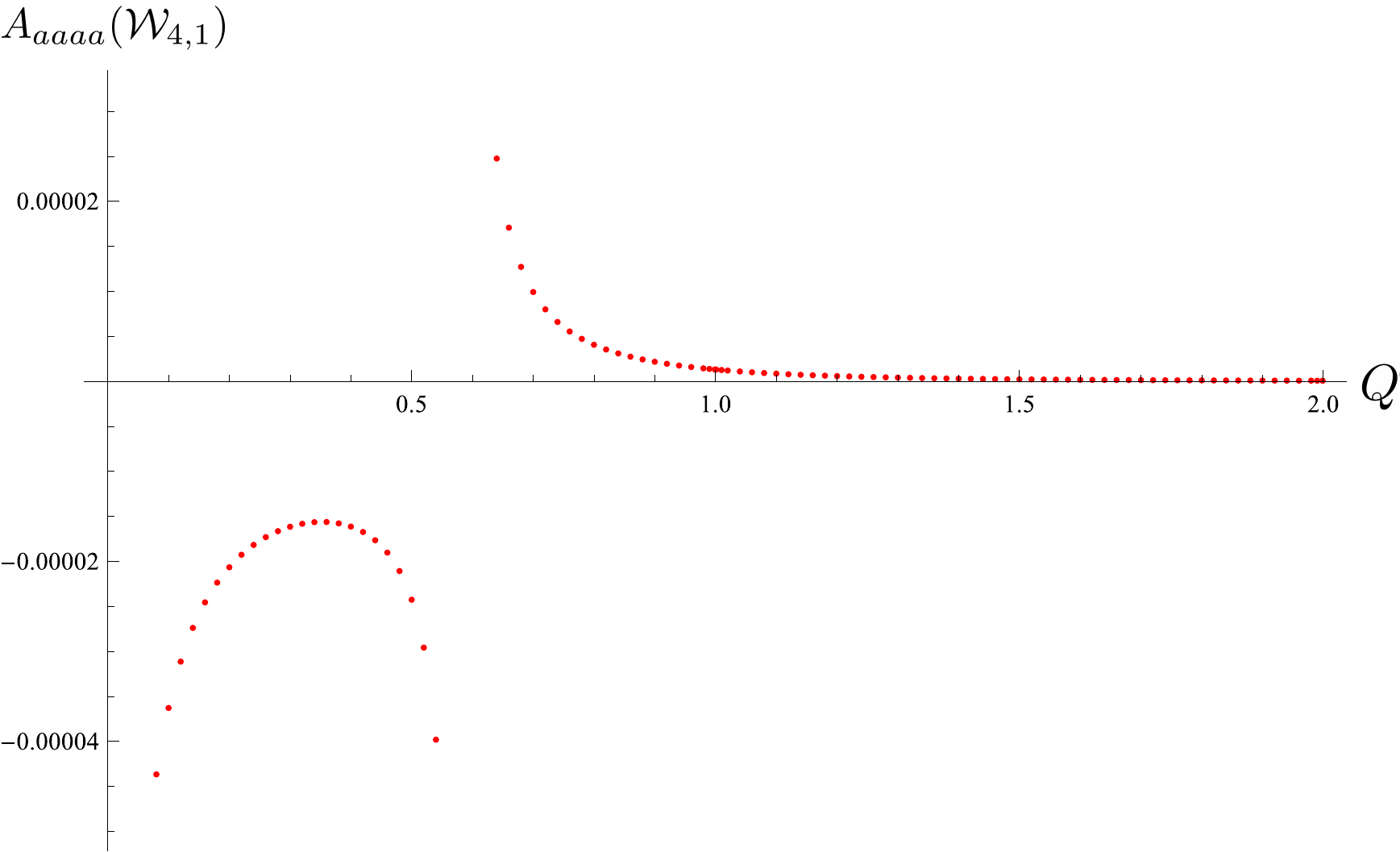}
		\end{subfigure}
		\begin{subfigure}{0.5\textwidth}
			\includegraphics[width=0.9\textwidth]{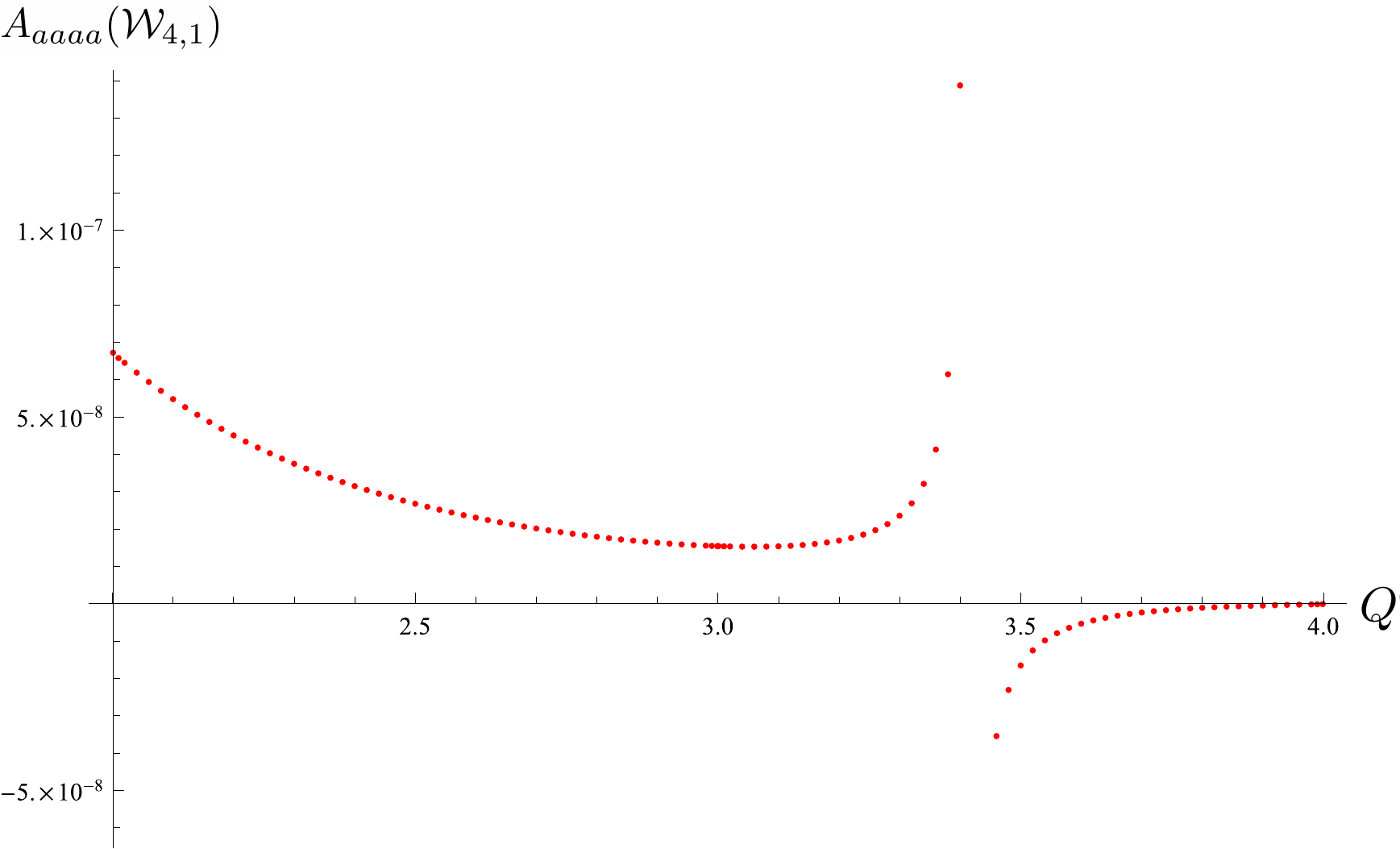}
		\end{subfigure}
	\end{centering}
	\caption{The bootstrap result of the amplitude $A_{aaaa}(\mathcal{W}_{4,1})$ and its detailed pole structure in the regions $0<Q<2$ and $2<Q<4$.}
	\label{aaaa40}
\end{figure}

The amplitude $A_{aaaa}(\mathcal{W}_{4,1})$ is shown in figure \ref{aaaa40}. It has simple poles at:
\begin{subequations}\label{aaaa41poles}
	\begin{eqnarray}
	&&Q=4\cos^2 \! \left( \frac{3\pi}{8} \right) =0.585786\ldots \;,\label{aaaa41poles2}\\
	&&Q=4\cos^2 \! \left( \frac{\pi}{8} \right) =3.414213\ldots \;,\label{aaaa41poles3}
	\end{eqnarray}
\end{subequations}
of which we also plot the details in the zoomed-in regions of $0<Q<2$ and $2<Q<4$ in the bottom part of the figure.

\subsubsection{$A_{abab}$}
In $P_{abab}$, up to $j=4$, we have the following amplitudes
\begin{equation}
\begin{aligned}
A_{abab}(\mathcal{W}_{2,1}),\;\;&A_{abab}(\mathcal{W}_{2,-1}),\\
A_{abab}(\mathcal{W}_{4,1}),\;\;&A_{abab}(\mathcal{W}_{4,-1}),\\
A_{abab}(\mathcal{W}_{4,i}),\;\;&A_{abab}(\mathcal{W}_{4,-i}).
\end{aligned}
\end{equation}
The second amplitude $A_{abab}(\mathcal{W}_{2,-1})$ was obtained analytically in \eqref{abab2m1ana}, and for the modules $\mathcal{W}_{2,1}$, $\mathcal{W}_{4,1}$, $\mathcal{W}_{4,-1}$ the corresponding amplitudes are related to the $A_{aaaa}$ through $\mathsf{R}_{\bar{\alpha}}$ in \eqref{ratioforbootstrap}. This was in fact used as input in the bootstrap for the final results we present here. However for completeness we plot all these amplitudes below.

\begin{figure}[H]
	\begin{centering}
		\begin{subfigure}[h]{0.5\textwidth}
			\includegraphics[width=0.9\textwidth]{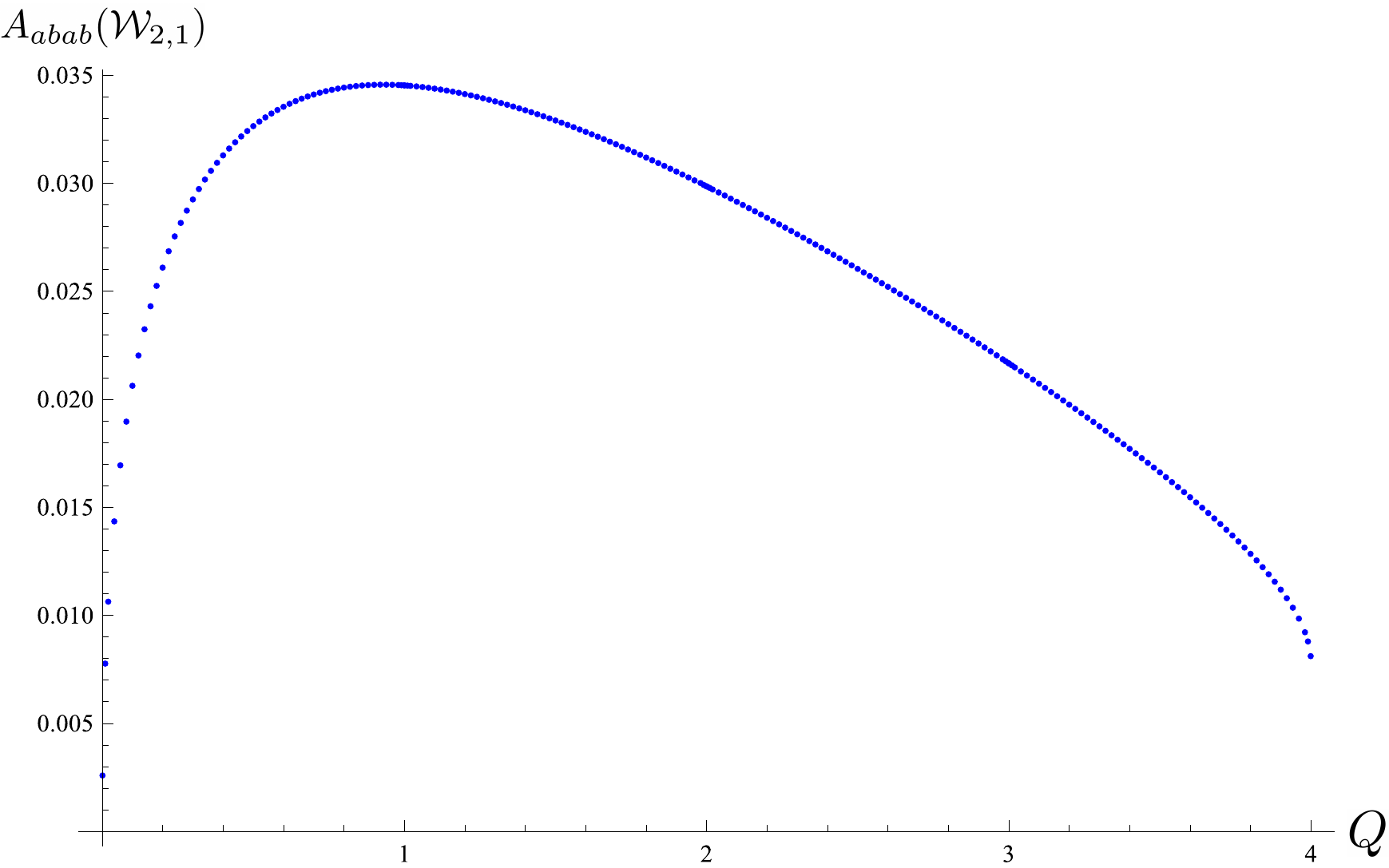}
			\caption{}
			\label{abab21}
		\end{subfigure}
		\begin{subfigure}[h]{0.5\textwidth}
			\includegraphics[width=0.9\textwidth]{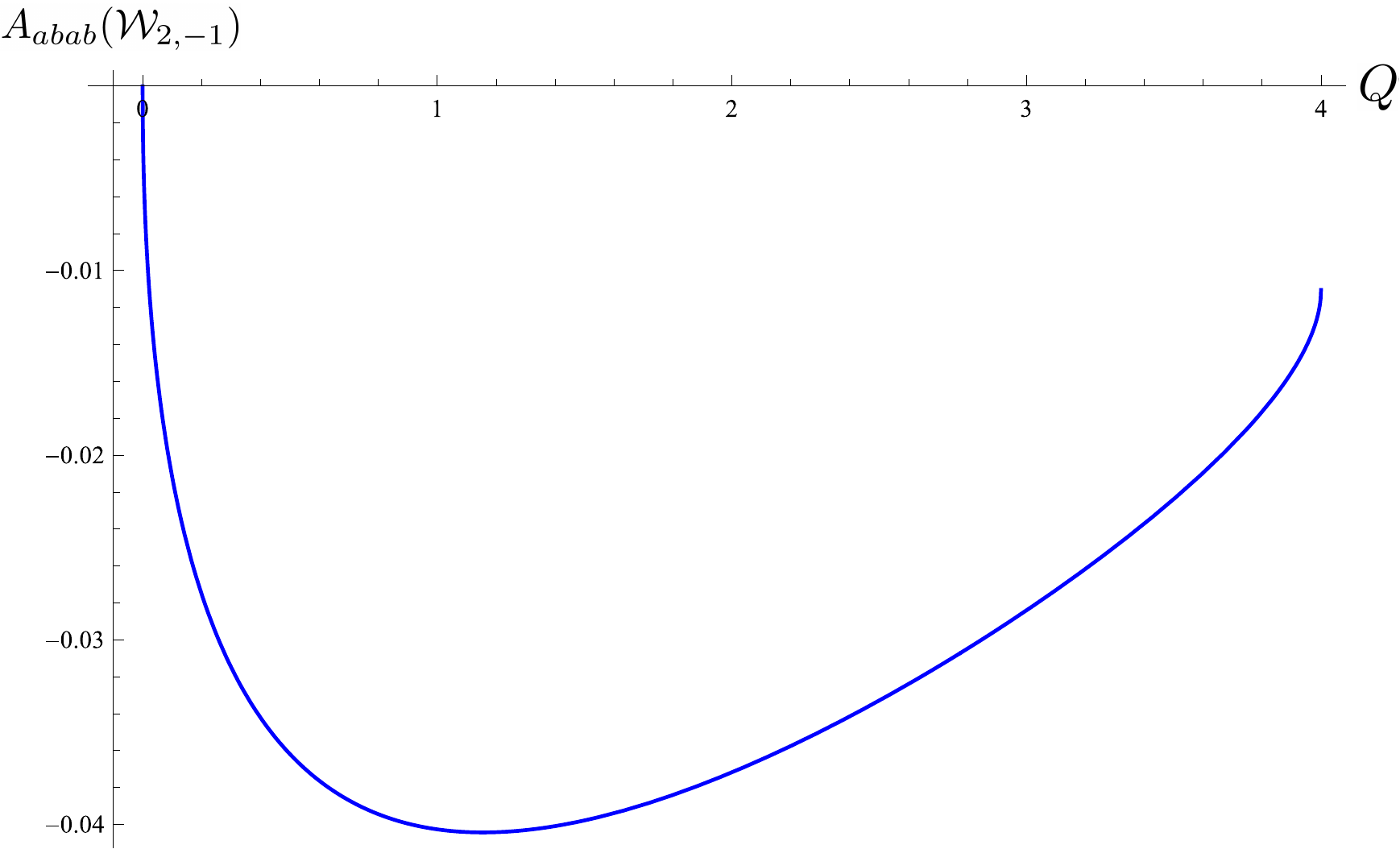}
			\caption{}
			\label{}
		\end{subfigure}
	\end{centering}
	\caption{The bootstrapped $A_{abab}(\mathcal{W}_{2,1})$ on the left and the analytic $A_{abab}(\mathcal{W}_{2,-1})$ on the right.}
	\label{abab212m1}
\end{figure}

The amplitudes $A_{abab}(\mathcal{W}_{2,1})$ and $A_{abab}(\mathcal{W}_{2,-1})$ are shown in figure \ref{abab212m1}. They are smooth with no singularities in the whole range $0<Q<4$.

\begin{figure}[H]
	\begin{centering}
		\begin{subfigure}[h]{0.5\textwidth}
			\includegraphics[width=0.9\textwidth]{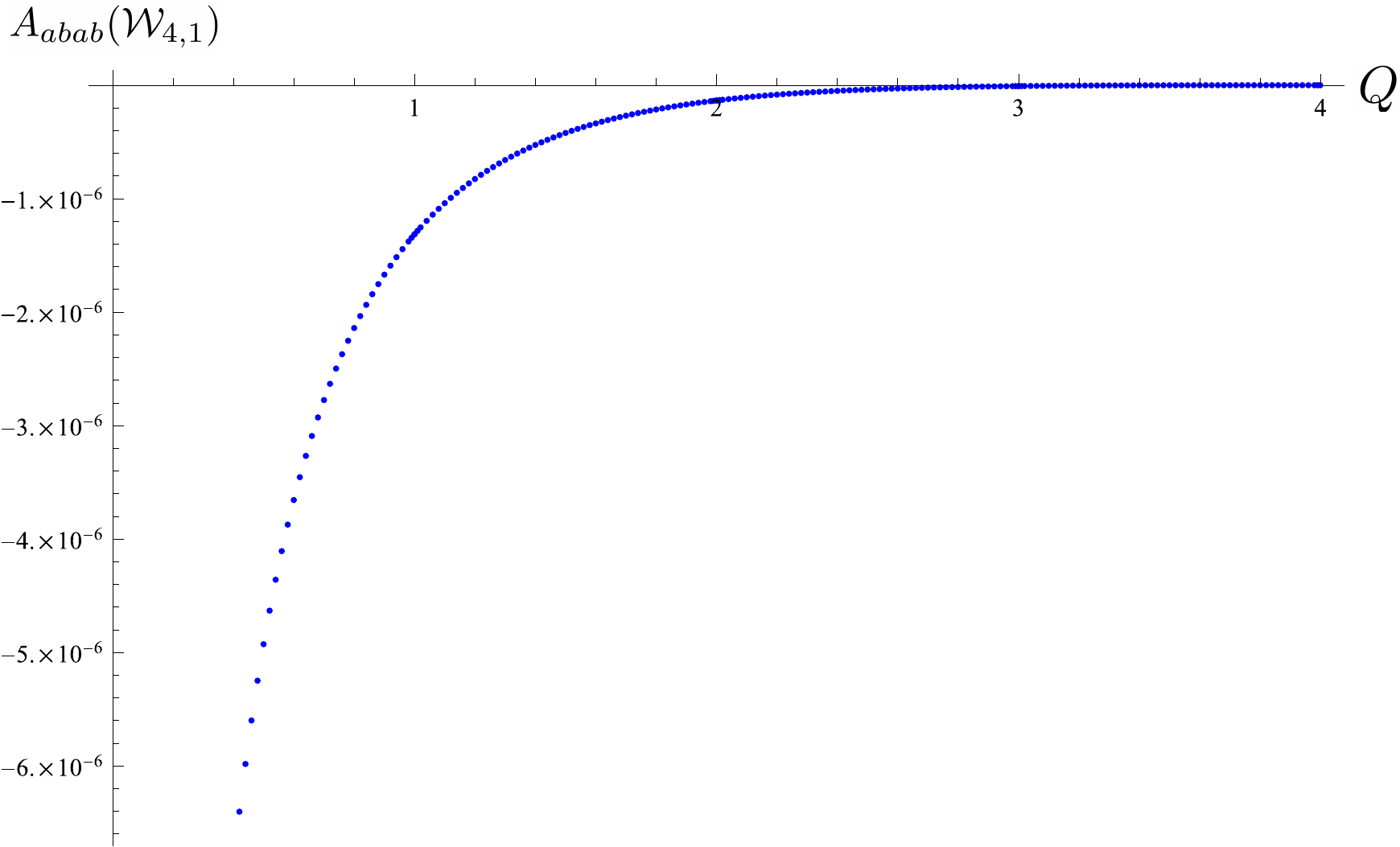}
			\caption{}
			\label{}
		\end{subfigure}
		\begin{subfigure}[h]{0.5\textwidth}
			\includegraphics[width=0.9\textwidth]{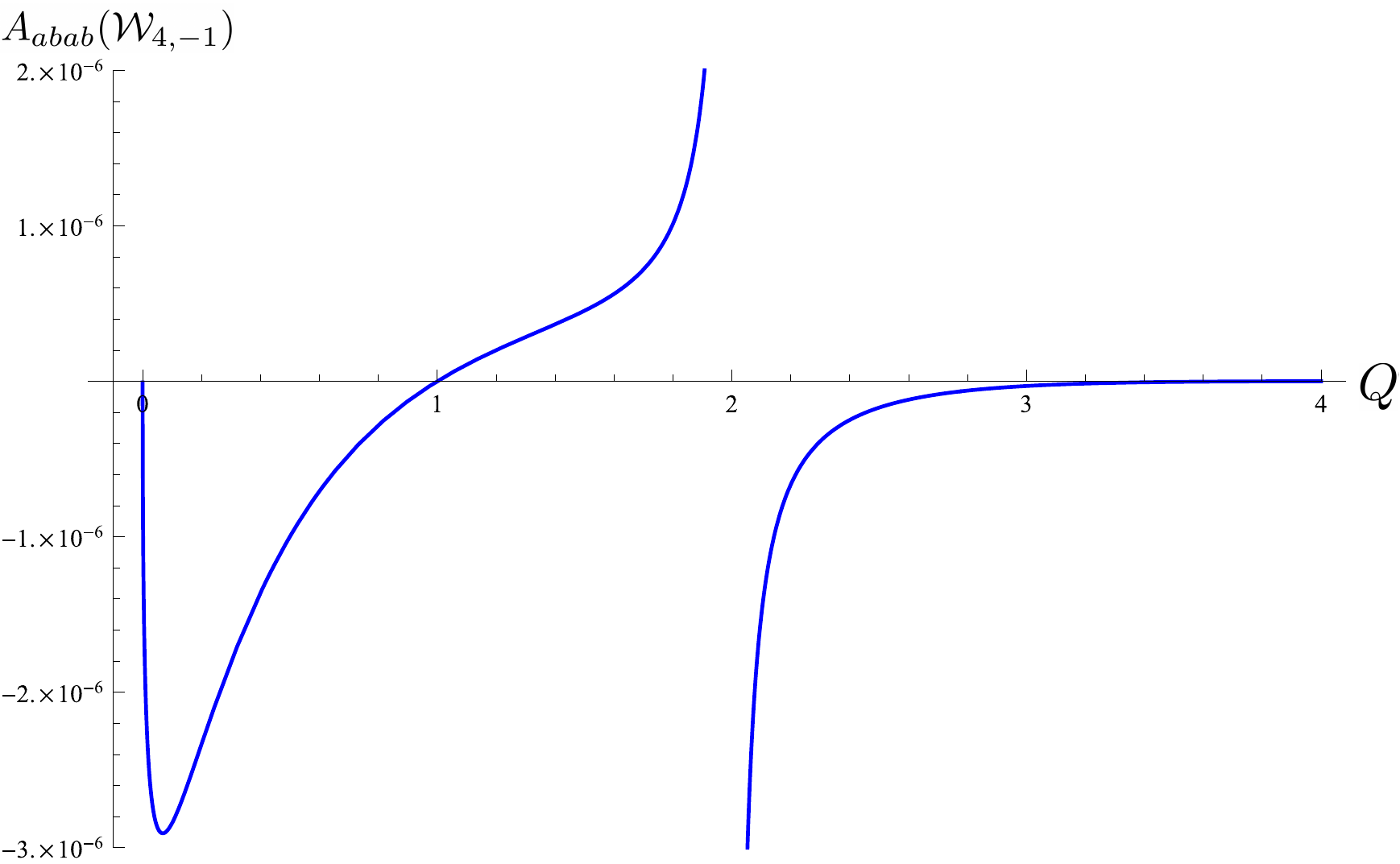}
			\caption{}
			\label{}
		\end{subfigure}
	\end{centering}
	\caption{The bootstrapped $A_{abab}(\mathcal{W}_{4,1})$ on the left and the analytic $A_{abab}(\mathcal{W}_{4,-1})$ on the right.}
	\label{abab404m1}
\end{figure}

The numerical $A_{abab}(\mathcal{W}_{4,1})$ and the analytic $A_{abab}(\mathcal{W}_{4,-1})$ are plotted in figure \ref{abab404m1}, where the latter are obtained using \eqref{A4m1ana} and \eqref{Ralphabar4m1}. Notice that the amplitude $A_{abab}(\mathcal{W}_{4,1})$ is smooth for $0<Q<4$, due to the cancellation of the zeros of $\mathsf{R}_{\bar{\alpha}}(\mathcal{W}_{4,1})$ in \eqref{Ralphabar41} with the poles in $A_{aaaa}(\mathcal{W}_{4,1})$, which further confirms that the singularities at \eqref{aaaa41poles2} and  \eqref{aaaa41poles3} in $A_{aaaa}(\mathcal{W}_{4,1})$ appear as simple poles.

\begin{figure}[H]
	\begin{centering}
		\begin{subfigure}{0.5\textwidth}
			\includegraphics[width=0.9\textwidth]{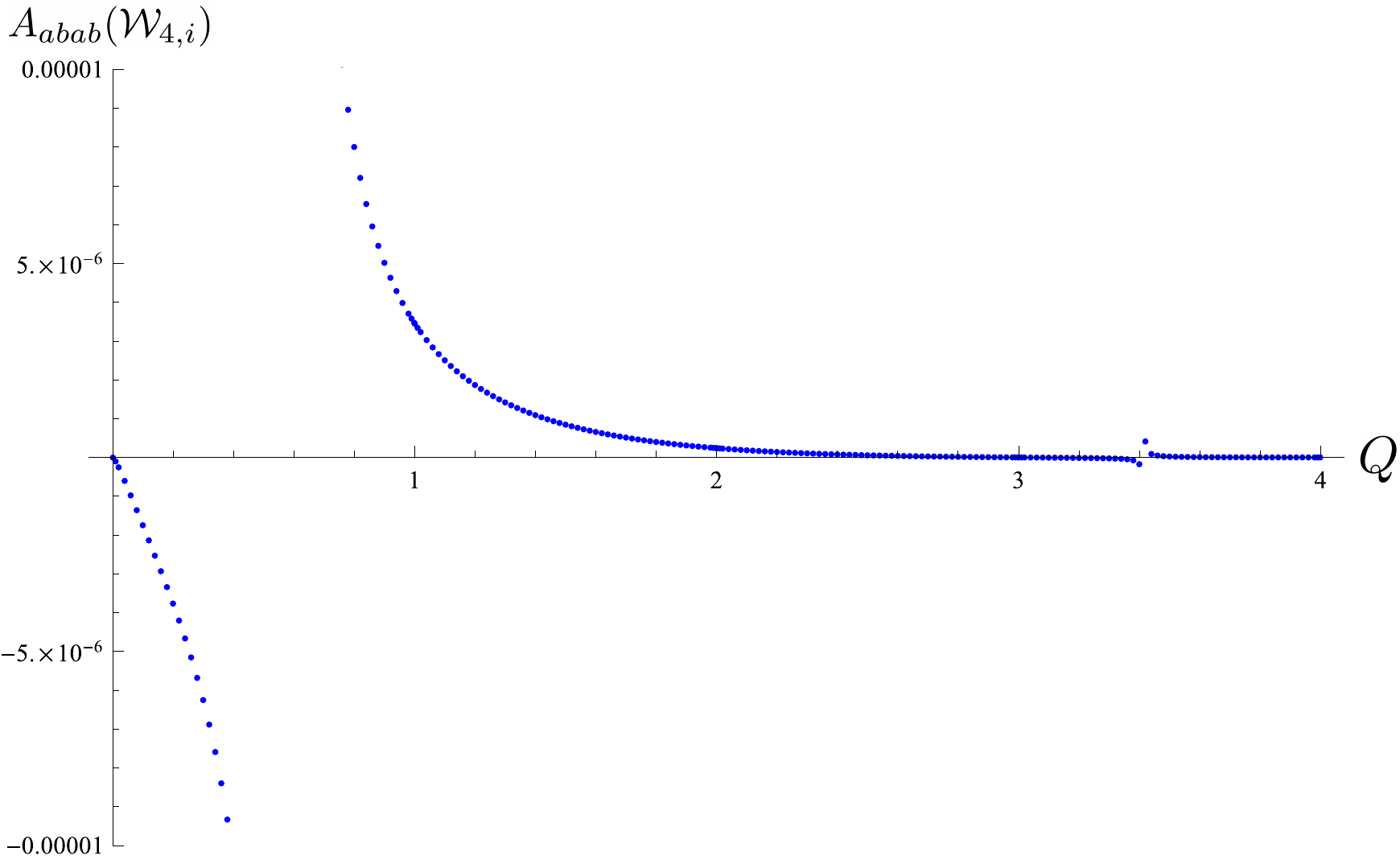}
		\end{subfigure}\\
		
		\begin{subfigure}{0.5\textwidth}
			\includegraphics[width=0.9\textwidth]{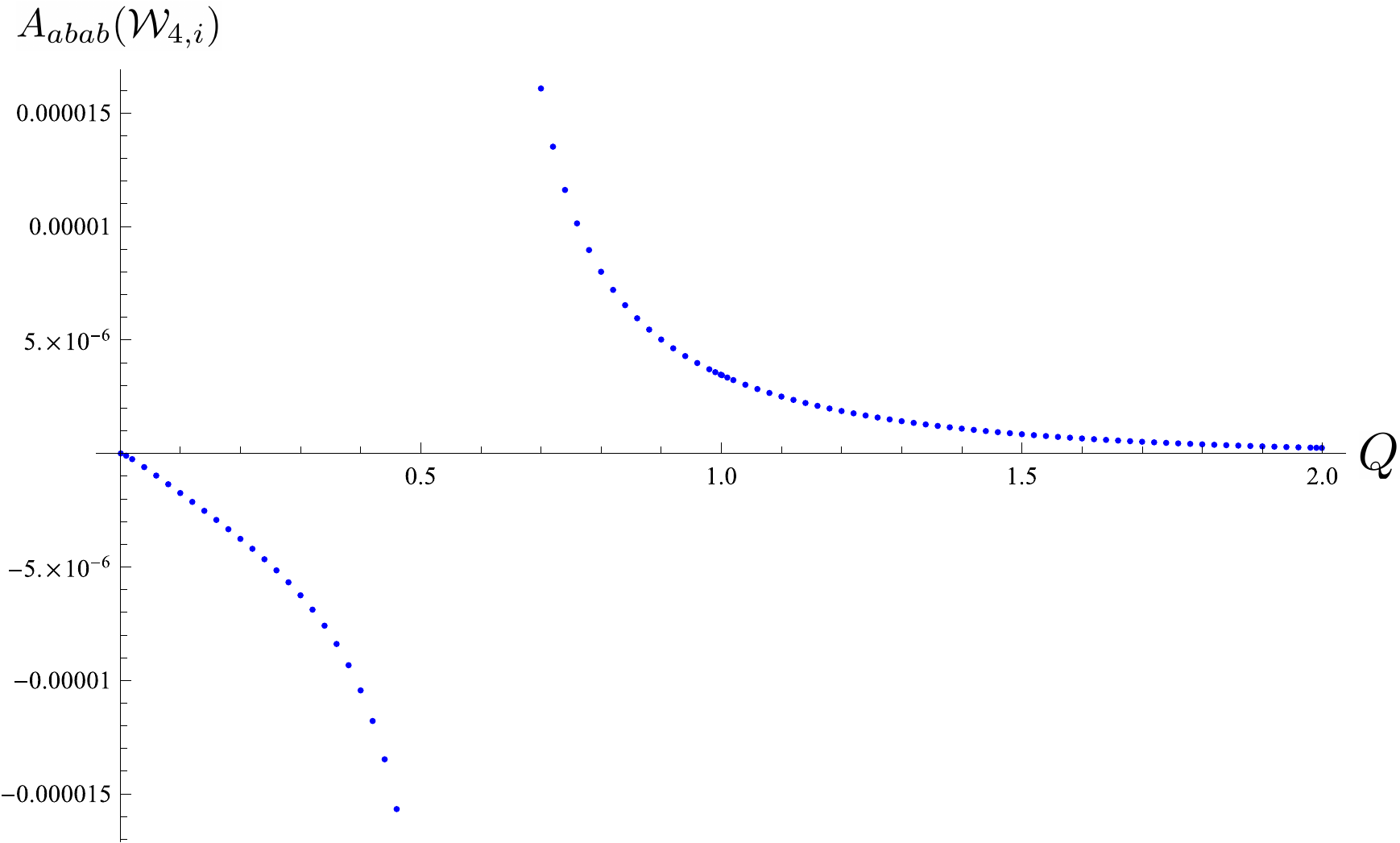}
		\end{subfigure}
		\begin{subfigure}{0.5\textwidth}
			\includegraphics[width=0.9\textwidth]{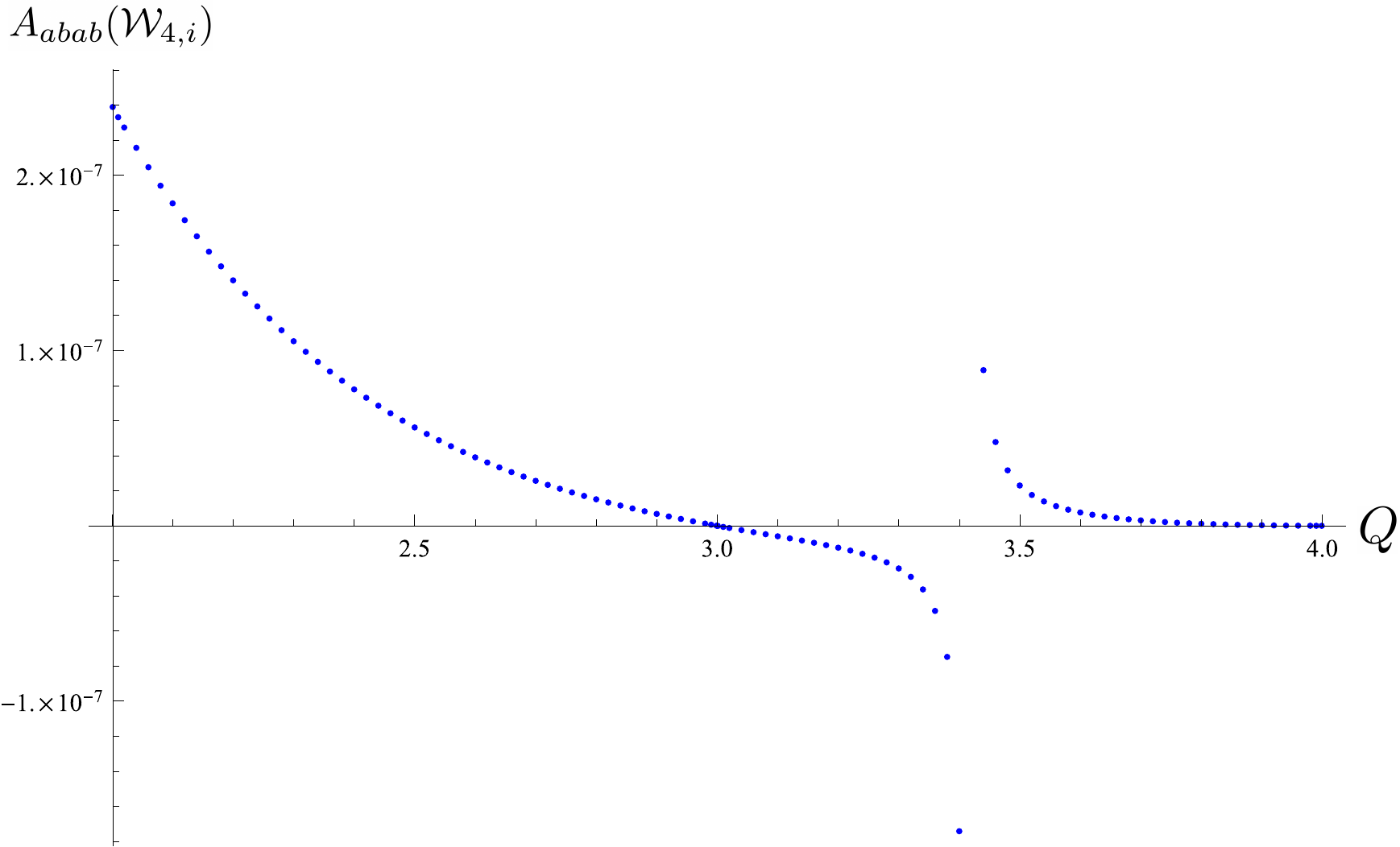}
		\end{subfigure}
	\end{centering}
	\caption{The bootstrap result of the amplitude $A_{abab}(\mathcal{W}_{4,i})$ and its detailed pole structures in the regions $0<Q<2$ and $2<Q<4$.}
	\label{abab4pi}
\end{figure}

\begin{figure}[H]
	\begin{centering}
		
		\begin{subfigure}{0.5\textwidth}
			\includegraphics[width=0.9\textwidth]{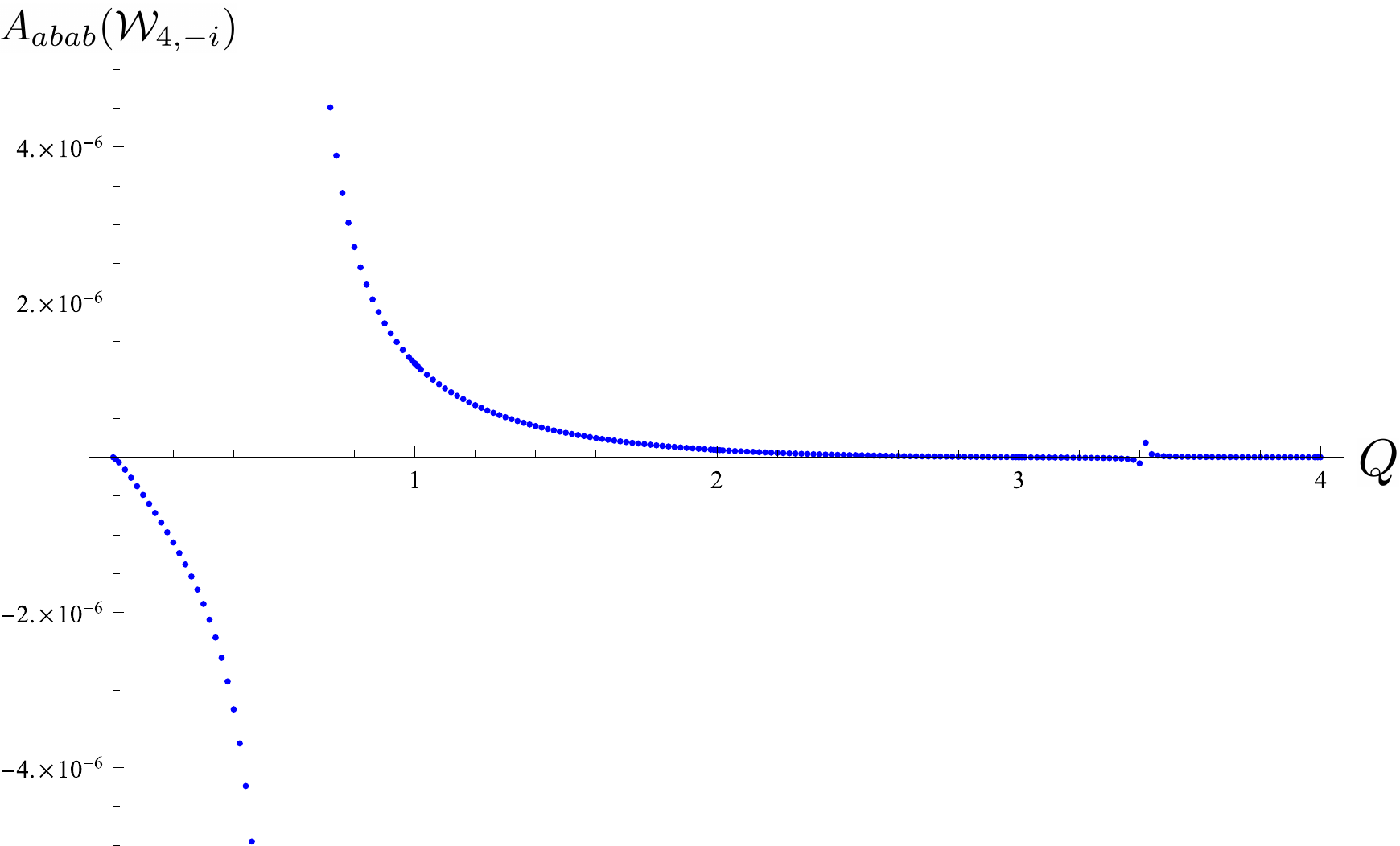}
		\end{subfigure}\\
		
		\begin{subfigure}{0.5\textwidth}
			\includegraphics[width=0.9\textwidth]{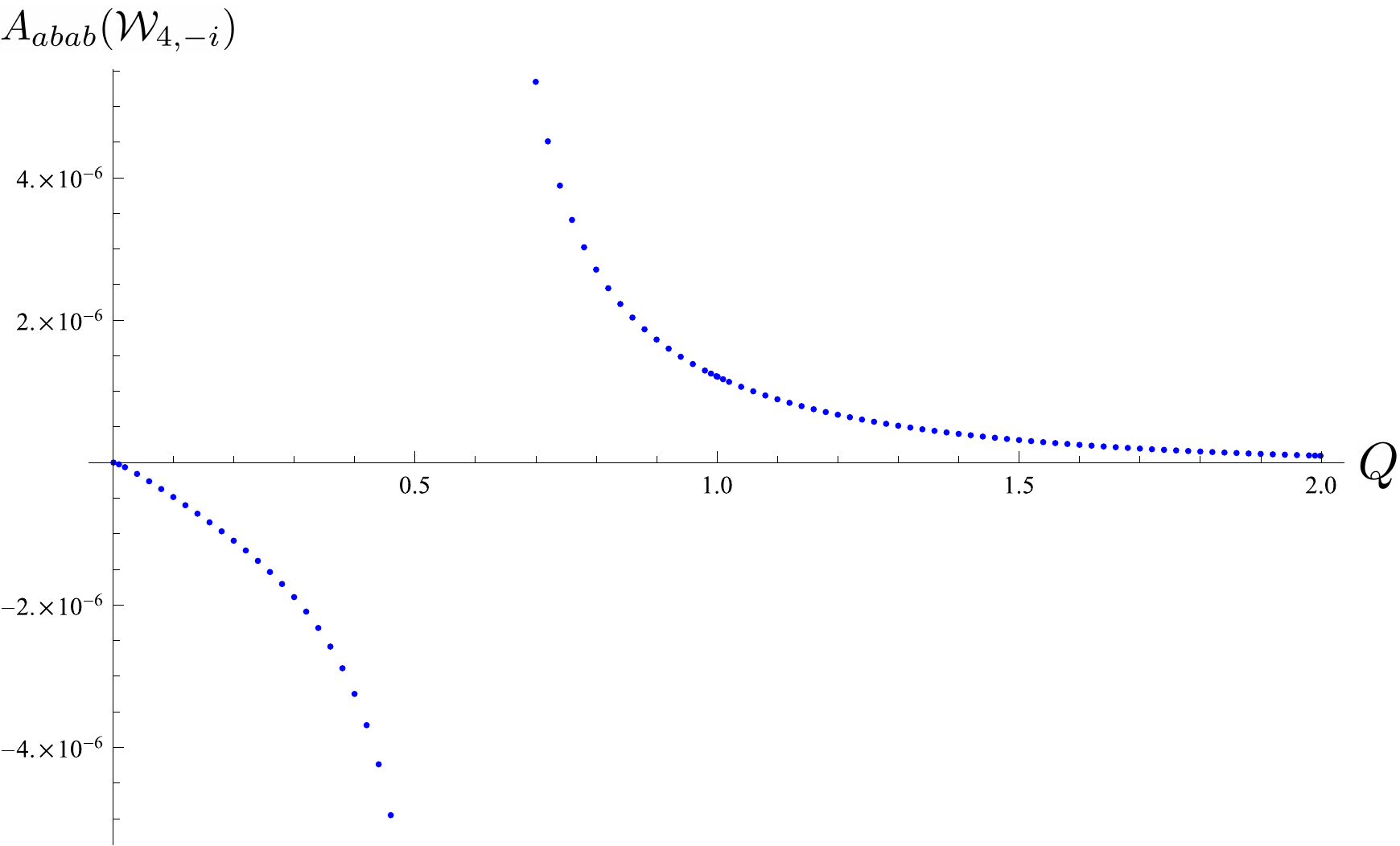}
		\end{subfigure}
		\begin{subfigure}{0.5\textwidth}
			\includegraphics[width=0.9\textwidth]{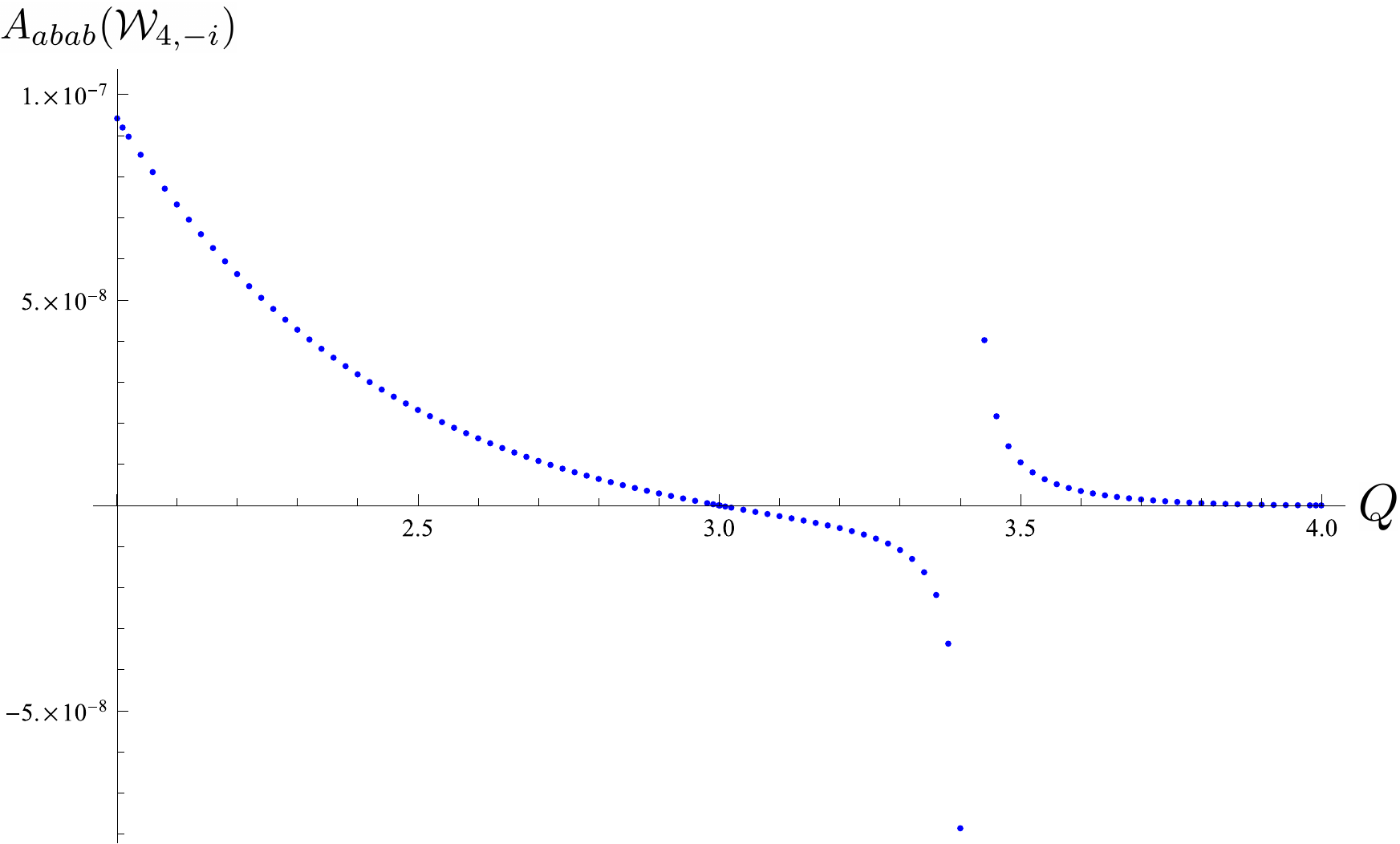}
		\end{subfigure}
	\end{centering}
	\caption{The bootstrap result of the amplitude $A_{abab}(\mathcal{W}_{4,-i})$ and its detailed pole structures in the regions $0<Q<2$ and $2<Q<4$.}
	\label{abab4mi}
\end{figure}

The modules $\mathcal{W}_{4,i}$ and $\mathcal{W}_{4,-i}$ only appear in $P_{abab}$ (and $P_{abba}$ with the same amplitude but the opposite sign) and hence were obtained purely through the numerical bootstrap. Recall that they are in fact related by \eqref{pm1pm}. The amplitudes display poles at
\begin{subequations}\label{abab4i1poles}
	\begin{eqnarray}
	&&Q=4\cos^2 \! \left( \frac{3\pi}{8} \right) =0.585786 \ldots \;,\\
	&&Q=4\cos^2 \! \left( \frac{\pi}{8} \right) =3.414213 \ldots \;.
	\end{eqnarray}
\end{subequations}
These poles were in fact already observed in \cite{Jacobsen:2018pti} for $A(h_{\frac{1}{4},4},h_{\frac{1}{4},-4})$ (i.e., $A_{abab}(\mathcal{W}_{4,i})$ in the present notation) which we will analyze in more details in the next section.
The results are plotted in figures \ref{abab4pi} and \ref{abab4mi} together with their detailed pole structures in the bottom parts of those figures.
 
\subsubsection{$A_{aabb}$}

\begin{figure}[H]
	\begin{centering}
	\begin{subfigure}{0.5\textwidth}
	\includegraphics[width=0.9\textwidth]{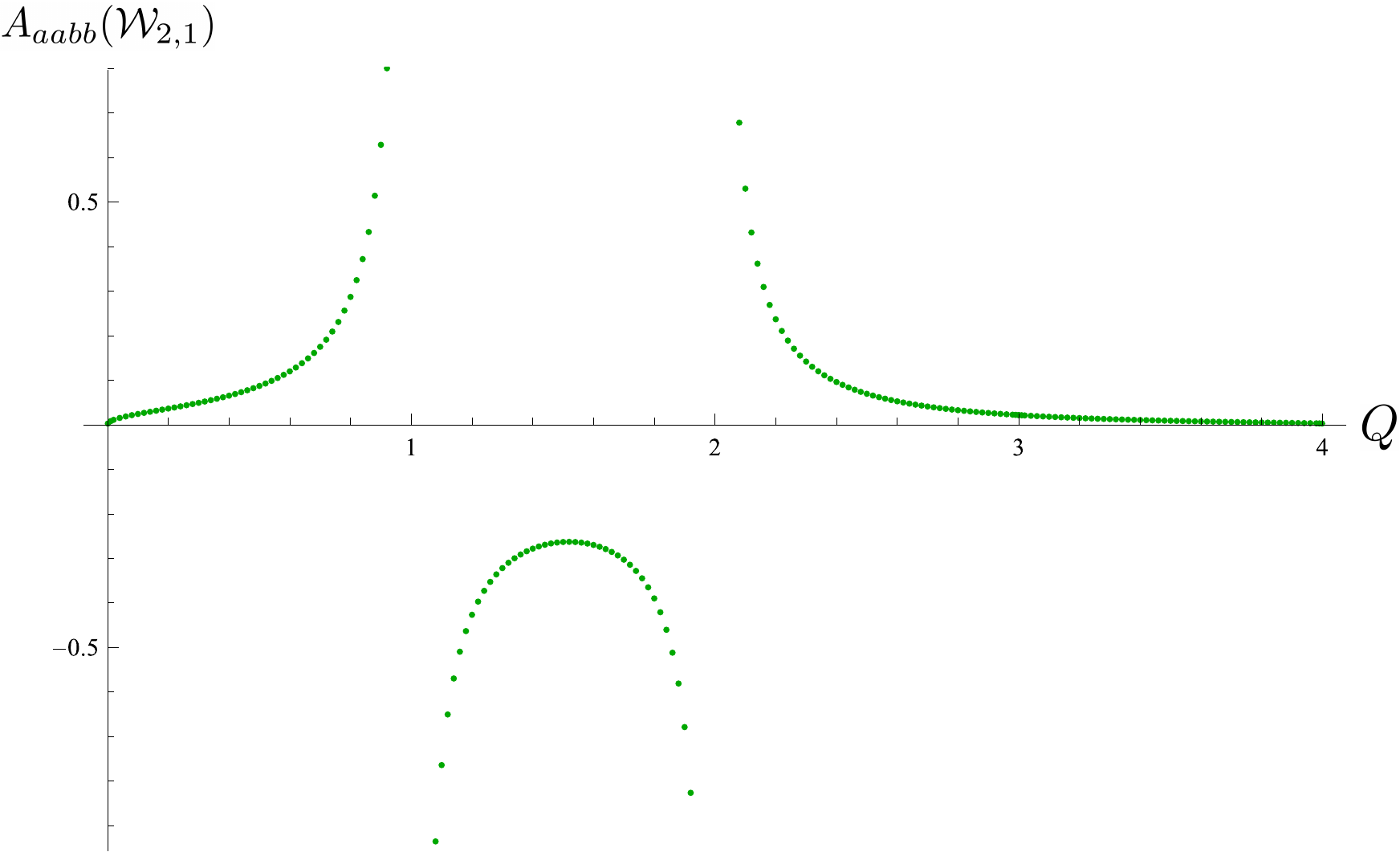}
	\caption{}
	\label{aabb21}
    \end{subfigure}
	\begin{subfigure}{0.5\textwidth}
	\includegraphics[width=0.9\textwidth]{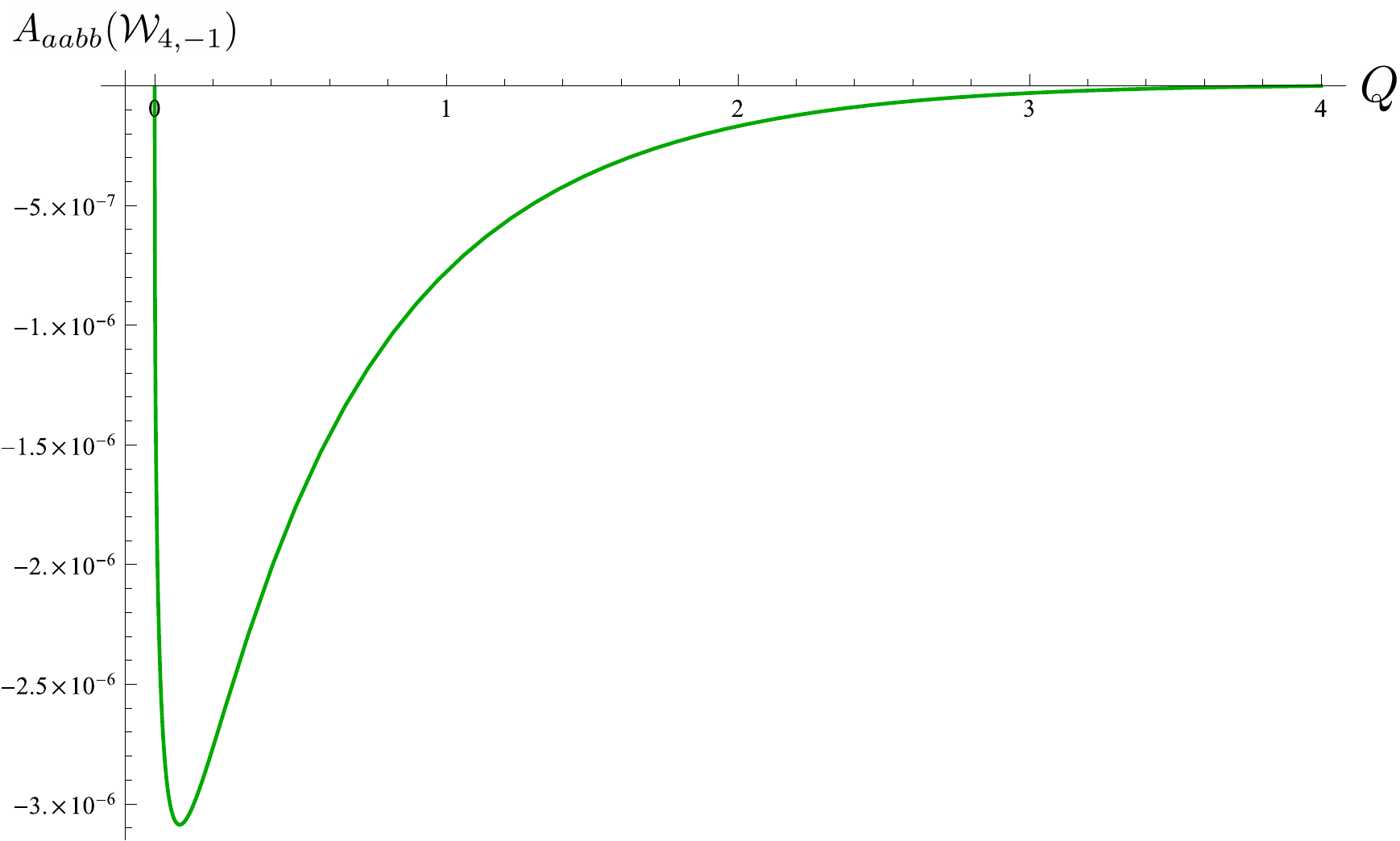}
	\caption{}
	\label{}
	\end{subfigure}
	\end{centering}
	\caption{The bootstrapped $A_{aabb}(\mathcal{W}_{2,1})$ on the left and the analytic $A_{aabb}(\mathcal{W}_{4,-1})$ on the right.}
	\label{aabb214m1}
\end{figure}

\begin{figure}[t]
	\begin{centering}
		
	\begin{subfigure}{0.5\textwidth}
	\includegraphics[width=\textwidth]{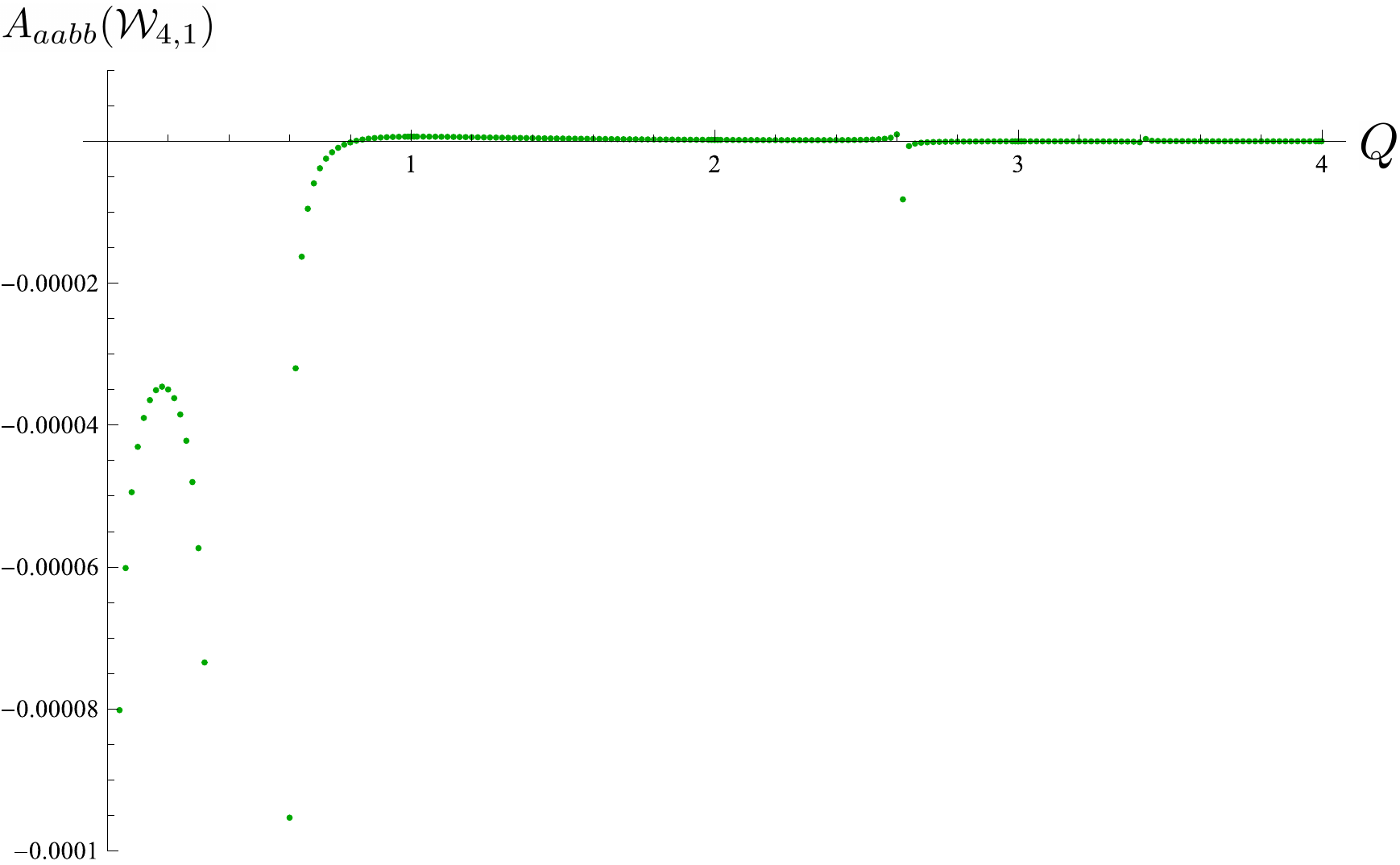}
	\end{subfigure}\\
		
	\begin{subfigure}{0.5\textwidth}
		\includegraphics[width=0.9\textwidth]{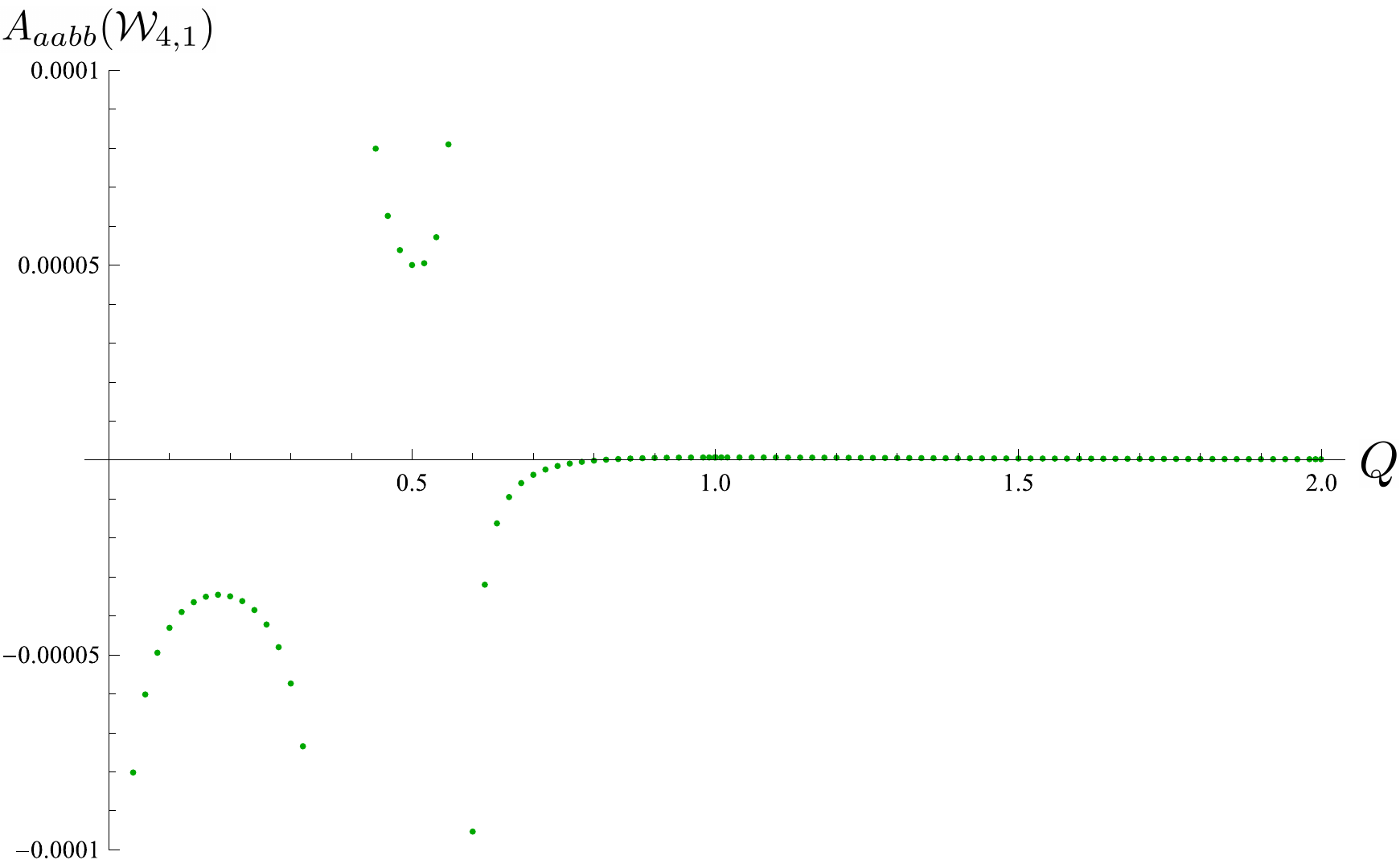}
    \end{subfigure}
	\begin{subfigure}{0.5\textwidth}
		\includegraphics[width=0.9\textwidth]{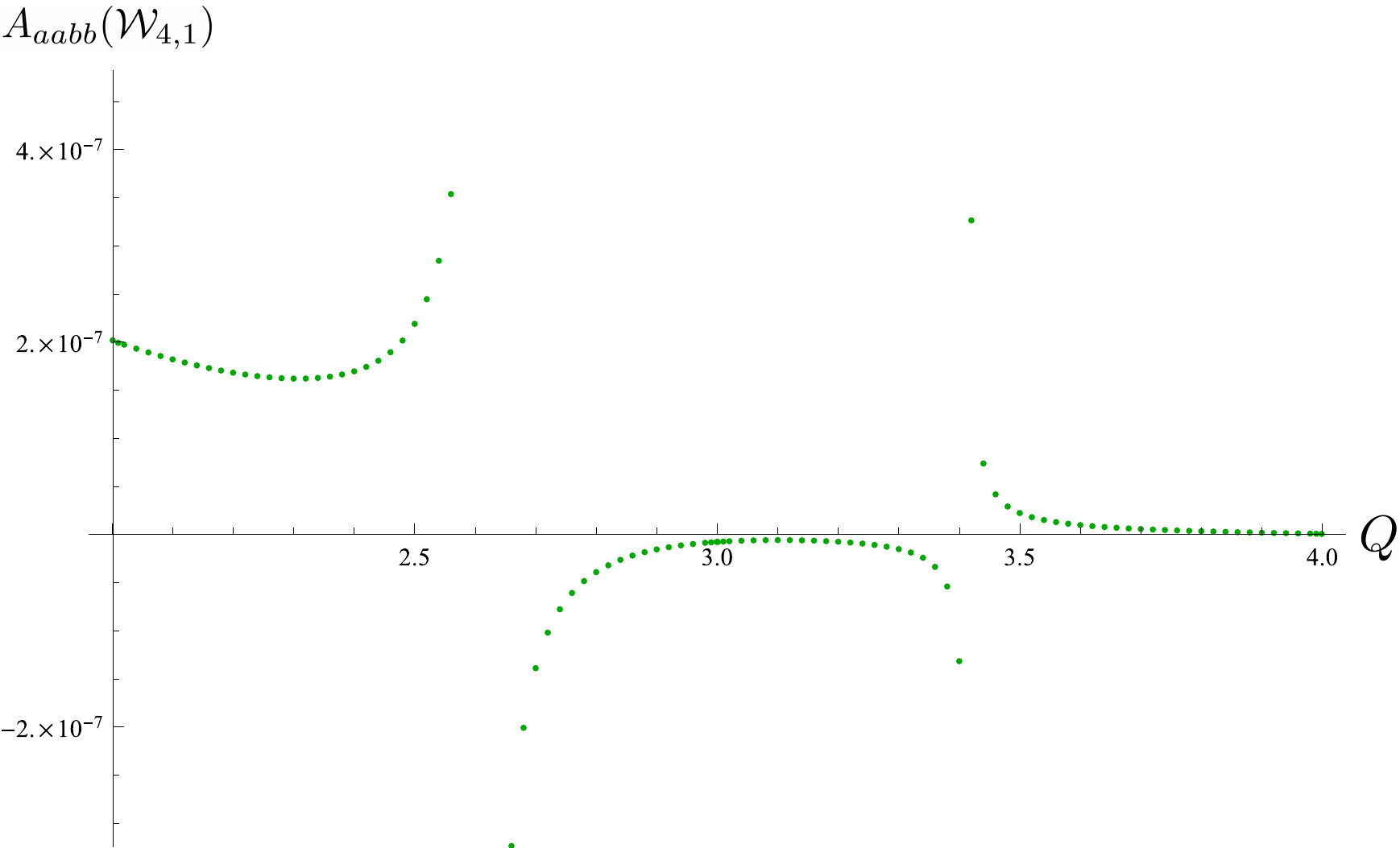}
	\end{subfigure}
	\end{centering}
	\caption{The bootstrap result of the amplitude $A_{aabb}(\mathcal{W}_{4,1})$ and its detailed pole structures in the regions $0<Q<2$ and $2<Q<4$.}
	\label{aabb41pole}
\end{figure}

In the probability $P_{aabb}$, we have the following amplitudes
\begin{equation}
\begin{aligned}
&A_{aabb}(\overline{\mathcal{W}}_{0,\q^2}),\;\;A_{aabb}(\mathcal{W}_{0,-1}),\\
&A_{aabb}(\mathcal{W}_{2,1}),\;\;\\
&A_{aabb}(\mathcal{W}_{4,-1}),\;\;A_{aabb}(\mathcal{W}_{4,1}).
\end{aligned}
\end{equation}
While the first amplitude provides the normalization \eqref{Anorm}, we plot last three in figures \ref{aabb214m1} and \ref{aabb41pole}. (Note here that $A_{aabb}(\mathcal{W}_{0,-1})$ is trivially related to $A_{aaaa}(\mathcal{W}_{0,-1})$ in figure \ref{aaaa0m1} by a minus sign as in \eqref{0m1alpha}.) The analytic structures of these amplitudes can be seen from that of $A_{aaaa}$ and $\mathsf{R}_{\alpha}$ from \eqref{Ralpha21}, \eqref{Ralpha4m1} and \eqref{Ralpha41}. In particular, $\mathsf{R}_{\alpha}(\mathcal{W}_{2,1})$ and $\mathsf{R}_{\alpha}(\mathcal{W}_{4,1})$ indicate new poles in $A_{aabb}(\mathcal{W}_{2,1})$ at $Q=1$ and
in $A_{aabb}(\mathcal{W}_{4,1})$ at
\begin{subequations}\label{aabbnewpoles}
	\begin{eqnarray}
	Q&=&4\cos^2 \! \left( \frac{2\pi}{5} \right) =0.381966 \ldots \;,\label{aabbnewpoles2}\\
	Q&=&4\cos^2 \! \left( \frac{\pi}{5} \right) =2.61803 \ldots \;.\label{aabbnewpoles3}
	\end{eqnarray}
\end{subequations}

\subsection{Singularities and exact amplitudes}\label{singexact}

As pointed out in \cite{Jacobsen:2018pti}, the proposal of \cite{Picco:2016ilr} cannot be the accurate description of the Potts geometrical correlations due to the appearance of divergences in $Q$ in their correlation functions, whereas the Potts probabilities are expected to be smooth functions in $Q$. The spectrum of \eqref{Pottsspectrum}, on the other hand, has the effect of canceling such unwanted singularities, as already studied in \cite{Jacobsen:2018pti} through an example. We now proceed further along this line to analyze in full detail the bootstrapped amplitudes that we have presented in the previous section. We will see that combining with the analytic amplitudes that we gave in section \ref{pp} and with the recursions that we established in section \ref{recursion}, this gives us exact amplitudes at special values of $Q$ corresponding to rational $\upbeta^2$ given by \eqref{upbetadef}. Such rational values of $\upbeta^2$ are currently not directly accessible to the numerical bootstrap.\footnote{The Zamolodchikov recursive formula for computing conformal blocks is singular at rational $\upbeta^2$ and therefore we do not bootstrap directly at the corresponding values of $Q$.} In the meantime, we will see the intricate interplay between the spectra involved in various Potts probabilities and the analytic structures in the amplitudes. This provides a CFT interpretation of some of the amplitude ratios in \eqref{ratioforbootstrap}, which were originally obtained as an observation in the lattice-model computations.


\subsubsection*{$A_{abab}(\mathcal{W}_{4,i})$}
In \cite{Jacobsen:2018pti}, it was argued that the leading field $(h_{\frac{1}{4},4},h_{\frac{1}{4},-4})$ is necessary in addition to the field $(h_{\frac{3}{2},2},h_{\frac{3}{2},-2})$ of the spectrum of \cite{Picco:2016ilr} in order for $P_{abab}$ to be a smooth function of $Q$. As our first case, we now make this analysis more precise and explain the poles in the amplitudes $A_{abab}(\mathcal{W}_{4,i})$ at \eqref{abab4i1poles}.

At $Q=4\cos^2 \! \left(\frac{3\pi}{8}\right)$, one finds a coincidence of conformal dimensions:
\begin{equation}
h_{\frac{1}{4},4}=\bar{h}_{1,2}, \quad \bar{h}_{\frac{1}{4},4}=\bar{h}_{\frac{3}{2},2}, \quad h_{\frac{3}{2},2}=h_{1,2}.
\end{equation}
The contribution of $(h_{\frac{3}{2},2},h_{\frac{3}{2},-2})$ in $P_{abab}$ therefore has a divergent term
\begin{equation}
A_{abab}(\mathcal{W}_{2,-1})\mathcal{R}_{\frac{3}{2},2}\text{Re}\left[\mathcal{F}_{h_{\frac{3}{2},2}}(z)\mathcal{F}_{h_{\frac{3}{2},-2}}(\bar{z})\right]=A_{abab}(\mathcal{W}_{2,-1})\mathcal{R}_{\frac{3}{2},2}\frac{\mathrm{R}_{1,2}}{h_{\frac{3}{2},2}-h_{1,2}}\text{Re}\left[\mathcal{F}_{h_{-1,2}}(z)\mathcal{F}_{{h}_{\frac{3}{2},-2}}(\bar{z})\right]+\ldots\,,
\end{equation}
where we have used \eqref{superA}, \eqref{superjm1}, \eqref{Rjm1} and \eqref{reg1}. The divergence is necessarily canceled by
\begin{equation}
A_{abab}(\mathcal{W}_{4,i})\text{Re}\left[\mathcal{F}_{h_{\frac{1}{4},4}}(z)\mathcal{F}_{h_{\frac{1}{4},-4}}(\bar{z})\right],
\end{equation}
which requires
\begin{equation}
A_{abab}(\mathcal{W}_{4,i})=-A_{abab}(\mathcal{W}_{2,-1})\mathcal{R}_{\frac{3}{2},2}\frac{\mathrm{R}_{1,2}}{h_{\frac{3}{2},2}-h_{1,2}}+O(1).
\end{equation}
Extracting the residue, we obtain at $Q=4\cos^2 \! \left( \frac{3\pi}{8} \right)$:
\begin{equation}\label{W4ires1}
\text{Res}\left[A_{abab}(\mathcal{W}_{4,i})\right]|_{Q=4\cos^2\frac{3\pi}{8}}=-A^L(\mathcal{W}_{0,-1})\frac{174607744311 \pi  \Gamma (-\frac{56}{5}) \Gamma
	(-\frac{3}{4}) \Gamma (\frac{1}{8})
	\Gamma (\frac{4}{5}) \Gamma
	(\frac{9}{8}) \Gamma
	(\frac{19}{10})}{5777653760000 \sqrt[5]{2} \Gamma
	(-\frac{34}{5}) \Gamma (-\frac{9}{8})
	\Gamma (-\frac{1}{8}) \Gamma
	(\frac{3}{4}) \Gamma (\frac{11}{10})
	\Gamma (\frac{11}{5})},\;\;
\end{equation}
where we have used the explicit expression \eqref{abab2m1ana} of $\mathcal{R}_{\frac{3}{2},2}$, and the expression of $A^L(\mathcal{W}_{0,-1})$ is given in \eqref{A0m1} in appendix \ref{AL}.

Similarly, at $Q=4\cos^2 \! \left(\frac{\pi}{8}\right)$, one finds
\begin{equation}
\bar{h}_{\frac{1}{4},4}=\bar{h}_{2,2}, \quad h_{\frac{1}{4},4}=\bar{h}_{\frac{3}{2},2}, \quad h_{\frac{3}{2},2}=h_{2,2}.
\end{equation}
A completely parallel calculation to the above leads to the exact result
\begin{equation}
A_{abab}(\mathcal{W}_{4,i})=-A_{abab}(\mathcal{W}_{2,-1})\mathcal{R}_{\frac{3}{2},2}\frac{\mathrm{R}_{2,2}}{h_{\frac{3}{2},2}-h_{2,2}}+O(1)
\end{equation}
and explicitly:
\begin{equation}\label{W4ires2}
\text{Res}\left[A_{abab}(\mathcal{W}_{4,i})\right]|_{Q=4\cos^2\frac{\pi}{8}}=A^L(\mathcal{W}_{0,-1})\frac{1932805 \pi  \Gamma (-\frac{5}{4}) \Gamma
	(\frac{5}{14}) \Gamma (\frac{11}{8})^2
	\Gamma (\frac{25}{14}) \Gamma
	(\frac{33}{14})}{501377302265856 \sqrt[7]{2} \Gamma
	(-\frac{19}{14}) \Gamma (-\frac{3}{8})^2
	\Gamma (\frac{9}{14}) \Gamma
	(\frac{17}{14}) \Gamma (\frac{5}{4})}.
\end{equation}

\begin{figure}[H]
	\begin{centering}
		\begin{subfigure}{0.5\textwidth}
			\includegraphics[width=0.95\textwidth]{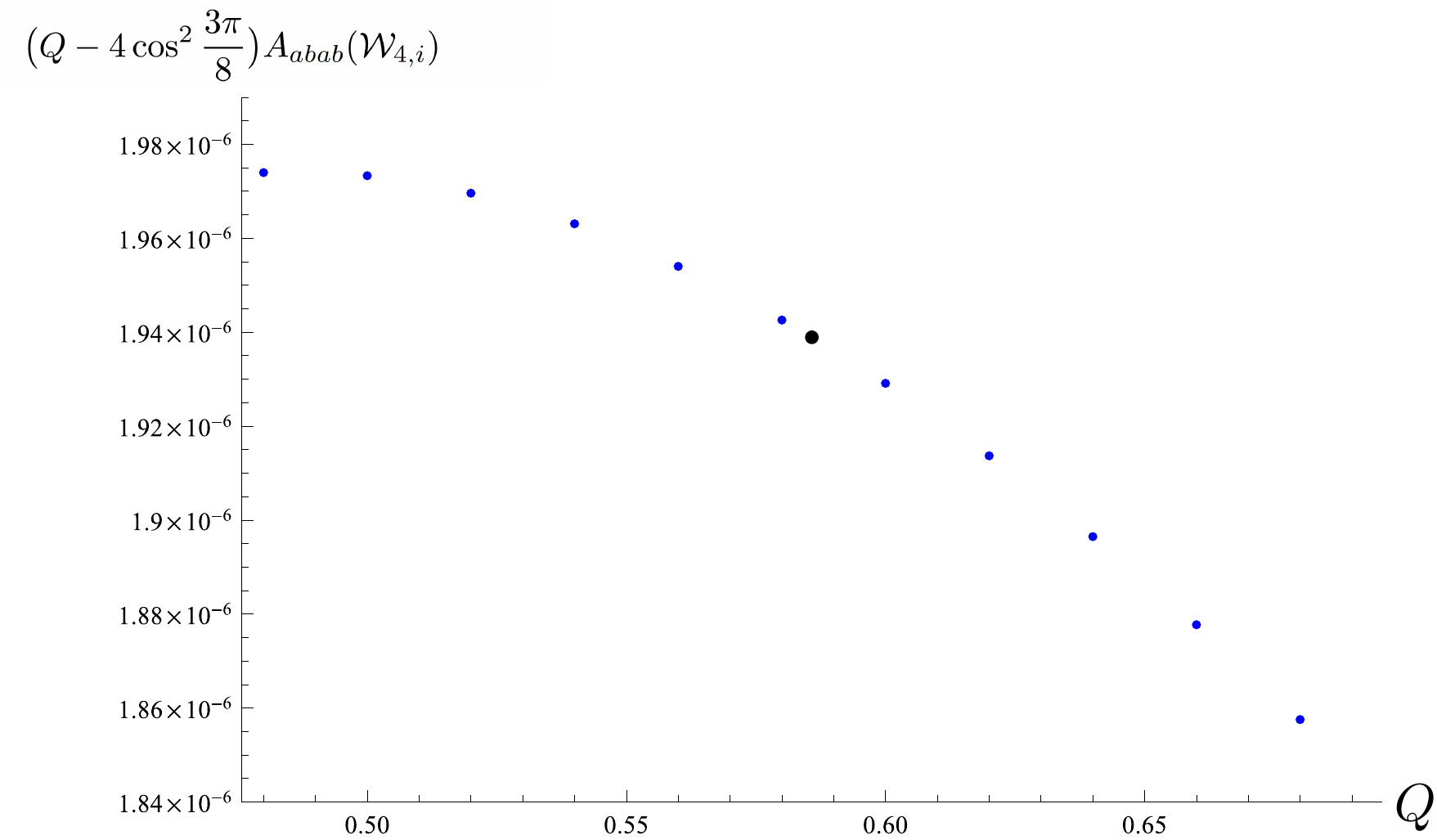}
		\end{subfigure}
		\begin{subfigure}{0.5\textwidth}
			\includegraphics[width=0.95\textwidth]{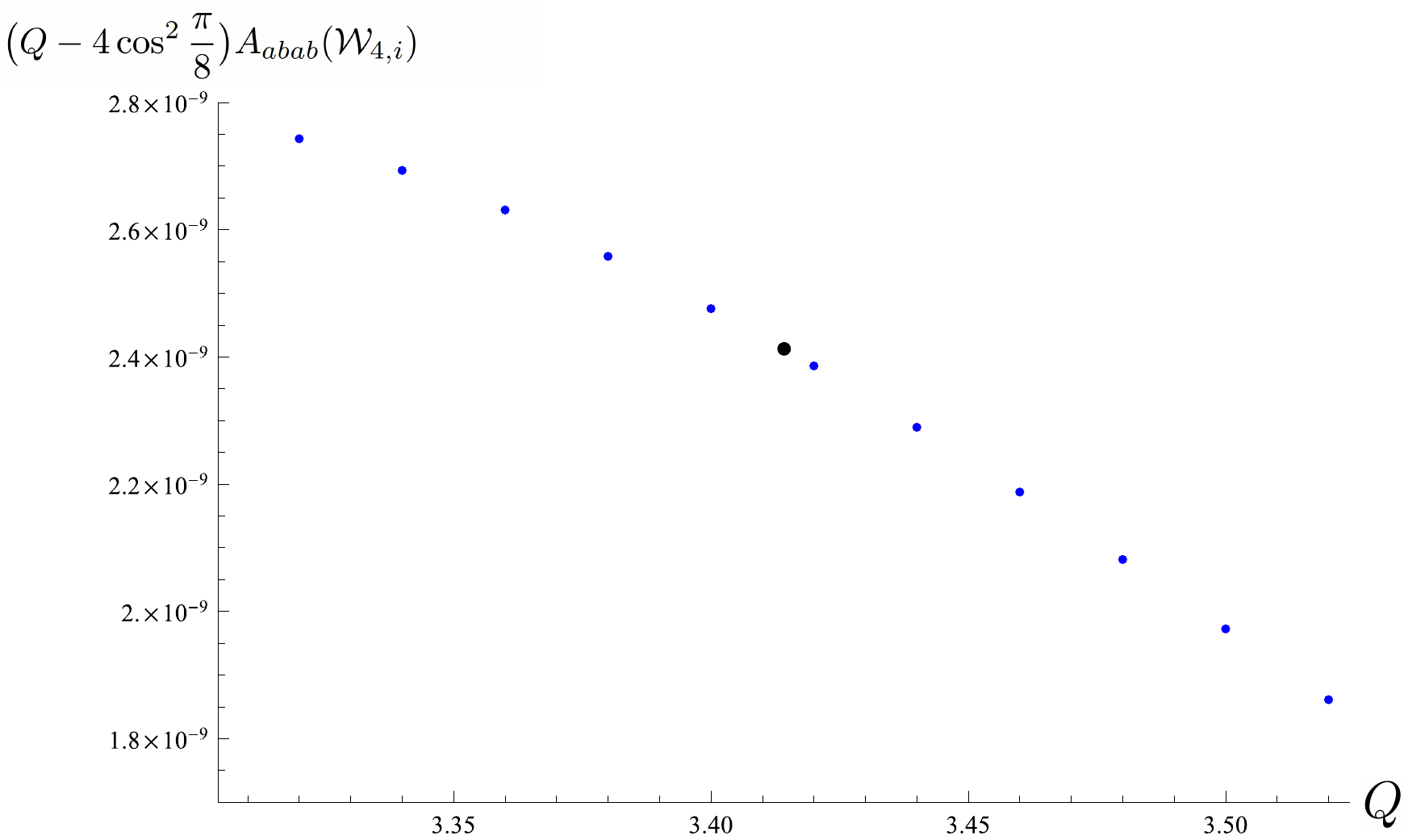}
		\end{subfigure}
	\end{centering}
	\caption{The residues of the amplitude $A_{abab}(\mathcal{W}_{4,i})$ at $Q=4\cos^2 \! \left( \frac{3\pi}{8} \right)$ (left) and $Q=4\cos^2 \! \left( \frac{\pi}{8} \right)$ (right) given by the exact expressions \eqref{W4ires1} and \eqref{W4ires2} are indicated with black dots. The slightly smaller {\color{blue}blue} dots are the bootstrap results in the nearby region.}
	\label{W4ires}
\end{figure}

In figure \ref{W4ires}, we plot \eqref{W4ires1} and \eqref{W4ires2} together with the bootstrap results in the respective regions of $Q$. As can be seen, the exact results interpolate smoothly between the numerical bootstrap results.

\bigskip

While the above analysis focuses on a single probability $P_{abab}$, in the following we consider the compa\-rison of analytic structures of the amplitudes in different probabilities which are explicitly related by \eqref{ratioforbootstrap}. We will focus on how such differences are related to the corresponding differences of the spectra in \eqref{Pottsspectrum}. This will give an analytic explanation of some of the ratios $\mathsf{R}$, as well as exact results on the amplitudes.

\subsubsection*{$\mathsf{R}_{\alpha}(\mathcal{W}_{2,1})$ and $\overline{\mathcal{W}}_{0,\q^2}$}
Consider now the amplitudes $A_{aabb}(\mathcal{W}_{2,1})$ in figure \ref{aabb21}. Compared to $A_{aaaa}(\mathcal{W}_{2,1})$, it has an extra pole at $Q=1$, as can be seen explicitly from the ratio $\mathsf{R}_{\alpha}(\mathcal{W}_{2,1})$ in \eqref{Ralpha21}. One naturally suspects that such difference in the analytic structure is directly related to the difference in the spectra of the two probabilities involved, the module $\overline{\mathcal{W}}_{0,\q^2}$ in this case. Indeed, at $Q=1$, one finds a collision of the conformal dimensions
\begin{equation}
h_{1,1}=\bar{h}_{1,1}=h_{1,2}.
\end{equation}
This means that the left and right conformal blocks for the identity field include the following divergent term:\footnote{We have in this section omitted the superscript of the conformal blocks indicating the channels. The arguments here apply to either one of the $s$- and $t$-channels whose blocks are related by \eqref{st}.}
\begin{subequations}\label{block11}
	\begin{eqnarray}
	\mathcal{F}_{h_{1,1}}(z)&=&\tilde{\mathcal{F}}_{h_{1,2}}(z)+\frac{\mathrm{R}_{1,2}}{h_{1,1}-h_{1,2}}\mathcal{F}_{h_{1,-2}}(z),\\
	\mathcal{F}_{\bar{h}_{1,1}}(\bar{z})&=&\tilde{\mathcal{F}}_{h_{1,2}}(\bar{z})+\frac{\mathrm{R}_{1,2}}{\bar{h}_{1,1}-h_{1,2}}\mathcal{F}_{h_{1,-2}}(\bar{z}).
	\end{eqnarray}
\end{subequations}
Note that the $\tilde{\mathcal{F}}$ still have divergences due to the coincidence of $(h_{1,1},\bar{h}_{1,1})$ with other fields, with however different $z,\bar{z}$-dependence. 
With the normalization $A_{aabb}(\overline{\mathcal{W}}_{0,\q^2})=1$ from \eqref{Anorm}, the identity field enters the $s$-channel of $P_{aabb}$ as
\begin{equation}\label{11blockQ1}
\mathcal{F}_{h_{1,1}}(z)\mathcal{F}_{\bar{h}_{1,1}}(\bar{z})=\frac{\mathrm{R}^2_{1,2}}{(h_{1,1}-h_{1,2})^2}\mathcal{F}_{h_{1,-2}}(z)\mathcal{F}_{h_{1,-2}}(\bar{z})+\frac{2\mathrm{R}_{1,2}}{h_{1,1}-h_{1,2}}\text{Re}\left[\tilde{\mathcal{F}}_{h_{1,2}}(z)\mathcal{F}_{h_{1,-2}}(\bar{z})\right]+ \ldots \,.
\end{equation}

First, notice that the double pole in the first term is canceled exactly within the block of $\overline{\mathcal{W}}_{0,\q^2}$. Due to the coincident dimensions
\begin{equation}
h_{3,1}=\bar{h}_{3,1}=h_{1,-2},
\end{equation}
the block $\mathbb{F}_{0,\q^2}$ from \eqref{superB0q} includes the term
\begin{equation}
\mathcal{R}_{3,1}\mathcal{F}_{h_{3,1}}(z)\mathcal{F}_{h_{3,1}}(\bar{z})=\mathcal{R}_{3,1}\mathcal{F}_{h_{1,-2}}(z)\mathcal{F}_{h_{1,-2}}(\bar{z}),
\end{equation}
where $\mathcal{R}_{3,1}$ has a double pole at $Q=1$ whose residue 
cancels the residue of $\frac{\mathrm{R}^2_{1,2}}{(h_{1,1}-h_{1,2})^2}$ exactly, as can be easily checked.

Now, in order to cancel the simple pole in the second term of \eqref{11blockQ1}, it is necessary for the amplitude of $(h_{1,2},h_{1,-2})$ at $Q=1$ to be of the form
\begin{equation}
A_{aabb}(h_{1,2},h_{1,-2})=-\frac{\mathrm{R}_{1,2}}{h_{1,1}-h_{1,2}}+O(1),
\end{equation}
where we recall the identification \eqref{AAA} to account for the factor of 2. Notice that the blocks in the second term of \eqref{11blockQ1} are precisely the regular part of the blocks for the field $(h_{1,2},h_{1,-2})$ after removing the pole, as described in \eqref{reg1}--\eqref{reg2}.\footnote{Even though our treatment of conformal blocks for fields with degenerate indices is not exact due to the logarithmic structure, we believe the regular part is accurate.}
We then deduce
\begin{equation}
A_{aabb}(\mathcal{W}_{2,1})=-\frac{\mathrm{R}_{1,2}}{(h_{1,1}-h_{1,2})R^N_{0,2}}+O(1) \,,
\end{equation}
where $R^N_{0,2}$ is given by the recursion \eqref{RshiftNdeg}. Now, using \eqref{Ralpha21}, we obtain the exact amplitude $A_{aaaa}(\mathcal{W}_{2,1})$ at $Q=1$:
\begin{equation}\label{Aaaaa21Q1}
\left. A_{aaaa}(\mathcal{W}_{2,1}) \right|_{Q=1}
=\frac{5\pi\Gamma(-\frac{5}{4})\Gamma(\frac{7}{4})}{144\sqrt{3}\Gamma(\frac{1}{4})\Gamma(\frac{5}{4})}\;.
\end{equation}
In figure \ref{aaaa20anaQ1}, we plot the value of \eqref{Aaaaa21Q1} together with the bootstrapped amplitude $A_{aaaa}(\mathcal{W}_{2,1})$ in the region around $Q=1$.

\bigskip

We have seen above that from the CFT point of view, the amplitude ratio $\mathsf{R}_{\alpha}(\mathcal{W}_{2,1})$ is necessary to introduce the pole at $Q=1$ in the amplitude $A_{aabb}(\mathcal{W}_{2,1})$, in order to cancel the simple pole generated by the conformal blocks of $\overline{\mathcal{W}}_{0,\q^2}$ appearing in the $s$-channel of $P_{aabb}$. This picture is quite generic as we shall now see in another example.

\subsubsection*{$\mathsf{R}_{\bar{\alpha}}(\mathcal{W}_{2,1})$ and $\mathcal{W}_{0,-1}$}

\begin{figure}[H]
	\begin{centering}
		\begin{subfigure}{0.5\textwidth}
			\includegraphics[width=0.95\textwidth]{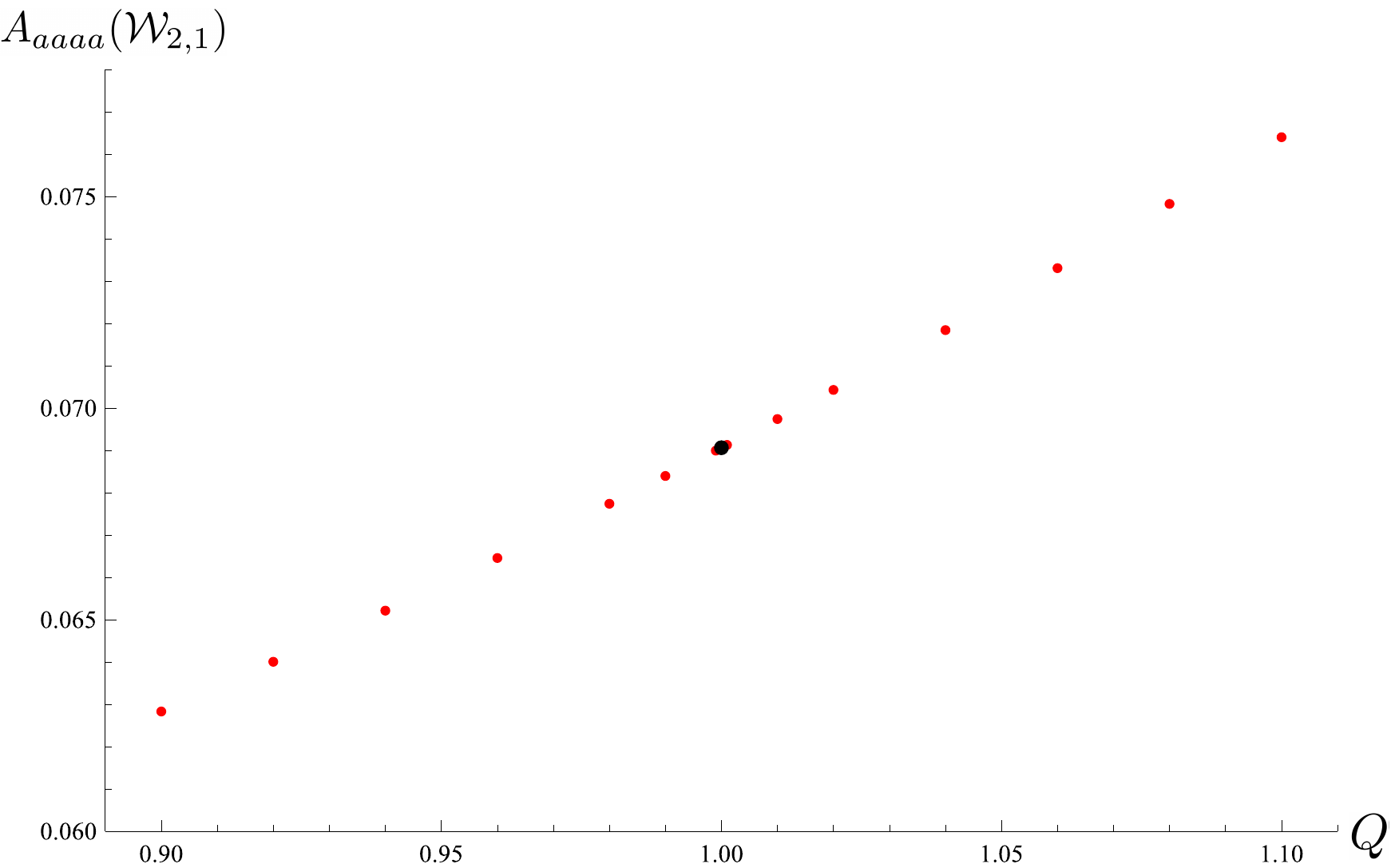}
			\caption{}
			\label{aaaa20anaQ1}
		\end{subfigure}
		\begin{subfigure}{0.5\textwidth}
			\includegraphics[width=0.95\textwidth]{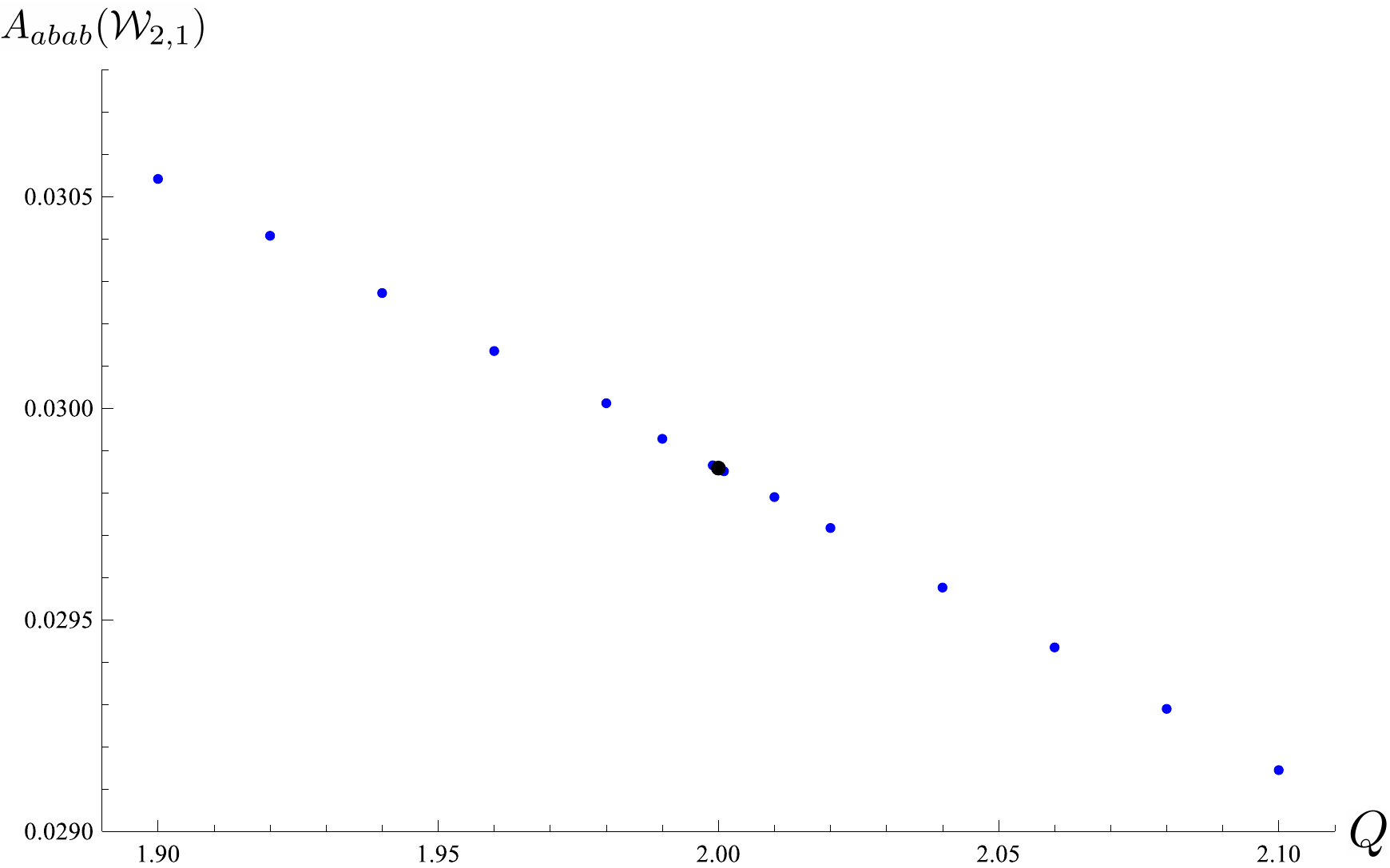}
			\caption{}
			\label{aaaa20anaQ2}
		\end{subfigure}
	\end{centering}
	\caption{The amplitudes $A_{aaaa}(\mathcal{W}_{2,1})$ at $Q=1$ (left) and $A_{abab}(\mathcal{W}_{2,1})$ at $Q=2$ (right). The {\color{red}red} and {\color{blue}blue} dots are the bootstrap results and the slightly bigger black dots are the exact expressions \eqref{Aaaaa21Q1} and \eqref{Aabab21Q2} obtained from the requirement of singularity cancellations.}
	\label{}
\end{figure}

From figures \ref{aaaa21} and \ref{abab21}, one can see that the amplitudes $A_{aaaa}(\mathcal{W}_{2,1})$ has a pole at $Q=2$ which is canceled by $\mathsf{R}_{\bar{\alpha}}(\mathcal{W}_{2,1})$ of \eqref{Rbar21} in $A_{abab}(\mathcal{W}_{2,1})$. This difference could easily be understood from the participation of the module $\mathcal{W}_{0,-1}$ in $P_{aaaa}$. At $Q=2$, one finds
\begin{equation}
h_{\frac{1}{2},0}=\bar{h}_{\frac{1}{2},0}=h_{1,2}=h_{2,2}
\end{equation}
leading to the following contribution to $P_{aaaa}$:
\begin{equation}
\begin{aligned}
\mathcal{F}_{h_{\frac{1}{2},0}}(z)\mathcal{F}_{\bar{h}_{\frac{1}{2},0}}(\bar{z})=&\frac{\mathrm{R}^2_{1,2}}{(h_{\frac{1}{2},0}-h_{1,2})^2}\mathcal{F}_{h_{1,-2}}(z)\mathcal{F}_{h_{1,-2}}(\bar{z})+\frac{\mathrm{R}^2_{2,2}}{(h_{\frac{1}{2},0}-h_{2,2})^2}\mathcal{F}_{h_{2,-2}}(z)\mathcal{F}_{h_{2,-2}}(\bar{z})\\
&+\frac{2\mathrm{R}_{1,2}}{h_{\frac{1}{2},0}-h_{1,2}}\text{Re}\left[\tilde{\mathcal{F}}_{h_{1,2}}(z)\mathcal{F}_{h_{1,-2}}(\bar{z})\right]+\frac{2\mathrm{R}_{2,2}}{h_{\frac{1}{2},0}-h_{2,2}}\text{Re}\left[\tilde{\mathcal{F}}_{h_{2,2}}(z)\mathcal{F}_{h_{2,-2}}(\bar{z})\right]+ \ldots
\end{aligned}
\end{equation}
with an overall amplitude $A_{aaaa}(\mathcal{W}_{0,-1})$. The two double poles are again canceled exactly within the block $\mathbb{F}_{0,-1}$ by the terms
\begin{equation}
\mathcal{R}_{\frac{5}{2},0}\mathcal{F}_{h_{\frac{5}{2},0}}(z)\mathcal{F}_{h_{\frac{5}{2},0}}(\bar{z})+\mathcal{R}_{\frac{7}{2},0}\mathcal{F}_{h_{\frac{7}{2},0}}(z)\mathcal{F}_{h_{\frac{7}{2},0}}(\bar{z})
\end{equation}
due to the coincident dimensions
\begin{equation}
h_{\frac{5}{2},0}=\bar{h}_{\frac{5}{2},0}=h_{1,-2}, \quad h_{\frac{7}{2},0}=\bar{h}_{\frac{7}{2},0}=h_{2,-2},
\end{equation}
as can be easily checked. To cancel the simple poles one needs the amplitudes for $(h_{1,2}, h_{1,-2})$ and $(h_{2,2},h_{2,-2})$ to be
\begin{equation}
-A_{aaaa}(\mathcal{W}_{0,-1})\frac{\mathrm{R}_{1,2}}{h_{\frac{1}{2},0}-h_{1,2}}\;, \quad -A_{aaaa}(\mathcal{W}_{0,-1})\frac{\mathrm{R}_{2,2}}{h_{\frac{1}{2},0}-h_{2,2}}.
\end{equation}
This on one hand requires the recursion
\begin{equation}
R^N_{1,2}=\frac{A(h_{2,2},h_{2,-2})}{A(h_{1,2},h_{1,-2})}=-\frac{\mathrm{R}_{2,2}}{\mathrm{R}_{1,2}} \,,
\end{equation}
which can indeed be shown to be true. On the other hand, one finds that at $Q=2$:
\begin{equation}
A_{aaaa}(\mathcal{W}_{2,1})=-A_{aaaa}(\mathcal{W}_{0,-1})\frac{\mathrm{R}_{1,2}}{(h_{\frac{1}{2},0}-h_{1,2})R^N_{0,2}}+O(1).
\end{equation}
Using \eqref{Rbar21}, this gives the exact value of $A_{abab}(\mathcal{W}_{2,1})$ at $Q=2$:
\begin{equation}\label{Aabab21Q2}
\left. A_{abab}(\mathcal{W}_{2,1}) \right|_{Q=2}=A^L(\mathcal{W}_{0,-1})\frac{21\pi\Gamma(-\frac{7}{6})\Gamma(\frac{5}{3})}{2048\sqrt[3]{2} \Gamma(\frac{1}{6}) \,,\Gamma(\frac{4}{3})},
\end{equation}
where the expression of $A^{L}(\mathcal{W}_{0,-1})$ is given in \eqref{A0m1}. In figure \ref{aaaa20anaQ2}, we show this exact amplitude at $Q=2$ together with the bootstrapped amplitudes in the region around $Q=2$.

\bigskip

We have seen in the above how singularities in the amplitudes cancel the divergences in the conformal blocks at special values of $Q$. In the last part of this section, we shall see another type of divergences which arises from the $\mathcal{R}$ in the construction of the interchiral conformal blocks in section \ref{recursion} (which ultimately comes from the recursions) and how it leads to singularities in the amplitudes, and also provides exact results.

\subsubsection*{Canceling divergences from $\mathcal{R}$}
As we have seen in figure \ref{aabb41pole}, the ratio $\mathsf{R}_{\alpha}(\mathcal{W}_{4,1})$ in \eqref{Ralpha41} introduces poles in $A_{aabb}(\mathcal{W}_{4,1})$ at $Q^2-3Q+1=0$, viz., those given in \eqref{aabbnewpoles}. This is naturally
due to the module $\overline{\mathcal{W}}_{0,\q^2}$. At $Q=4\cos^2 \! \left({\frac{2\pi}{5}}\right)$, one finds the coincidence of dimensions between the leading field in $\mathcal{W}_{4,1}$ with the diagonal field $(h_{3,1},h_{3,1})$ in $\overline{\mathcal{W}}_{0,\q^2}$:
\begin{equation}
h_{0,4}=\bar{h}_{0,4}=h_{3,1}=\bar{h}_{3,1}.
\end{equation}
Meanwhile recall that the contribution of $(h_{3,1},h_{3,1})$ is given by
\begin{equation}
\mathcal{R}_{3,1}\mathcal{F}_{h_{3,1}}(z)\mathcal{F}_{h_{3,1}}(\bar{z}),
\end{equation}
and that $\mathcal{R}_{3,1}$ as defined in \eqref{Rd} has a simple pole.
To cancel this divergence in the $s$-channel of $P_{aabb}$ (or the $t$-channel of $P_{abba}$), we need
\begin{equation}
A_{aabb}(\mathcal{W}_{4,1})=-\mathcal{R}_{3,1}+O(1),
\end{equation}
or using \eqref{Ralpha41}:
\begin{equation}
A_{aaaa}(\mathcal{W}_{4,1})=-\frac{\mathcal{R}_{3,1}}{\mathsf{R}_{\alpha}(\mathcal{W}_{4,1})}.
\end{equation}
This gives the exact amplitude at $Q=4\cos^2 \! \left({\frac{2\pi}{5}}\right)$:
\begin{equation}\label{aaaa41B}
\left. A_{aaaa}(\mathcal{W}_{4,1})\right|_{Q=4\cos^2 \! \left(\frac{2\pi}{5}\right)}=\frac{9 \sqrt{ (5+\sqrt{5})} \pi  \Gamma (-\frac{10}{3}) \Gamma
	(-\frac{4}{3}) \Gamma (\frac{5}{6})^2 \Gamma
	(\frac{5}{3})^3}{256\sqrt{10} \Gamma (-\frac{2}{3})^3 \Gamma
	(\frac{1}{6})^2 \Gamma (\frac{7}{3}) \Gamma (\frac{10}{3})}.
\end{equation}

Similarly at $Q=4\cos^2 \! \left(\frac{\pi}{5}\right)$, one has instead
\begin{equation}
h_{0,4}=\bar{h}_{0,4}=h_{4,1}=\bar{h}_{4,1}.
\end{equation}
and therefore
\begin{equation}
A_{aabb}(\mathcal{W}_{4,1})=-\mathcal{R}_{4,1}+O(1).
\end{equation}
This means that
\begin{equation}
A_{aaaa}(\mathcal{W}_{4,1})=-\frac{\mathcal{R}_{4,1}}{\mathsf{R}_{\alpha}(\mathcal{W}_{4,1})},
\end{equation}
which is explicitly given by
\begin{equation}\label{aaaa41A}
\left. A_{aaaa}(\mathcal{W}_{4,1})\right|_{Q=4\cos^2 \! \left({\frac{\pi}{5}}\right)}=\frac{\sqrt{(5-\sqrt{5})} \pi  \Gamma (-\frac{11}{4}) \Gamma
	(-\frac{7}{4}) \Gamma (\frac{5}{8})^2 \Gamma (\frac{5}{4})^3
	\Gamma (\frac{15}{8})^4}{10\sqrt{10} \Gamma (-\frac{7}{8})^4 \Gamma
	(-\frac{1}{4})^3 \Gamma (\frac{3}{8})^2 \Gamma(\frac{11}{4})
	\Gamma (\frac{15}{4})}.
\end{equation}

\smallskip

One can carry out the same analysis on the poles of $A_{aaaa}(\mathcal{W}_{4,1})$ and $A_{abab}(\mathcal{W}_{4,1})$ at $Q^2-4Q+2=0$ which disappear in $A_{aabb}(\mathcal{W}_{4,1})$ due to the ratio $\mathsf{R}_{\bar{\alpha}}(\mathcal{W}_{4,1})$ from \eqref{Ralphabar41}. This can be understood from the module $\mathcal{W}_{0,-1}$ with the divergences in $\mathcal{R}_{\frac{7}{2},0}$ and $\mathcal{R}_{\frac{5}{2},0}$. We do not repeat the details here but give the exact results from the cancellation of these divergences:


\begin{subequations} \label{abab41AB}
\begin{eqnarray}
\left. A_{abab}(\mathcal{W}_{4,1}) \right|_{Q=4\cos^2 \! (\frac{3\pi}{8})} \!\!\! &=& \!\!\! -A^L(\mathcal{W}_{0,-1})\frac{45 (2+\sqrt{2}) \pi  \Gamma (-\frac{12}{5}) \Gamma
	(-\frac{7}{5}) \Gamma (-\frac{4}{5}) \Gamma (\frac{9}{10})^2
	\Gamma (\frac{17}{10})^4}{16384 \Gamma (-\frac{7}{10})^4 \Gamma
	(\frac{1}{10})^2 \Gamma (\frac{9}{5}) \Gamma (\frac{12}{5})
	\Gamma (\frac{17}{5})},\label{abab41A}\\
\left. A_{abab}(\mathcal{W}_{4,1}) \right|_{Q=4\cos^2 \! (\frac{\pi}{8})} \!\!\! &=&\!\!\! A^L(\mathcal{W}_{0,-1})\frac{823543 (\sqrt{2}-2) \Gamma
	(-\frac{6}{7}) \Gamma (-\frac{3}{7})
	\Gamma (\frac{1}{7}) \Gamma
	(\frac{11}{14})^2 \Gamma
	(\frac{19}{14})^3 \Gamma
	(\frac{27}{14})^2}{6871947673600 \sqrt[7]{2} \Gamma
	(-\frac{13}{14})^2 \Gamma
	(-\frac{5}{14})^2 \Gamma (\frac{3}{14})
	\Gamma (\frac{6}{7}) \Gamma
	(\frac{10}{7}) \Gamma (\frac{17}{7})}, \quad \quad \quad \label{abab41B}
\end{eqnarray}
\end{subequations}
where again the $A^L(\mathcal{W}_{0,-1})$ is given in \eqref{A0m1}.

\smallskip

\begin{figure}[H]
	\begin{centering}
		\begin{subfigure}{0.5\textwidth}
			\includegraphics[width=0.95\textwidth]{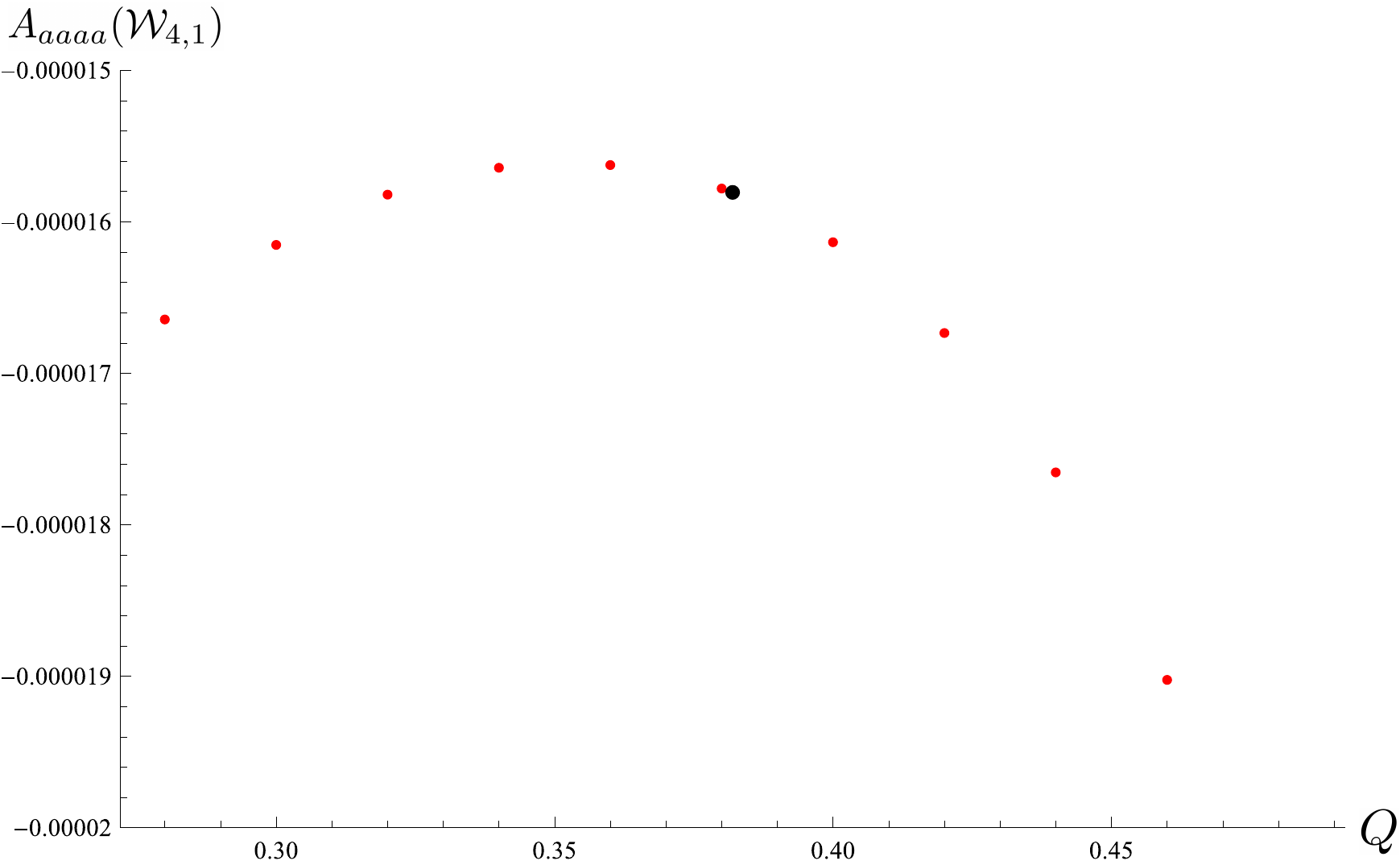}
		\end{subfigure}
		\begin{subfigure}{0.5\textwidth}
			\includegraphics[width=0.95\textwidth]{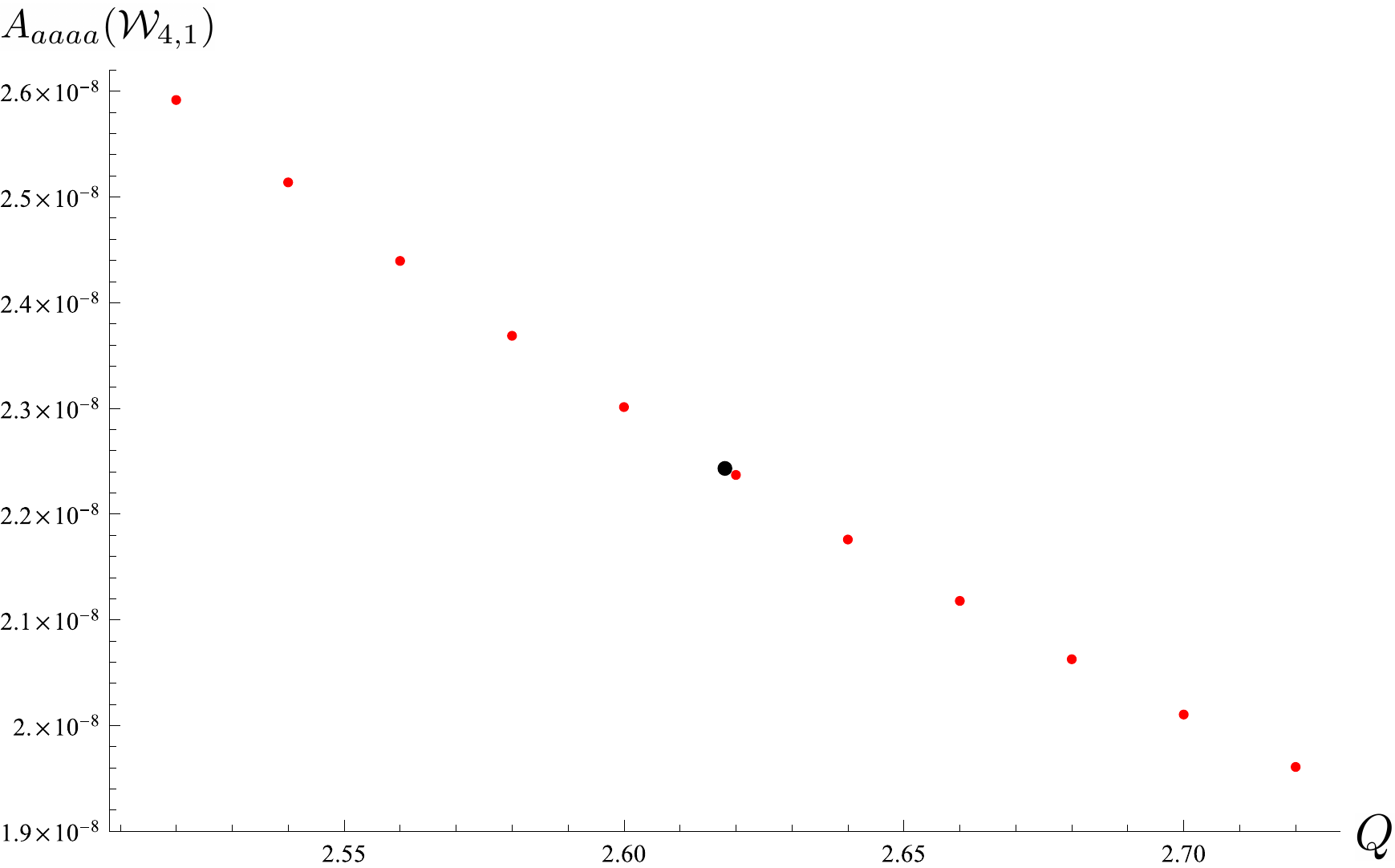}
		\end{subfigure}
	\end{centering}
	\caption{The amplitudes $A_{aaaa}(\mathcal{W}_{4,1})$ in the regions around $Q=4\cos^2 \! \left(\frac{2\pi}{5}\right)$ (left) and $Q=4\cos^2 \! \left(\frac{\pi}{5}\right)$ (right). The {\color{red}red} dots are the bootstrap results and the black dots are the exact expressions \eqref{aaaa41B} and \eqref{aaaa41A}.}
	\label{aaaa41analyticQ}
\end{figure}

\begin{figure}[H]
	\begin{centering}
		\begin{subfigure}{0.5\textwidth}
			\includegraphics[width=0.95\textwidth]{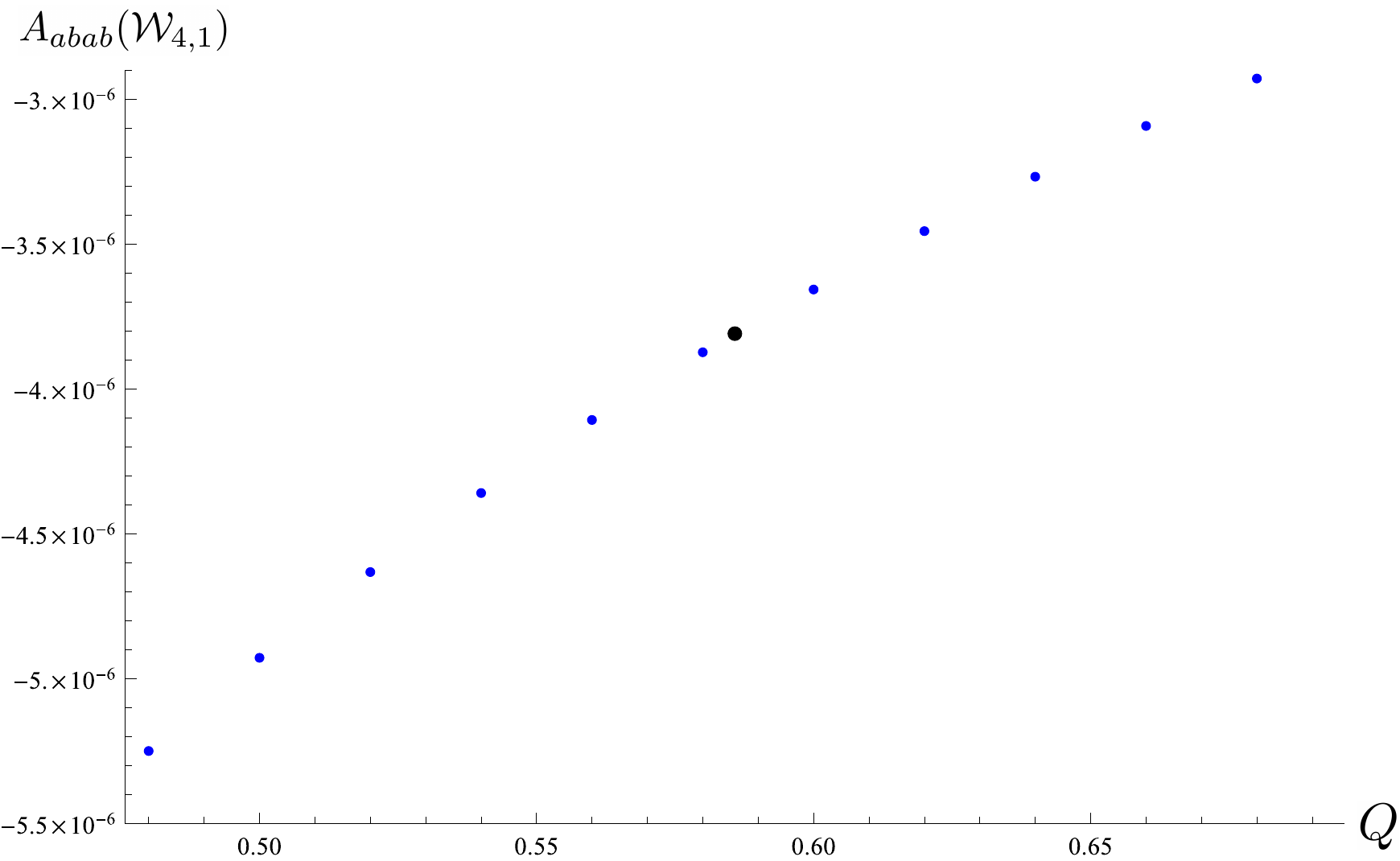}
		\end{subfigure}
		\begin{subfigure}{0.5\textwidth}
			\includegraphics[width=0.95\textwidth]{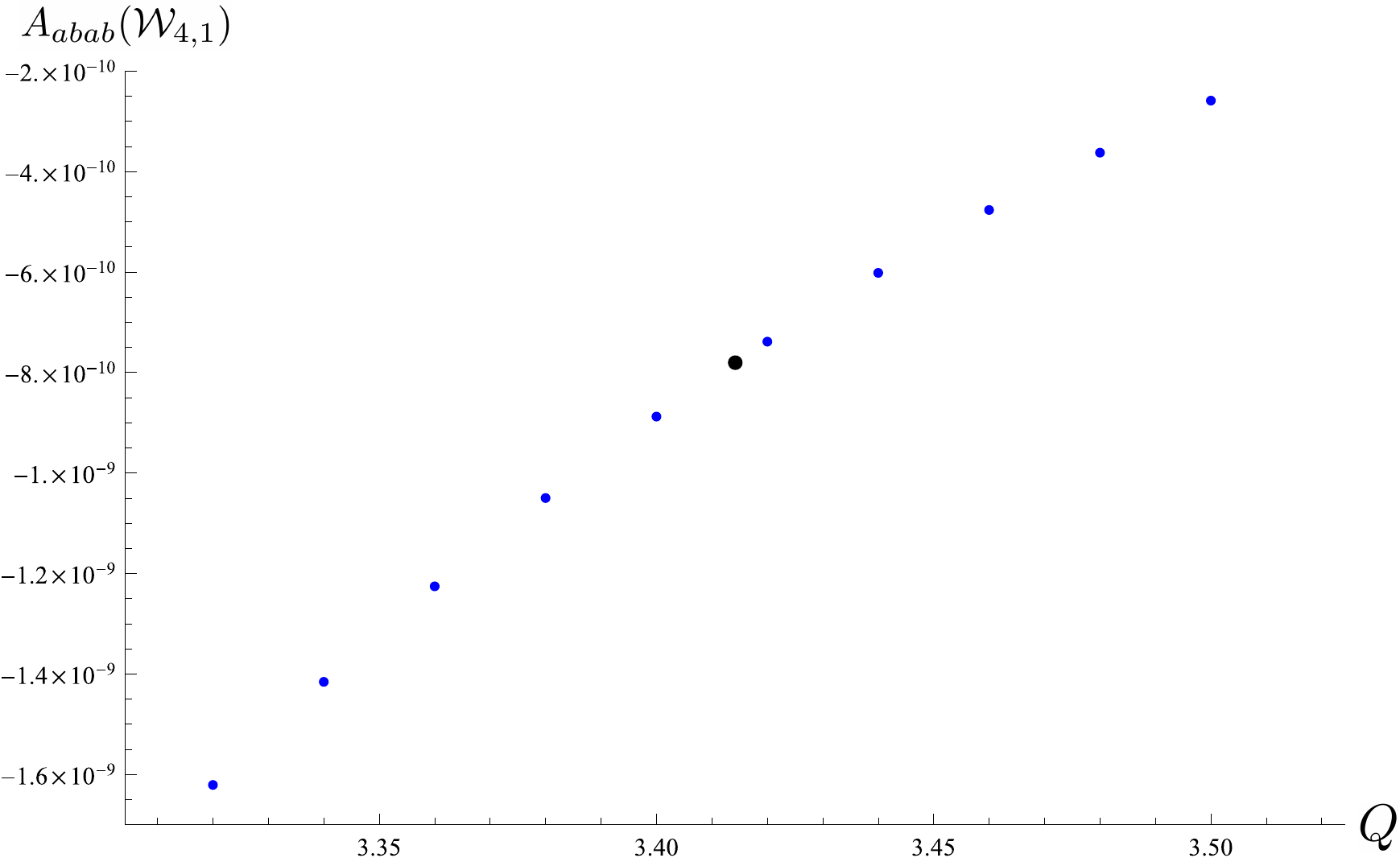}
		\end{subfigure}
	\end{centering}
	\caption{The amplitudes $A_{abab}(\mathcal{W}_{4,1})$ in the regions around $Q=4\cos^2 \! \left(\frac{3\pi}{8}\right)$ (left) and $Q=4\cos^2 \! \left(\frac{\pi}{8}\right)$ (right). The {\color{blue}blue} dots are the bootstrap results and the black dots are the exact expressions \eqref{abab41A} and \eqref{abab41B}.}
	\label{abab41analyticQ}
\end{figure}

In figures \ref{aaaa41analyticQ} and \ref{abab41analyticQ}, we plot the analytic expressions \eqref{aaaa41A}, \eqref{aaaa41B}, \eqref{abab41A} and \eqref{abab41B} together with the bootstrap results.

\subsection{Comparisons}\label{com}

The amplitudes that we have obtained from the bootstrap in section \ref{bootresults} can be compared with a few existing partial results. Note that while such comparisons provide some sanity checks on the bootstrapped amplitudes, there has not been a complete determination of the amplitudes in the Potts probabilities up to this level before our work. In the following we will discuss the comparison of the bootstrap results with numerical transfer-matrix computations in the lattice model \cite{Jacobsen:2018pti}, and with the non-diagonal Liouville theory of \cite{Picco:2016ilr} which provides an approximate description.

\subsubsection{Lattice}
The approach of computing the amplitudes on the lattice is described in details in \cite{Jacobsen:2018pti} where a few examples were also given.%
\footnote{Specifically we are using the scalar product method exposed in section 4.3.2 of \cite{Jacobsen:2018pti}, for which ample technical details
were given in appendix A.2 of that paper. We were generally able to obtain finite-size results for cylinders of circumferences $\mathsf{L}=5,6,\ldots,\mathsf{L}_{\rm max}$, with a maximal size $\mathsf{L}_{\rm max}=11$ for the amplitudes corresponding to the lowest-lying eigenvalues in the transfer matrix spectrum, and $\mathsf{L}_{\rm max}=10$ for higher-lying cases. Extrapolations to the scaling limit $\mathsf{L}\to\infty$ were done separately for even and odd sizes, using the tricks given in section 4.3.3 of \cite{Jacobsen:2018pti}. Indicative error bars were estimated from the difference between the extrapolations through even and odd sizes.}
Here we apply this lattice approach to obtain the amplitudes associated with the primary fields in the four probabilities within the range $0<Q<4$. On one hand, this provides a check on the Potts solution to the bootstrap we obtained above, namely the amplitudes $A(\mathcal{W})$ associated with entire ATL modules, which we show in this section. On the other hand, the lattice results for the sub-leading primaries in a ATL module also serve as basic checks on the interchiral conformal blocks as established in section \ref{recursion}, by means of a comparison of the lattice results with the recursions \eqref{Rshift1N} and \eqref{Rshift1D}. We leave this latter issue to appendix \ref{basiccheck}.

We now consider the lattice comparisons for the bootstrapped amplitudes
\begin{equation}\label{Acompforlattice}
A_{aaaa}(\mathcal{W}_{2,1}),\;A_{aaaa}(\mathcal{W}_{4,1}),\;A_{abab}(\mathcal{W}_{4,i}),\;A_{abab}(\mathcal{W}_{4,-i}).
\end{equation}
Note that due to the normalization in the lattice computation (see below), we do not have the lowest amplitude in each probability for the lattice results. On the other hand, since the bootstrap has imposed the amplitude ratios \eqref{ratioforbootstrap}, which were obtained originally from lattice computations \cite{He:2020mhb}, it suffices to consider the comparisons of the amplitudes \eqref{Acompforlattice} and ignore the ones related to them through \eqref{ratioforbootstrap}.

\begin{figure}[H]
	\centering
	\includegraphics[width=0.6\textwidth]{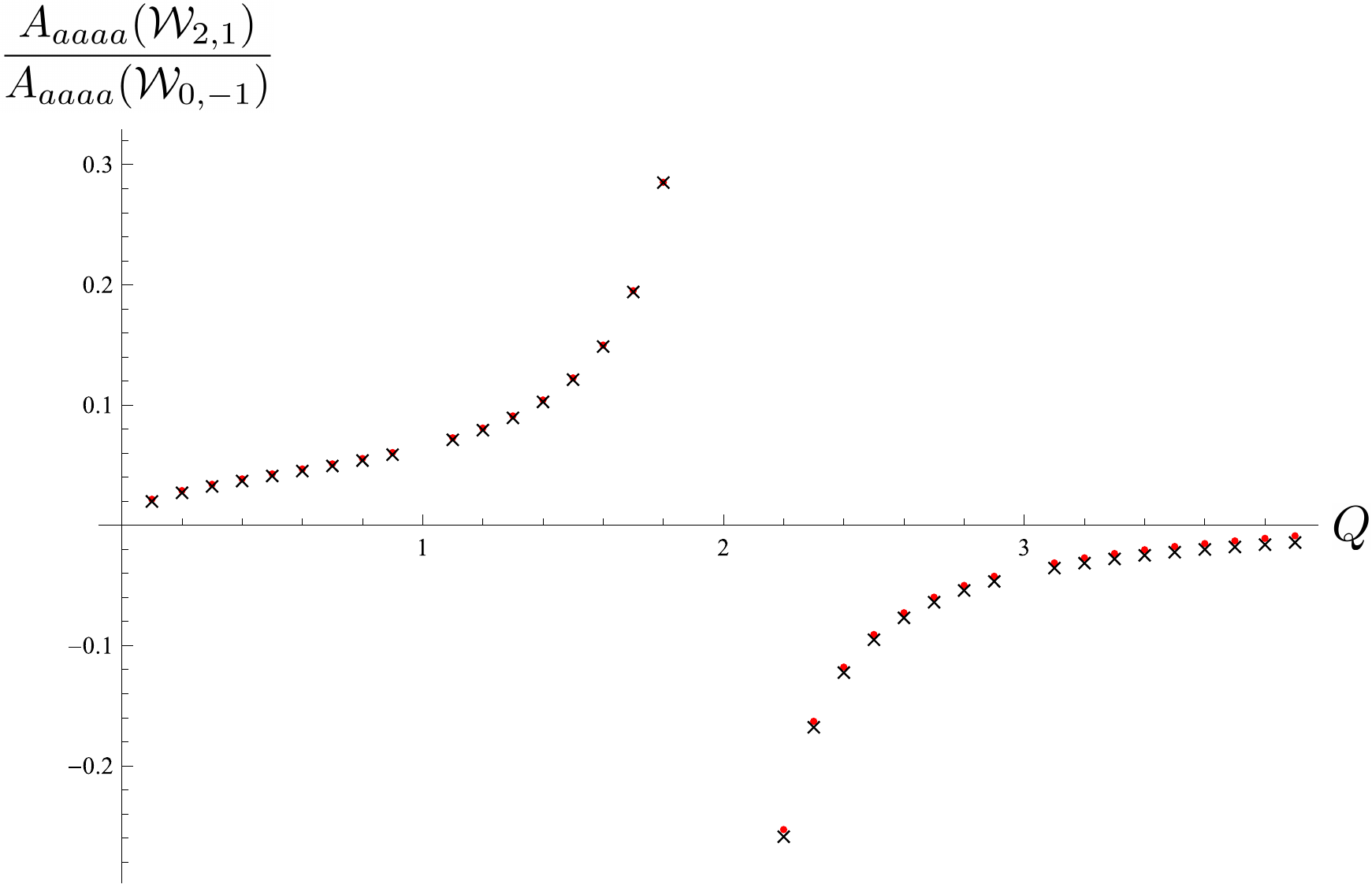}
	\caption{The amplitude $A_{aaaa}(\mathcal{W}_{2,1})$ normalized with the leading amplitude $A_{aaaa}(\mathcal{W}_{0,-1})$ in $P_{aaaa}$. Comparison of the lattice results (indicated with {\footnotesize$\times$}) and bootstrap results (indicated with {\footnotesize\color{red}$\bullet$}).}
	\label{aaaa20lattice}
\end{figure}

\begin{figure}[H]
	\centering
	\includegraphics[width=\textwidth]{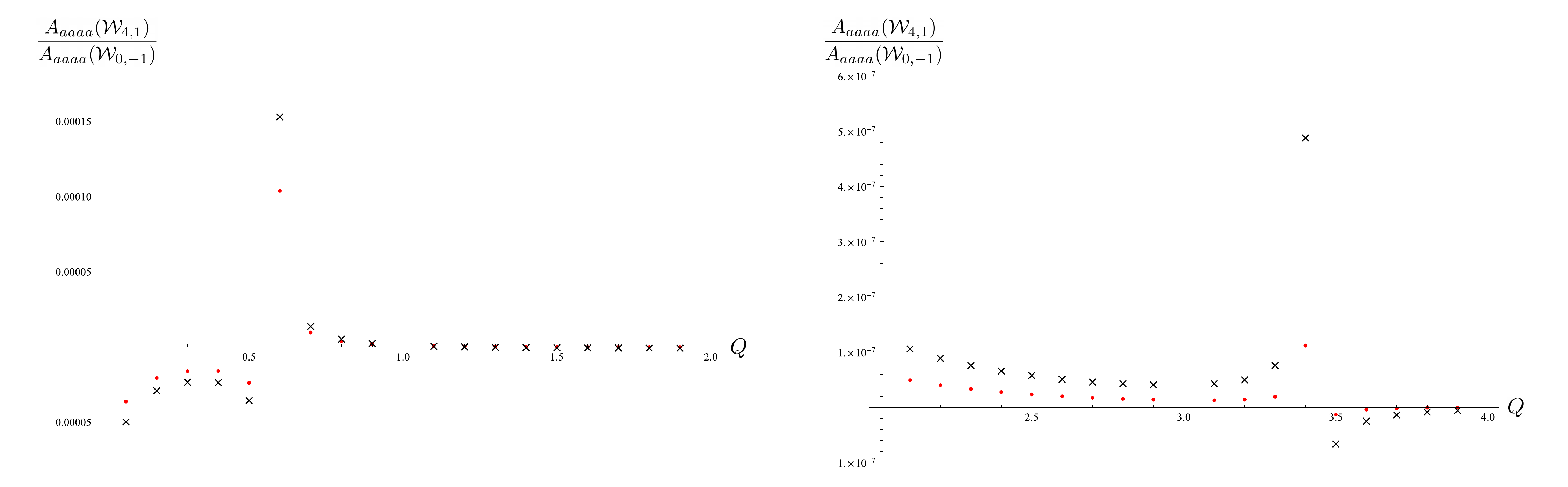}
	\caption{The amplitude $A_{aaaa}(\mathcal{W}_{4,1})$ normalized with the leading amplitude $A_{aaaa}(\mathcal{W}_{0,-1})$ in $P_{aaaa}$. Comparison of the lattice results (indicated with {\footnotesize$\times$}) and bootstrap results (indicated with {\footnotesize\color{red}$\bullet$}) in the regions $0< Q< 2$ and $2< Q< 4$.}
	\label{aaaa40lattice}
\end{figure}

\begin{figure}[H]
	\centering
	\includegraphics[width=\textwidth]{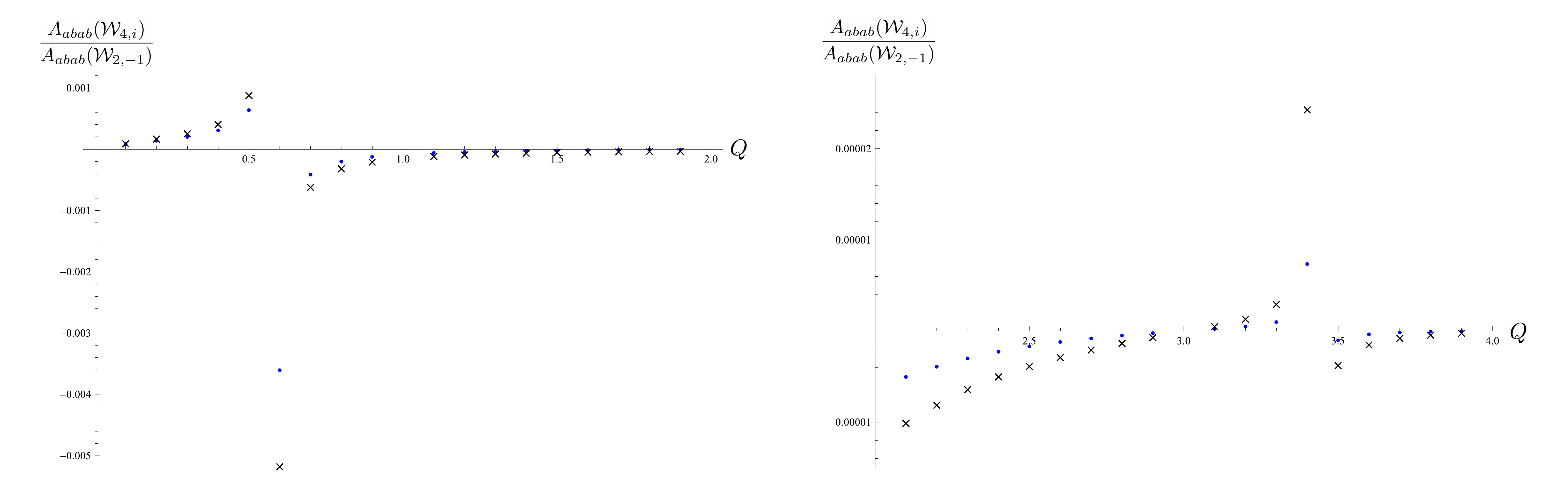}
	\caption{The amplitude $A_{abab}(\mathcal{W}_{4,i})$ normalized with the leading amplitude $A_{abab}(\mathcal{W}_{0,-1})$ in $P_{abab}-P_{abba}$. Comparison of the lattice results (indicated with {\footnotesize$\times$}) and bootstrap results (indicated with {\footnotesize\color{blue}$\bullet$}) in the regions $0< Q< 2$ and $2< Q< 4$.}
	\label{abab4quatlattice}
\end{figure}

\begin{figure}[H]
	\centering
	\includegraphics[width=\textwidth]{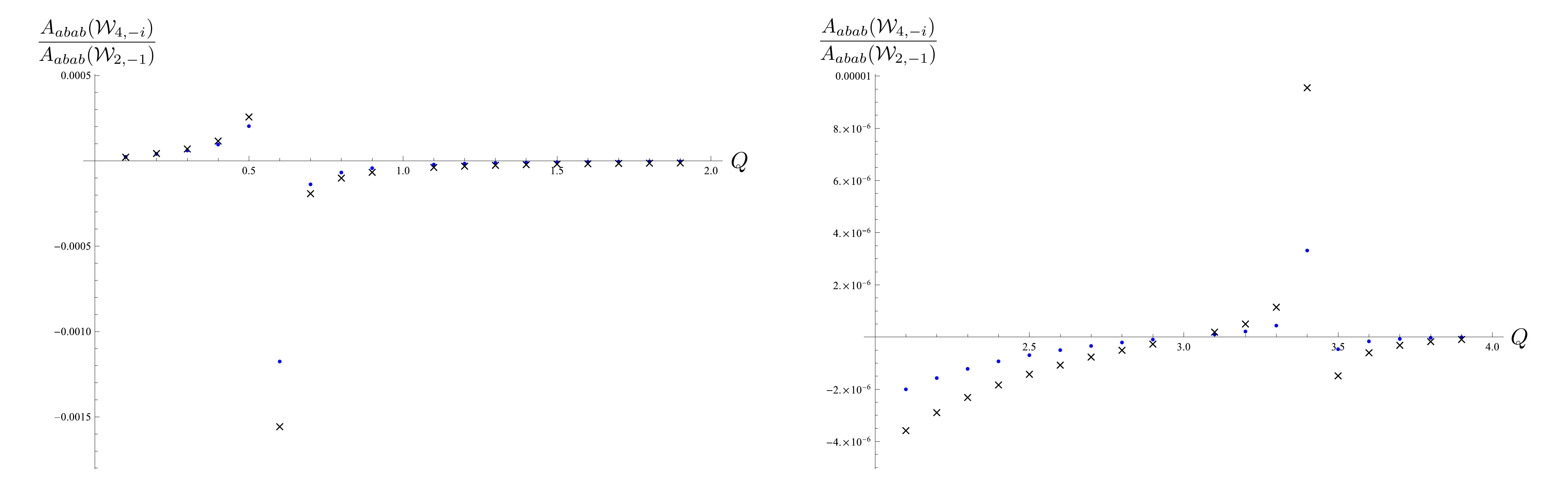}
	\caption{The amplitude $A_{abab}(\mathcal{W}_{4,-i})$ normalized with the leading amplitude $A_{abab}(\mathcal{W}_{0,-1})$ in $P_{abab}-P_{abba}$. Comparison of the lattice results (indicated with {\footnotesize$\times$}) and bootstrap results (indicated with {\footnotesize\color{blue}$\bullet$}) in the regions $0< Q< 2$ and $2< Q< 4$.}
	\label{abab43quatlattice}
\end{figure}

In the lattice computation, one needs to choose a normalization for each probability which we have chosen to be the amplitude of the field with the lowest dimension. In particular, as described in \cite{Jacobsen:2018pti,He:2020mhb}, for the probabilities $P_{abab}$ and $P_{abba}$, we focus on the symmetric and antisymmetric combinations
\begin{equation}
P_S=P_{abab}+P_{abba}, \quad P_{A}=P_{abab}-P_{abba} \,,
\end{equation}
which consist respectively of modules with even and odd spins, due to \eqref{Aeven} and \eqref{Aodd}. 
This way, we obtain the following results on the lattice:
\begin{equation}
\frac{A_{aaaa}(\mathcal{W}_{2,1})}{A_{aaaa}(\mathcal{W}_{0,-1})},\; \frac{A_{aaaa}(\mathcal{W}_{4,1})}{A_{aaaa}(\mathcal{W}_{0,-1})},\; \frac{A_{abab}(\mathcal{W}_{4,i})}{A_{abab}(\mathcal{W}_{2,-1})},\; \frac{A_{abab}(\mathcal{W}_{4,-i})}{A_{abab}(\mathcal{W}_{2,-1})}
\end{equation}
and plot them in figures \ref{aaaa20lattice}, \ref{aaaa40lattice}, \ref{abab4quatlattice} and \ref{abab43quatlattice} together with the corresponding bootstrap results. In each of these, the bootstrap and the lattice results agree on the analytic structures (the location of poles and zeros), the order of magnitudes (which vary considerably with the amplitude being considered) and the generic behavior as a function of $Q$ (sign, monotonicity, local extrema). The difference in the actual values is likely largely due to the finite-size effect of the lattice computations.
In particular, for each parity of the lattice size $\mathsf{L}$ we have only three points at our disposal, which is a rather precarious situation for performing reliable extrapolations.%
\footnote{Concretely, we extrapolated to the limit $\mathsf{L} \to \infty$ using a second-order polynomial in $1/\mathsf{L}$, which might not always be sufficient due to the amount of curvature observed in the data.}
Overall, given these constraints, we find the agreement on the general features of the curves highly satisfactory, while the detailed comparison of the actual values ranges from excellent (for the lowest-lying amplitudes) to acceptable (for the higher-lying ones).

\subsubsection{Non-diagonal Liouville theory}
As claimed by the authors of \cite{Picco:2019dkm}, the spectrum $\mathcal{S}_{\mathbb{Z}+\frac{1}{2},2\mathbb{Z}}$ applied to the four-point function \eqref{PRS4pt} provides an approximate description that becomes accurate at $Q=0,3,4$. This means that the difference between the spectrum \eqref{Pottsspectrum} and $\mathcal{S}_{\mathbb{Z}+\frac{1}{2},2\mathbb{Z}}$ vanishes at these values of $Q$ for the combination in \eqref{PRS4pt}, as can be easily checked using our results. This involves the modules $\mathcal{W}_{2,1}$, $\mathcal{W}_{4,1}$, $\mathcal{W}_{4,i}$ and $\mathcal{W}_{4,-i}$. 

For $\mathcal{W}_{2,1}$, using \eqref{Rbar21}, one has
\begin{equation}\label{21cancel}
A_{aaaa}(\mathcal{W}_{2,1})+\frac{2}{Q-2}A_{abab}(\mathcal{W}_{2,1})=0
\end{equation}
for all values of $Q$, which explains why \cite{Picco:2016ilr} gives a reasonable numerical approximation, since this is the next-to-leading contribution in the combination $P_{aaaa}+P_{abab}$ after the module $\mathcal{W}_{0,-1}$. On the other hand, as we have studied in \cite{He:2020mhb}, the expression \eqref{PRS4pt} is accurate up to diagrams involving two non-contractible loops, which is reflected in the amplitude identities for modules with $j=2$, i.e.,  eqs.~\eqref{abab2m1ana} above and \eqref{21cancel} here.

\begin{figure}[H]
	\centering
	\includegraphics[width=0.6\textwidth]{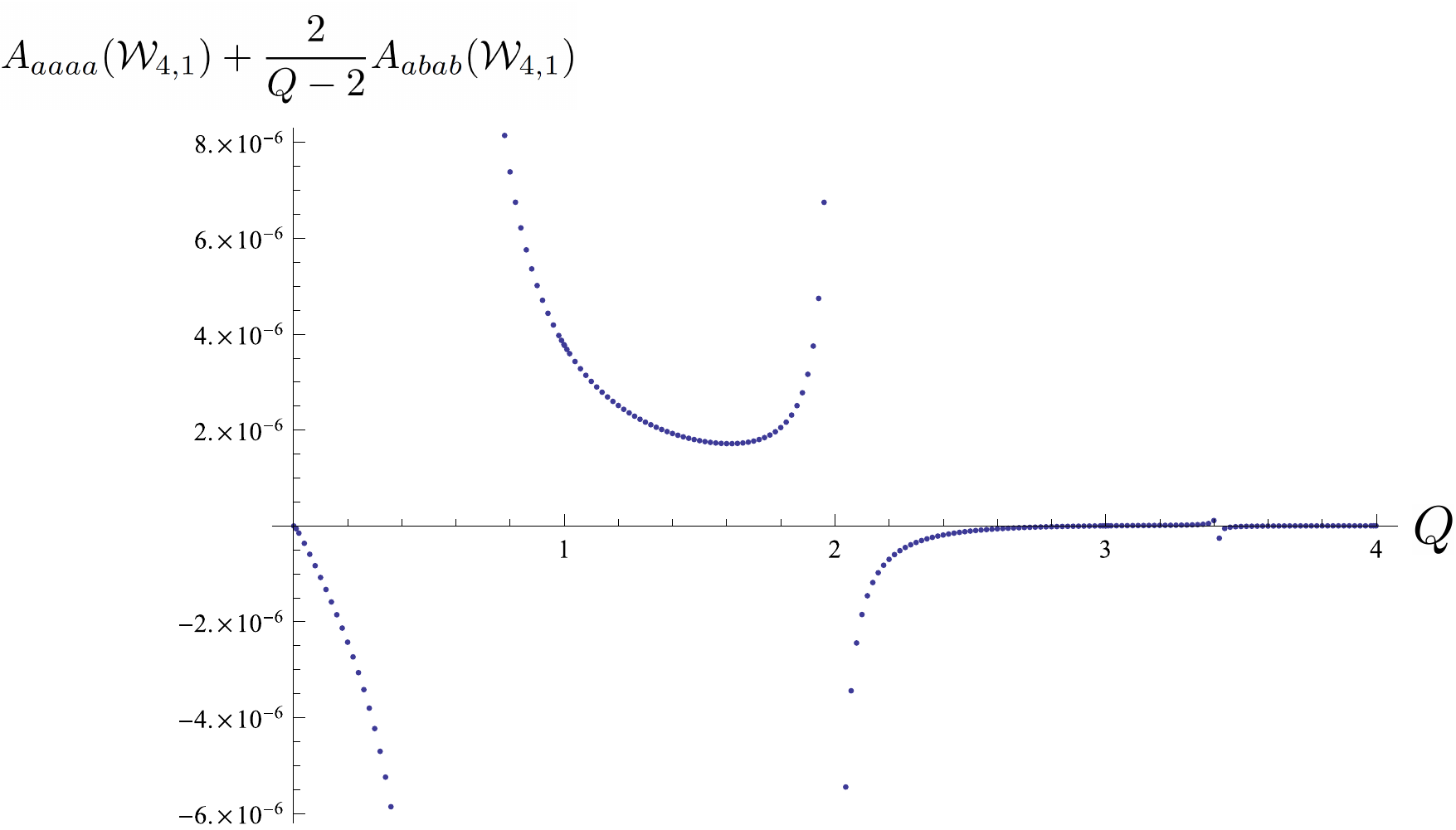}
	\caption{The amplitude of $\mathcal{W}_{4,1}$ in the combination $P_{aaaa}+\frac{2}{Q-2}P_{abab}$ (the right-hand side of \eqref{PRS4pt}) which is approximated by the four-point function of \cite{Picco:2016ilr} (the left-hand side of \eqref{PRS4pt}) whose spectrum does not include these fields. The values are generically small and vanishes at $Q=0,3,4$ where the approximation becomes exact.}
	\label{PRSabab40}
\end{figure}

For $\mathcal{W}_{4,1}$, using \eqref{Ralphabar41}, one obtains
\begin{equation}
A_{aaaa}(\mathcal{W}_{4,1})+\frac{2}{Q-2}A_{abab}(\mathcal{W}_{4,1})=-A_{aaaa}(\mathcal{W}_{4,1})\frac{Q^2(Q-3)(Q-4)}{2(Q-2)}
\end{equation}
and indeed, the module disappears in this combination exactly at $Q=0,3,4$. For generic values of $Q$, this does not vanish but the values are numerically small---except obviously for the regions near the poles at $Q=2$, $4\cos^2 \! \left(\frac{3\pi}{8}\right)$ and $4\cos^2 \! \left(\frac{\pi}{8}\right)$---as we show in figure \ref{PRSabab40}.
The situation is similar for $\mathcal{W}_{4,i}$ and $\mathcal{W}_{4,-i}$: since they do not appear in $P_{aaaa}$, the combination \eqref{PRS4pt} involves simply the amplitudes
\begin{equation}
\frac{2}{Q-2}A_{abab}(\mathcal{W}_{4,i}), \quad \frac{2}{Q-2}A_{abab}(\mathcal{W}_{4,-i})
\end{equation}
which---as is clearly seen from figures \ref{abab4pi} and \ref{abab4mi}---vanish at $Q=0,3,4$ and remain small for generic values of $Q$. In this case, they also give rise to poles at $Q=2$, $4\cos^2 \! \left(\frac{3\pi}{8}\right)$ and $4\cos^2 \! \left(\frac{\pi}{8}\right)$ which appear in the correlation functions \eqref{PRS4pt}.

\subsection{``Renormalized'' Liouville recursions}\label{renormL}
\begin{figure}[H]
\centering
	\includegraphics[width=0.6\textwidth]{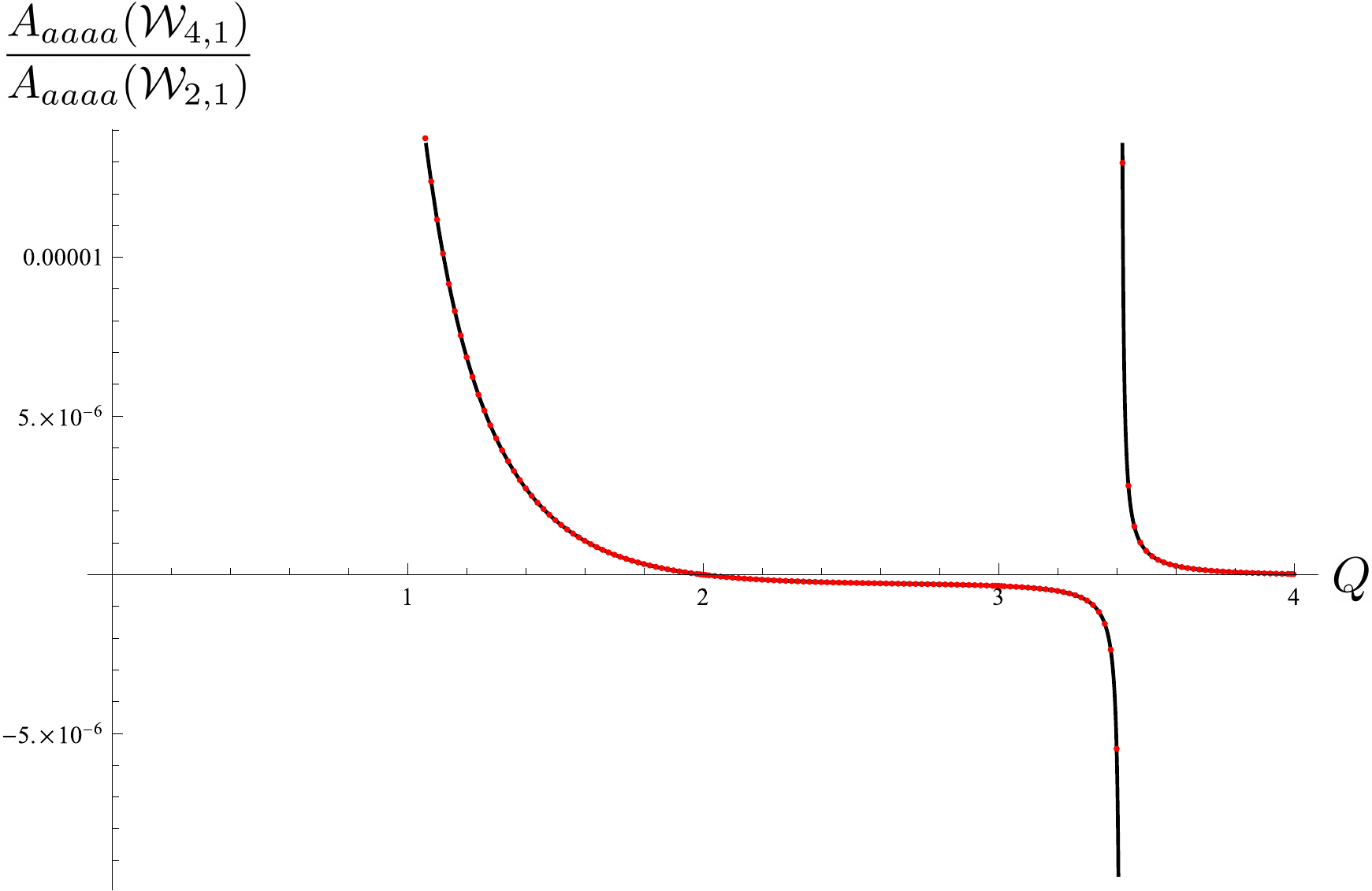}
	\caption{The renormalized Liouville recursion of $\frac{A_{aaaa}(\mathcal{W}_{4,1})}{A_{aaaa}(\mathcal{W}_{2,1})}$ from the bootstrap result (red dots) matches perfectly with the analytic expression on the right-hand side of \eqref{4020ratio} (black curve).}
	\label{aaaa4020ratio}
\end{figure}

\begin{figure}[H]
	\begin{centering}
		\begin{subfigure}{0.5\textwidth}
			\includegraphics[width=0.9\textwidth]{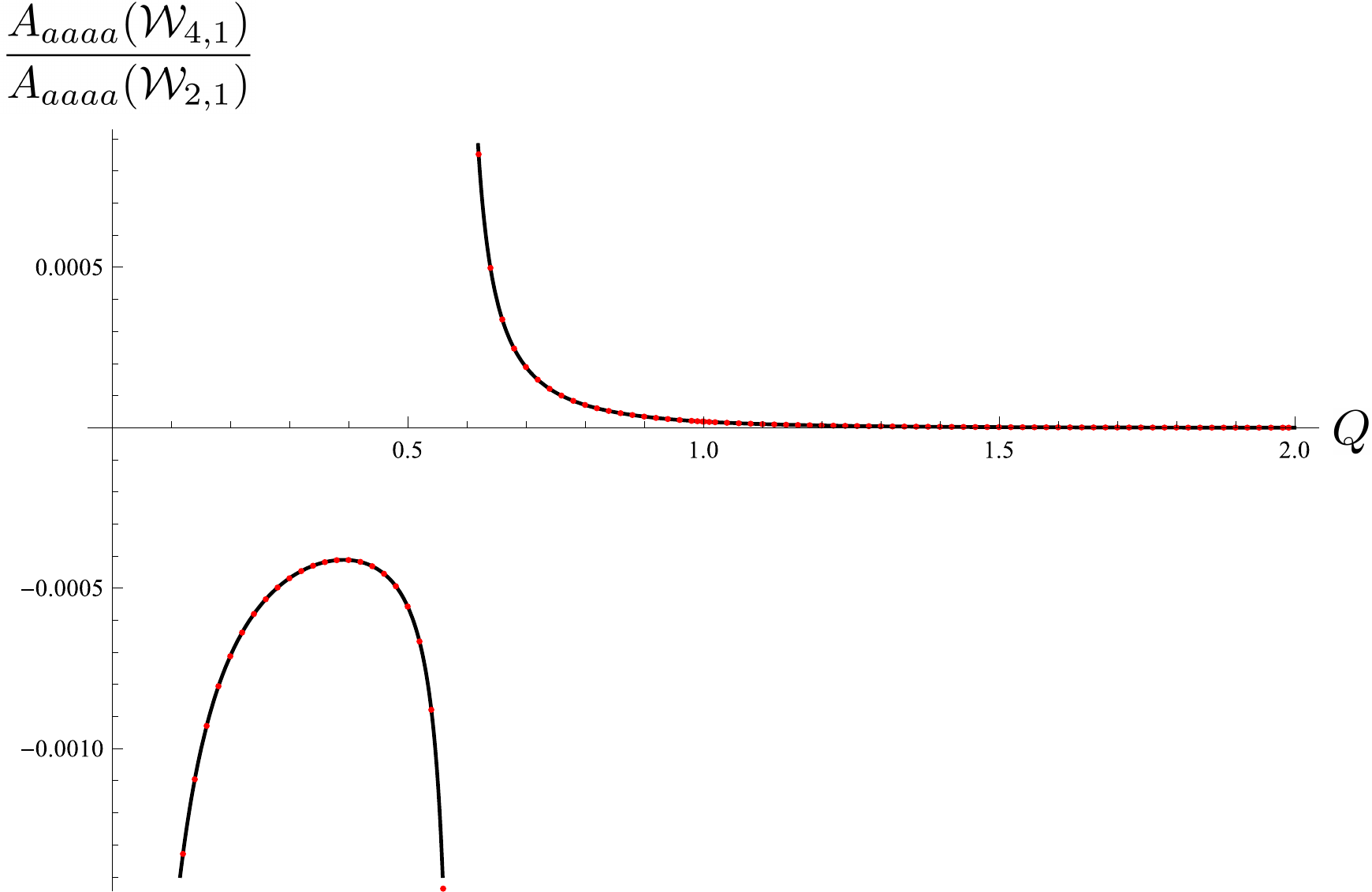}
		\end{subfigure}
		\begin{subfigure}{0.5\textwidth}
			\includegraphics[width=0.9\textwidth]{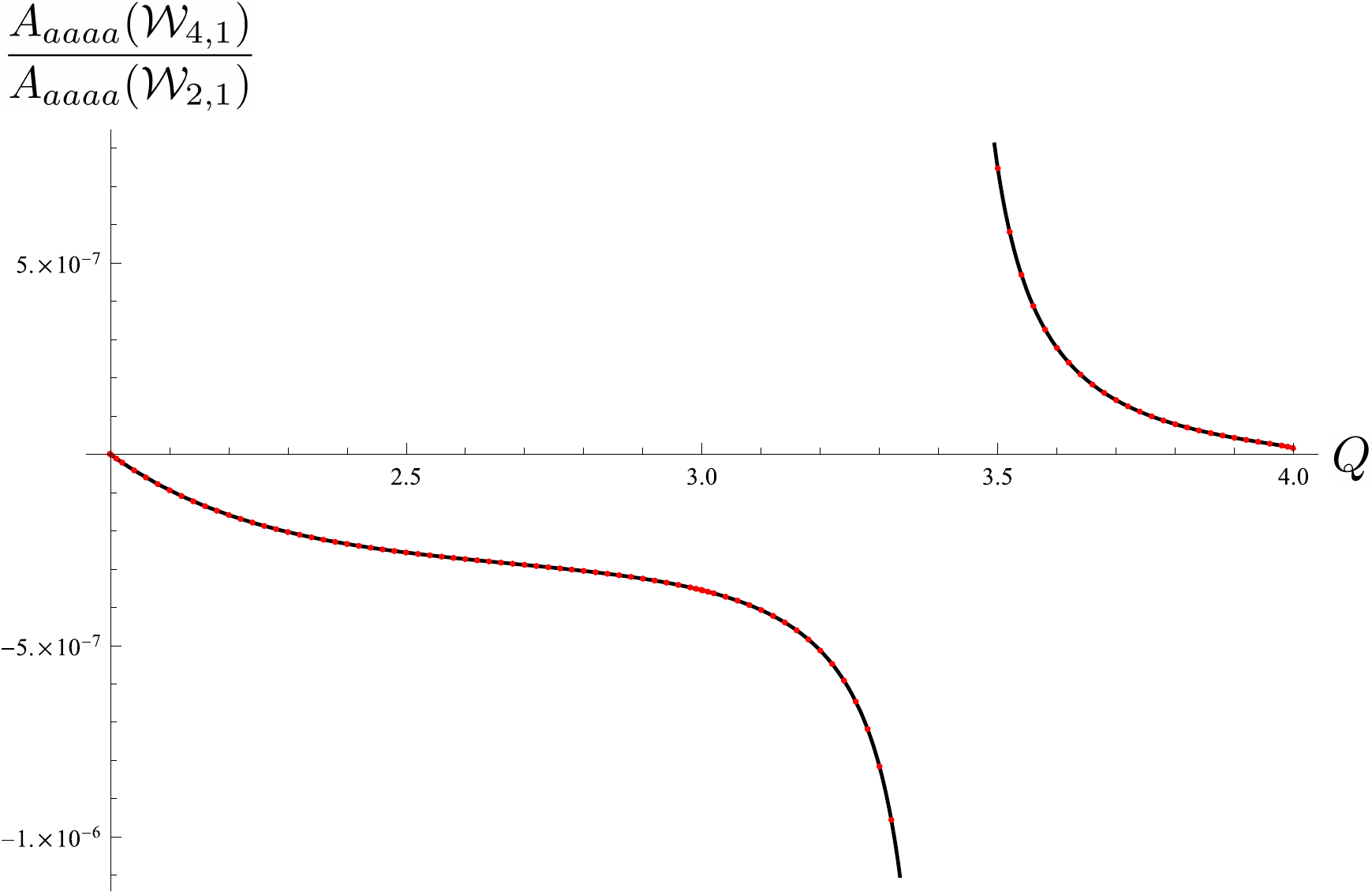}
		\end{subfigure}
	\end{centering}
	\caption{The agreement of the bootstrap result of $\frac{A_{aaaa}(\mathcal{W}_{4,1})}{A_{aaaa}(\mathcal{W}_{2,1})}$ with the right-hand side of \eqref{4020ratio} in the regions $0< Q<2$ and $2< Q<4$.}
	\label{aaaa4020ratio12}
\end{figure}

As we have mentioned in section \ref{recursion}, the field $\Phi^D_{1,2}$ in (non-diagonal) Liouville theory is degenerate, and this feature leads to the recursions $\frac{A^L(\mathcal{W}_{j+1,-1})}{A^L({\mathcal{W}_{j-1,-1}})}$. The explicit expressions were obtained in \cite{Estienne:2015sua,Migliaccio:2017dch} and we recall them in appendix \ref{AL}. In the case of the Potts model, the degeneracy of this field is absent and therefore the usual Liouville recursions for shifting the $j$-index do not hold any more. We see however in \eqref{PAusingL}, that this is replaced by a renormalized version in which the Liouville recursion is dressed by a factor consisting of ratios of polynomials in $Q$:
\begin{subequations}\label{renoLrecu}
\begin{eqnarray}
\frac{A_{aaaa}(\mathcal{W}_{4,-1})}{A_{aaaa}(\mathcal{W}_{0,-1})}&=&\frac{(Q-2)(Q^2-4Q+2)}{Q(Q-3)^2}\frac{A^L(\mathcal{W}_{4,-1})}{A^L(\mathcal{W}_{0,-1})},\\
\frac{A_{abab}(\mathcal{W}_{4,-1})}{A_{abab}(\mathcal{W}_{2,-1})}&=&\frac{(Q-1)(Q-4)(Q^2-4Q+2)}{2Q(Q-3)^2}\frac{A^L(\mathcal{W}_{4,-1})}{A^L(\mathcal{W}_{2,-1})}.
\end{eqnarray}
\end{subequations}
Interestingly, using the bootstrap results, we have managed to conjecture another renormalized Liouville recursion:
\begin{equation}\label{4020ratio}
\frac{A_{aaaa}(\mathcal{W}_{4,1})}{A_{aaaa}(\mathcal{W}_{2,1})}=\frac{(Q-2)^2}{(Q-1)^2(Q^2-4Q+2)}\frac{A^L(\mathcal{W}_{4,1})}{A^L(\mathcal{W}_{2,1})}.
\end{equation}
It is certainly remarkable that the precision of the numerical bootstrap results is sufficient for such a relation to be established.
Notice that despite of the fields in modules $\mathcal{W}_{4,1}$ and $\mathcal{W}_{2,1}$ being absent in the spectrum of (non-diagonal) Liouville theory, the recursion $\frac{A^L(\mathcal{W}_{4,1})}{A^L(\mathcal{W}_{2,1})}$ exists as a result of the degeneracy of $\Phi^D_{1,2}$ there. In the case of the Potts-model probabilities considered here, it is renormalized by a $Q$-dependent factor similar to \eqref{renoLrecu} which we have established analytically.
In figure \ref{aaaa4020ratio} and \ref{aaaa4020ratio12}, we plot the bootstrap results of \eqref{4020ratio} and the analytic expression on the right-hand side and they agree perfectly.

\bigskip

\section{Conclusions}

Our results fully  confirms the correctness of the spectrum for the Potts-model four-point functions proposed in  \cite{Jacobsen:2018pti}, and, for all practical purposes, solve the bootstrap problem and determine accurately the leading amplitudes, hence carrying to its term the program initiated in 
\cite{Picco:2016ilr}. To be fair, our treatment of conformal blocks arising from fields in the modules $\mathcal{W}_{j,1}$ with degenerate conformal weights is a bit unsatisfactory, as we did not take into account the likely presence of logarithmic terms ($\ln(z\bar{z})$). We do not expect this, however, to affect the numerically determined amplitudes significantly, as witnessed by the excellent agreement with data from lattice calculations.
Nonetheless, we hope to revisit this question in our next paper. 

In the course of this work, we have also unearthed a lot of structure that remains to be understood. The degeneracy of fields with weight $h_{r,1}$, for $r\in\mathbb{N}^*$, arising in  $\overline{\mathcal{W}}_{0,\q^2}$ led naturally to the existence of interchiral conformal blocks, a structure deeply related with the underlying affine Temperley-Lieb algebra. This begs further study of the continuum limit of this algebra, which is more than the product of left and right Virasoro algebras, and was postulated in \cite{Gainutdinov:2012nq} to be described by an {\sl interchiral algebra}. The results of this paper should make possible the construction of this algebra beyond the case $c=-2$ discussed in \cite{Gainutdinov:2012nq}: we also plan to come back to this question soon.

Certainly the most fascinating result of our work is the existence of ``renormalized'' Liouville recursions, hinting at a structure in the Potts model that would replace the degeneracy of fields $\Phi_{12}^D$ familiar in Liouville theory. This ``structure'' manifests itself by an infinite series of rational functions of $Q$ (see, e.g., \eqref{renoLrecu} and \eqref{4020ratio}), whose origin remains largely mysterious to us. Understanding these functions would likely require a deeper study of the algebraic structure of the models on the lattice, and result in the full analytic determination of the correlation functions in the Potts model and in particular, the analytic expressions for the amplitudes. This question belongs as well to our list of ongoing investigations.

In conclusion, it is worth mentioning that a fully similar approach would lead to geometrical correlation functions in the $O(n)$ model (involving ``polymer lines'' instead of clusters).


\section*{Acknowledgements} 

This work was supported by the ERC Advanced Grant NuQFT. We thank S.~Ribault for many useful discussions and for detailed comments on the manuscript. We are also grateful to J.~Cardy and A.~Zamolodchikov for their kind encouragements throughout this work, and L.~Gr\"ans-Samuelsson for discussions and collaborations on related topics.

\newpage

\appendix

\section{$\alpha$, $\beta$, $\gamma$ from \cite{He:2020mhb}}\label{ratiosappend}
In \cite{He:2020mhb}, we have stated several facts regarding how the eigenvalues of the lattice transfer matrix (which in the continuum limit become the fields in the CFT) contribute to various quantities with different geometric content. Here we recall the definitions and expressions which are used in the main text for deriving further consequences on the Potts model.

One of the main results in \cite{He:2020mhb} is that the ratios of the amplitudes of these eigenvalues in different probabilities, different sub-diagrams contributing to the probabilities, and the diagrams with modified loop weights, depend only on the corresponding ATL modules and $Q$, which allows one to define the following quantities:
\begin{equation}
\begin{aligned}
\alpha_{j,\mathrm{z}^2}\equiv{A_{aabb}\over A_{aaaa}}(\AStTL{j}{\mathrm{z}^2}),\;\;&
\bar{\alpha}_{j,\mathrm{z}^2}\equiv{A_{abab}+A_{abba}\over A_{aaaa}}(\AStTL{j}{\mathrm{z}^2}),\;\;\\
\beta^{(k)}_{j,\mathrm{z}^2}\equiv{A_{abab}^{(k)}\over A_{abab}^{(2)}}(\AStTL{j}{\mathrm{z}^2}),\label{betadef}\;\;&
\gamma^{(\mathsf{a})}_{j,\mathrm{z}^2}\equiv{A^{(\mathsf{a})}_{aabb}\over A_{aabb}}(\AStTL{j}{\mathrm{z}^2}).
\end{aligned}
\end{equation}
Here $A^{(k)}$ refers to the amplitude of a module in sub-diagrams with a fixed number $k$ (necessarily even) of non-contractible loops between the marked FK clusters, and note that a module $\mathcal{W}_{j,\mathrm{z}^2}$ has at most $k=j$. By definition, $\beta^{(2)}_{j,\mathrm{z}^2}=1$. $A^{(\mathsf{a})}$ refers to the amplitude of a module when the non-contractible loop is given weight
\begin{equation}
n_{\mathsf{a}}=\q^{\mathsf{a}}+\q^{-\mathsf{a}},
\end{equation}
with $\mathsf{a}=1$ in the case of the Potts model. 
In section \ref{aratios}, we have used $\alpha_{j,\mathrm{z}^2},\bar{\alpha}_{j,\mathrm{z}^2},\beta^{(k)}_{j,\mathrm{z}^2}$ to define the ratios $\mathsf{R}_{\alpha},\mathsf{R}_{\bar{\alpha}},\mathsf{R}_{\beta}$. One can similarly define the ratio
\begin{equation}
\mathsf{R}_{\gamma}(\mathcal{W}_{j,\mathrm{z}^2})=
\frac{\tilde{A}_{aabb}}{A_{aabb}}(\mathcal{W}_{j,\mathrm{z}^2})=\sum_{\mathsf{a}=1\;\text{odd}}^{p-1}(-1)^{\frac{\mathsf{a}-1}{2}}\gamma^{(\mathsf{a})}_{j,\mathrm{z}^2},
\end{equation}
where $p$ is from the minimal models labeled by $\mathcal{M}(p,q)$ before taking the irrational limit (see \cite{He:2020mhb} for more details). 

We have then \cite{He:2020mhb}:
\begin{subequations}\label{alphabeta}
\begin{eqnarray}
\alpha_{0,-1} &=& -1\,,\\
\alpha_{2,1} &=& {1\over 1-Q} \,,\\
\bar{\alpha}_{2,1} &=& 2-Q\,,\\
\alpha_{4,-1} &=& {2-Q\over 2} \,, \\
\bar{\alpha}_{4,-1} &=& {(Q-1)(Q-4)\over 2}\,,\\
\alpha_{4,1} &=& -\frac{Q^5-7 Q^4+15 Q^3-10 Q^2+4 Q-2}{2 (Q^2-3Q+1)}\,, \\
\bar{\alpha}_{4,1} &=&-\frac{ (Q^2 - 4 Q+2) (Q^2 - 3 Q -2)}{2}\,.
\end{eqnarray}
\end{subequations}
In addition, we have
\begin{equation} \label{betas}
\beta^{(4)}_{4,-1}=-\frac{Q(Q-2)}{3Q-4}.
\end{equation}
The explicit expressions of $\beta^{(4)}_{4,1}, \beta^{(4)}_{4,\pm i}$ and $\gamma^{(\mathsf{a})}_{j,\mathrm{z}^2}$ are also given \cite{He:2020mhb}, but since they are irrelevant in this paper we do not repeat them here.

\section{More details on the numerical bootstrap}\label{bootstrapapp}
\begin{figure}[t!]
	\centering
	\includegraphics[width=0.9\textwidth]{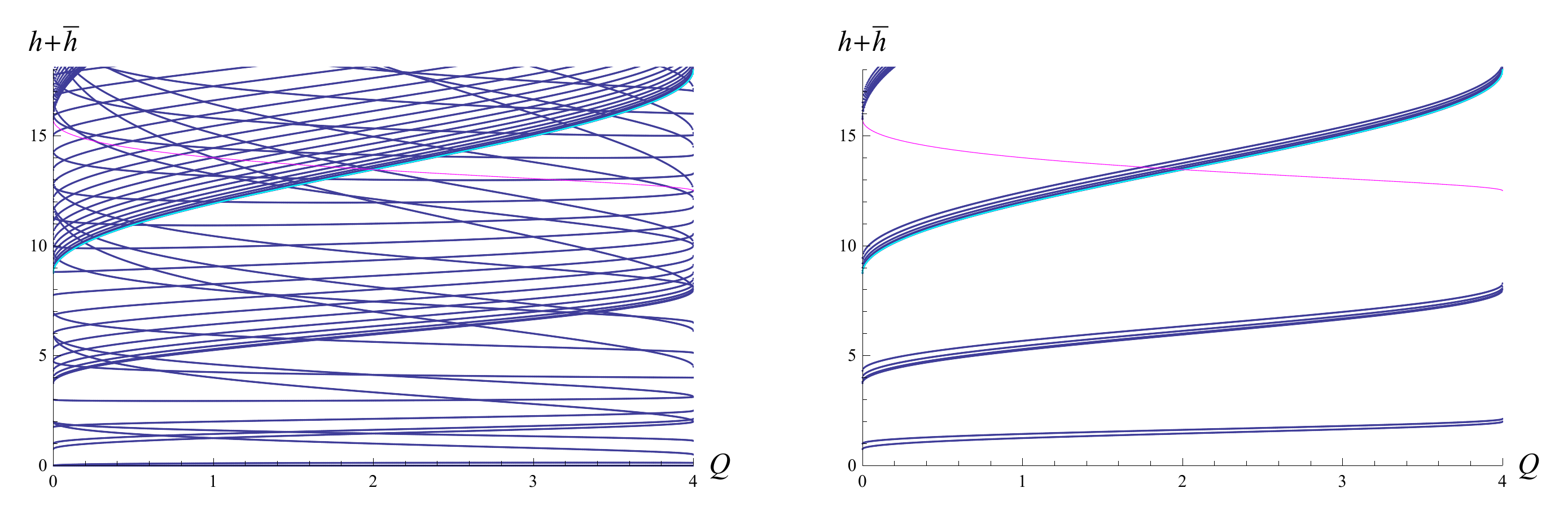}
	\caption{The truncation of the Potts spectrum \eqref{Pottsspectrum} in the interchiral bootstrap. On the left we plot the total dimensions $h+\bar{h}$ for the primary fields in the spectrum. After organizing them in terms of ATL modules, we plot on the right the total dimension of the leading fields in the ATL modules whose amplitudes are taken as the overall amplitude $A(\mathcal{W})$ in the interchiral conformal block expansions of the geometrical correlations and need to be determined by the bootstrap. We truncate at $j=6$ and plotted the total dimension of $(r,s)=(0,6)$ in {\color{cyan}cyan}. However we include the diagonal null descendant $(r,s)^D=(3,-2)$ whose dimension is plotted in {\color{magenta}magenta}. Notice that its total dimension is larger than the truncation for $0\le Q<2$.}
	\label{spectrunc}
\end{figure}

Our numerical bootstrap follows the general philosophy proposed in \cite{Picco:2016ilr} but adapted to our bootstrap program with the interchiral conformal block construction. Namely we solve the interchiral bootstrap equations \eqref{superP} as a linear system for the amplitudes $\eqref{ATLamp}$ with the coefficients given by the blocks $\mathbb{F}^{(s,t)}$ evaluated at a set of points $\{z_i\}$ in the region where the $s$- and $t$-channel conformal blocks converge fast.\footnote{The $q$-expansion in the Zamolodchikov recursive formula is convergent everywhere on the $z$-plane except at $z=1,\infty$ for the $s$-channel and at $z=0,\infty$ for the $t$-channel. This speeds up the convergence within $\{z_i||z_i|<1\}\cap\{z_i||z_i-1|<1\}$ where we evaluate the blocks.} See appendix A.2 of \cite{Ribault:2015sxa} for a rewriting of the recursive formula \eqref{CBZamo} convenient for numerical implementation. We then re-sum the conformal blocks into the interchiral conformal blocks using the analytic recursions \eqref{Rshift1N} and \eqref{Rshift1D} and truncate the blocks according to the total conformal dimension of the primaries (see the caption of figure \ref{spectrunc}).
Solving the bootstrap equations a few times with different sets of points $\{\{z_i\}_\mathrm{m}\}$ gives a set of amplitudes $A_\mathrm{m}(\{z_i\})$. As pointed out in \cite{Picco:2016ilr}, since the amplitudes are supposed to be constants arising from the three-point structure constants in the fusion channels, they should not depend on $\{z_i\}_\mathrm{m}$ and therefore have a small variation within the numerical errors. After imposing various constraints to fix the solution to the Potts model, as described at the beginning of section \ref{bootresults},  we use this criteria of small variation as a check for the stability of the solution, up to the chosen truncation. This variation is defined through the quantity
\begin{equation}
\delta(A)=\frac{\sqrt{\frac{\sum_{\mathrm{m}=1}^\mathrm{M}(A_\mathrm{m}-\bar{A})^2}{\mathrm{M}-1}}}{\bar{A}}\;,
\end{equation}
where $\bar{A}$ is the average among the set of $\mathrm{m}=1,\ldots,\mathrm{M}$ results. Meanwhile, $\delta(A)$ also provides an estimate of the number of significant figures reliable of the average $\bar{A}$.

As we have discussed at the end of section \ref{CB}, since for the moment we do not know the exact regularization procedure for the pole terms in \eqref{CBH} for the modules $\mathcal{W}_{j,1}$ and therefore leave a free amplitude for the block $\mathcal{F}_{h_{m,-n}}(z)\mathcal{F}_{h_{m,-n}}(\bar{z})$, the latter amplitudes are not expected to be constants and will have unstable results. 
In addition, same as for the primary fields, we also impose the amplitude ratios \eqref{ratioforbootstrap} for these free amplitudes associated with the null descendants, for the modules $\mathcal{W}_{2,1}$ and $\mathcal{W}_{4,1}$ in particular. This is a reasonable constraint as in obtaining these ratios on the lattice in \cite{He:2020mhb}, it was observed that they hold for all fields in the same ATL module, regardless of primaries or descendants. Of course, as stated above, this has to be taken as a numerical approximation since the possible logarithm is not taken into consideration. As a result, the instability of these null descendant amplitudes does not influence our final results on the amplitudes that we have presented in section \ref{bootresults}.

The spectrum \eqref{Pottsspectrum} is truncated at a certain total conformal dimension $h+\bar{h}$ after which the primary fields are organized into ATL modules and enter the interchiral conformal block constructions. We have found that in order to obtain a stable bootstrap result on the amplitudes with $j\le4$ for the whole range of $0< Q<4$, it is good to truncate out the modules with $j\ge 6$.\footnote{By including the $j=6$ modules, the resulting amplitudes with $j\le 4$ on average do not exhibit a significant change, but they are less stable.} This being said, it is however necessary to include the conformal block of the diagonal field $(r,s)^D=(3,-2)$, i.e., the level-$6$ null descendant of $(r,s)=(3,2)$ whose block comes with a free amplitude, despite that its total dimension is {\em above} the dimension of the lowest $j=6$ field $(r,s)=(0,6)$---left out by the truncation---within the region $0< Q< 2$. We illustrate the truncation of the spectrum in figure \ref{spectrunc}.

As mentioned at the beginning of section \ref{bootresults}, in addition to the constraints \eqref{constraints}, we have also imposed the analytic ratios \eqref{analyticratios} for obtaining the final results we give in section \ref{bootresults}. Essentially, this is imposing the renormalized Liouville recursions we observed in section \ref{renormL} to fix the bootstrap solution to the Potts model. 
The input of $A_{aaaa}(\mathcal{W}_{4,-1})$ is particular important for obtaining stable amplitudes for $j=4$. This is likely due to a certain instability introduced by the ``naive" regularization we implement in the module $\mathcal{W}_{2,1}$.

\begin{figure}[H]
	\begin{centering}
		\begin{subfigure}{0.5\textwidth}
			\includegraphics[width=0.9\textwidth]{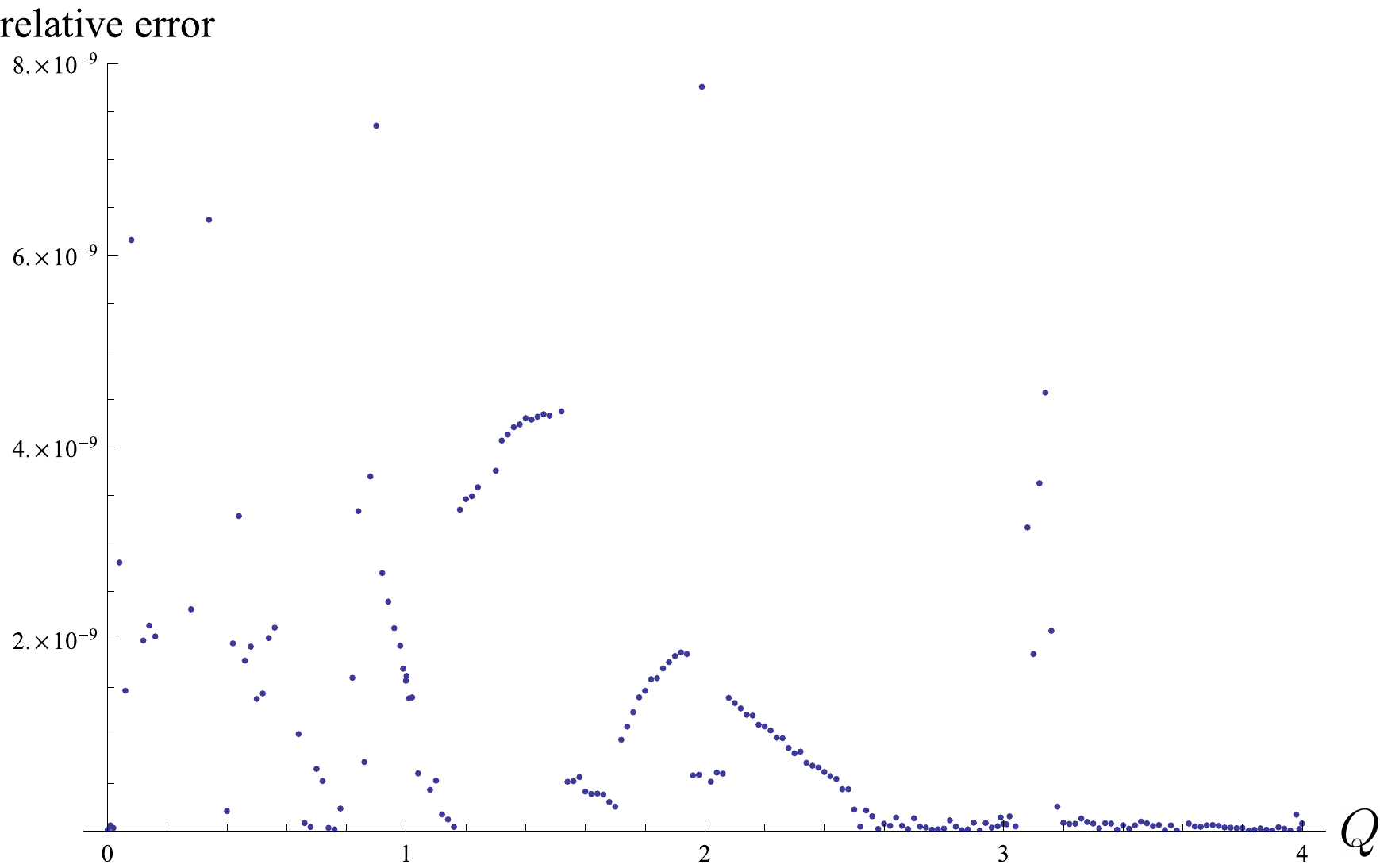}
		\end{subfigure}
		\begin{subfigure}{0.5\textwidth}
			\includegraphics[width=0.9\textwidth]{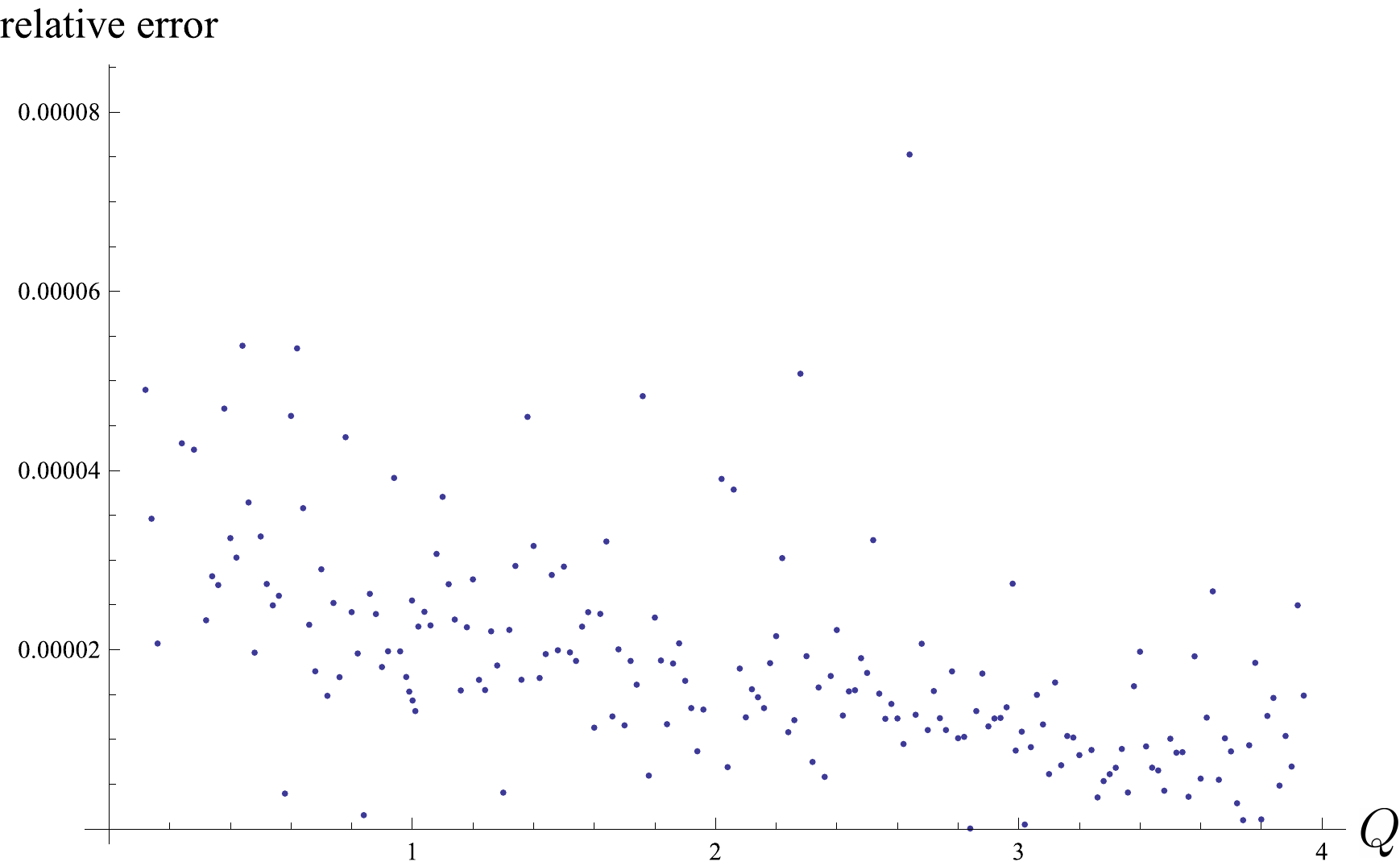}
		\end{subfigure}
	\end{centering}
    \caption{The relative errors in the comparisons of the bootstrap results with the analytic expressions of $A_{aaaa}(\mathcal{W}_{0,-1})$ in figure \ref{aaaa0m1} (left) and $\frac{A_{aaaa}(\mathcal{W}_{4,1})}{A_{aaaa}(\mathcal{W}_{2,1})}$ in figure \ref{aaaa4020ratio} (right). These give a measure of the accuracy in the bootstrapped amplitudes.}
	\label{errors}
\end{figure}

For the plots we give in section \ref{bootresults}, we have typically $\delta(A)\approx 10^{-5},10^{-6}$ except for a few less stable cases (near $Q=0$ for example) with $\delta(A)\approx 10^{-3},10^{-4}$. We present here the detail of a typical bootstrap result:
\begin{align}
\setlength{\arraycolsep}{6mm}
\renewcommand{\arraystretch}{1.2}
\begin{array}{c|c|c}
\mbox{amplitude} & \bar{A} & \delta(A) \\
\hline
A_{aaaa}(\mathcal{W}_{0,-1}) & 1.07789 & 3.16955\times 10^{-11}\\
A_{aaaa}(\mathcal{W}_{2,1}) & 0.137467 & 5.27509\times 10^{-6}\\
A_{aaaa}(\mathcal{W}_{4,1}) & 1.77935\times 10^{-7} & 1.0648\times 10^{-5} \\ 
A_{abab}(\mathcal{W}_{4,i}) & 6.7319\times 10^{-7} &  4.8053\times 10^{-6} \\
\end{array}
\end{align}
at $Q=1.56$ and the other amplitudes are obtained using the recursions \eqref{Rshift1N}, \eqref{Rshift1D} and the amplitude ratios \eqref{ratioforbootstrap}. As stated in \cite{Picco:2016ilr} and mentioned above, the $\delta(A)$ here gives an estimate of the accuracy in the bootstrap determination of the amplitudes $A$. This is also reflected in the few comparisons with the analytic expressions we give in figs. \ref{aaaa0m1}, \ref{aaaa4020ratio} above. We plot the relative error of these comparisons in figure \ref{errors}.

\subsection{Basic checks}\label{basiccheck}
In the bootstrap, we have re-summed the Virasoro conformal blocks into the interchiral conformal blocks using the recursions obtained from the degeneracy of field $\Phi^D_{2,1}$, as discussed in details in section \ref{recursion}. On the other hand, we have also imposed the relations \eqref{analyticratios} which, as discussed in section \ref{renormL}, is essentially a renormalized version of the Liouville recursions. It is actually a fun exercise to check these constraints with lattice computations or ``reduced bootstrap" where the constraints are loosened. We provide some results on such basic checks in this subsection.

\subsubsection*{$\frac{A_{abab}(\mathcal{W}_{2,-1})}{A_{aaaa}(\mathcal{W}_{0,-1})}$ and $\frac{A_{aaaa}(\mathcal{W}_{4,-1})}{A_{aaaa}(\mathcal{W}_{0,-1})}$}

\begin{figure}[H]
	\begin{centering}
	\begin{subfigure}{0.5\textwidth}
		\includegraphics[width=0.9\textwidth]{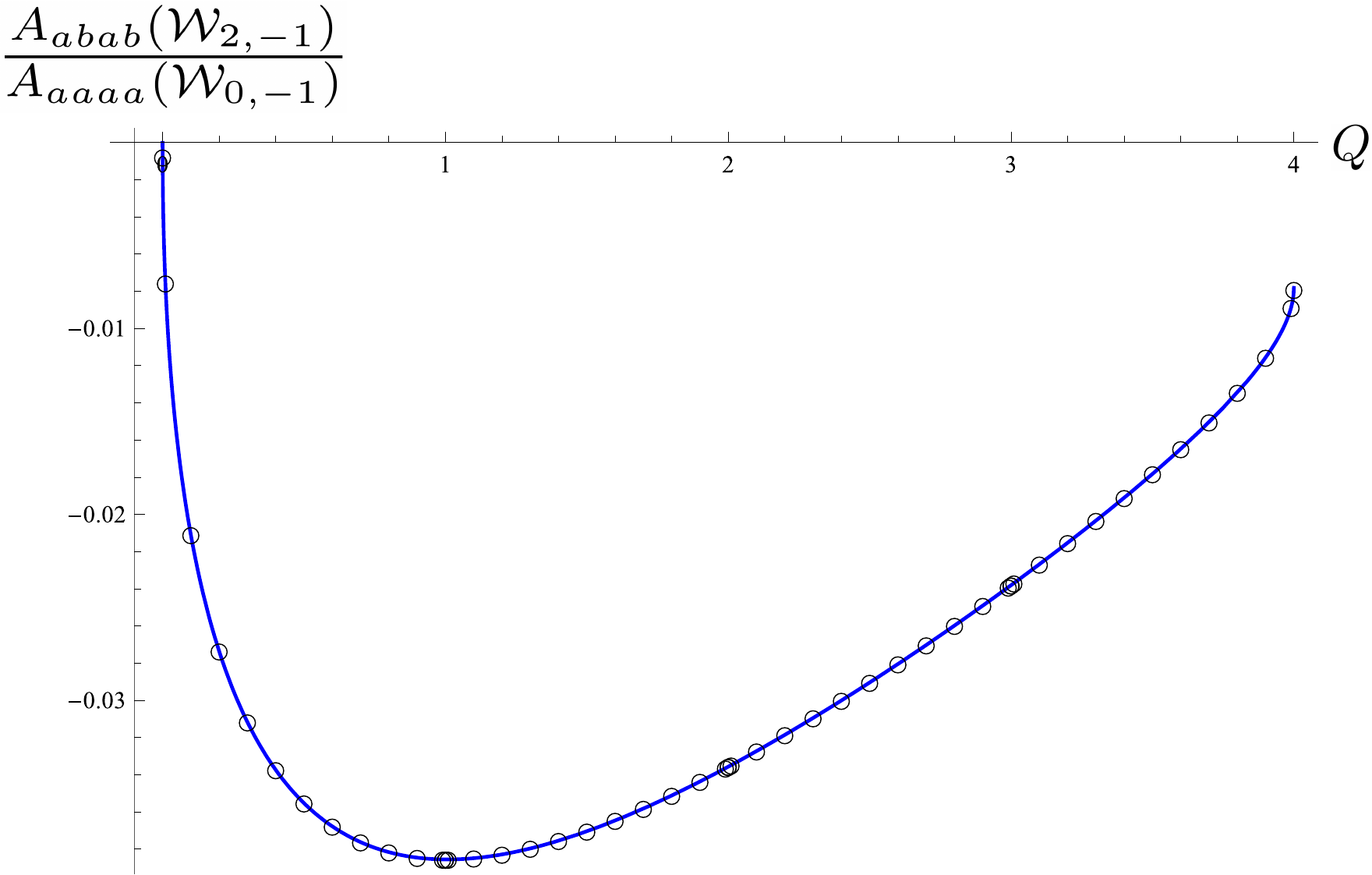}
	\end{subfigure}
	\begin{subfigure}{0.5\textwidth}
		\includegraphics[width=0.9\textwidth]{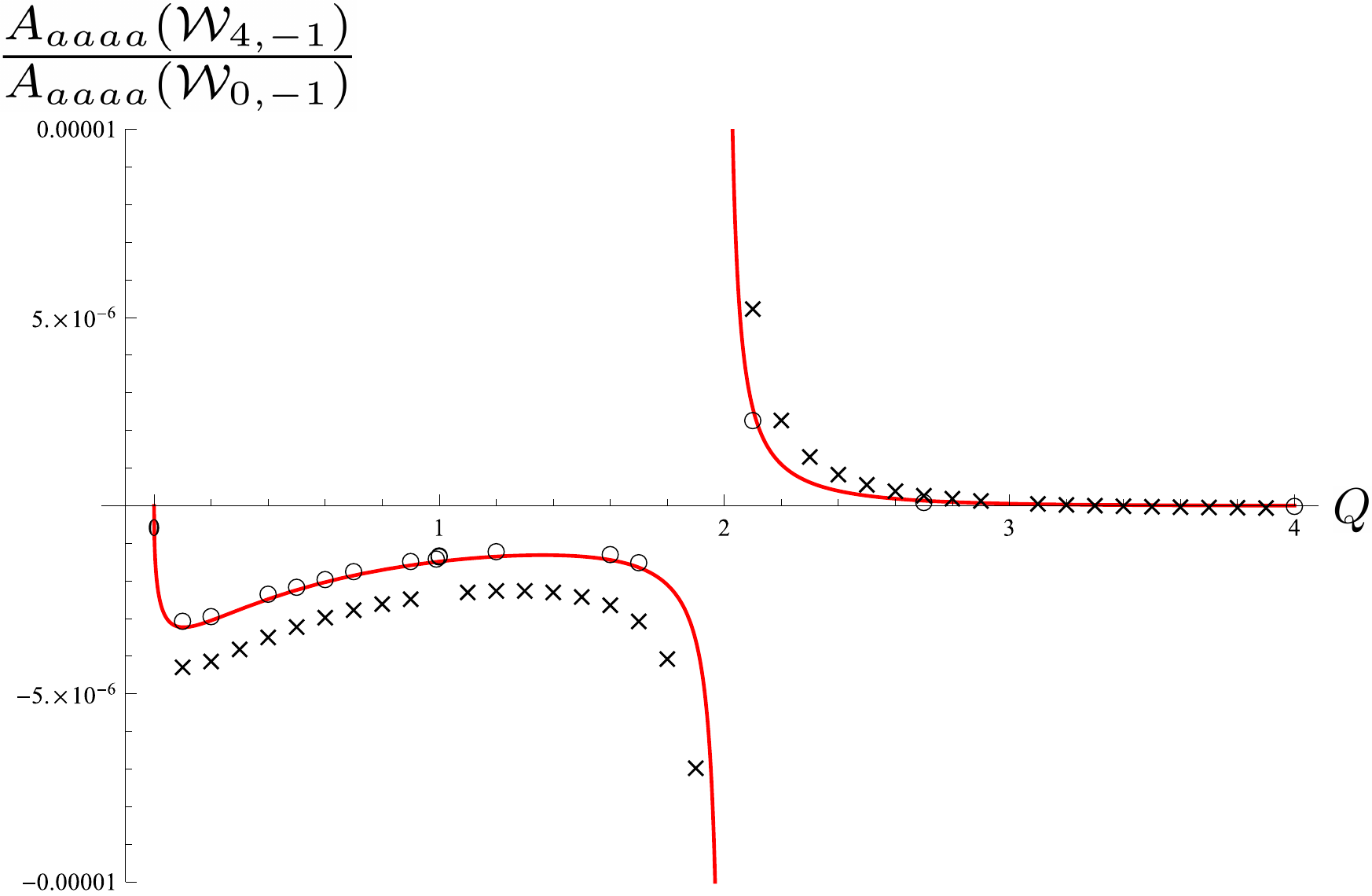}
	\end{subfigure}
\end{centering}
\caption{The analytic results of $\frac{A_{abab}(\mathcal{W}_{2,-1})}{A_{aaaa}(\mathcal{W}_{0,-1})}$ and $\frac{A_{aaaa}(\mathcal{W}_{4,-1})}{A_{aaaa}(\mathcal{W}_{0,-1})}$ compared with lattice (indicated with {\footnotesize$\times$}) and reduced bootstrap results (indicated with {\footnotesize$\Circle$}).}
\label{Aanalattice}
\end{figure}

In figure \ref{Aanalattice}, we plot the analytic results of $\frac{A_{abab}(\mathcal{W}_{2,-1})}{A_{aaaa}(\mathcal{W}_{0,-1})}$ and $\frac{A_{aaaa}(\mathcal{W}_{4,-1})}{A_{aaaa}(\mathcal{W}_{0,-1})}$ obtained in section \ref{pp} (see also eq.~\eqref{renoLrecu}) compared with lattice-computation and ``reduced bootstrap" results. In particular, we bootstrapped $\frac{A_{abab}(\mathcal{W}_{2,-1})}{A_{aaaa}(\mathcal{W}_{0,-1})}$ using eqs.~\eqref{superaaaa} and \eqref{superabab} with constraints \eqref{Rabab}, whose result is quite stable with $\delta(A)\lesssim 10^{-6}$ in general. We do not have the lattice results for this ratio, since it involves the leading amplitudes in $P_{aaaa}$ and $P_{abab}-P_{abba}$ which are used as normalization for the other amplitudes in the same probabilities. For the second ratio $\frac{A_{aaaa}(\mathcal{W}_{4,-1})}{A_{aaaa}(\mathcal{W}_{0,-1})}$, the bootstrap result is much less stable, with $\delta(A)\approx 10^{-2}$ at most. This is obtained by bootstrapping eq.~\eqref{superaaaa} alone. In the plot we show at a few values of $Q$ with $\delta(A)\approx 10^{-2}$ and also the lattice results. As can be seen, the bootstrap results match the analytical values more accurately than the lattice results which have finite-size errors.

\subsubsection*{Recursions \eqref{Rshift1N}}

\begin{figure}[H]
	\centering
	\includegraphics[width=\textwidth]{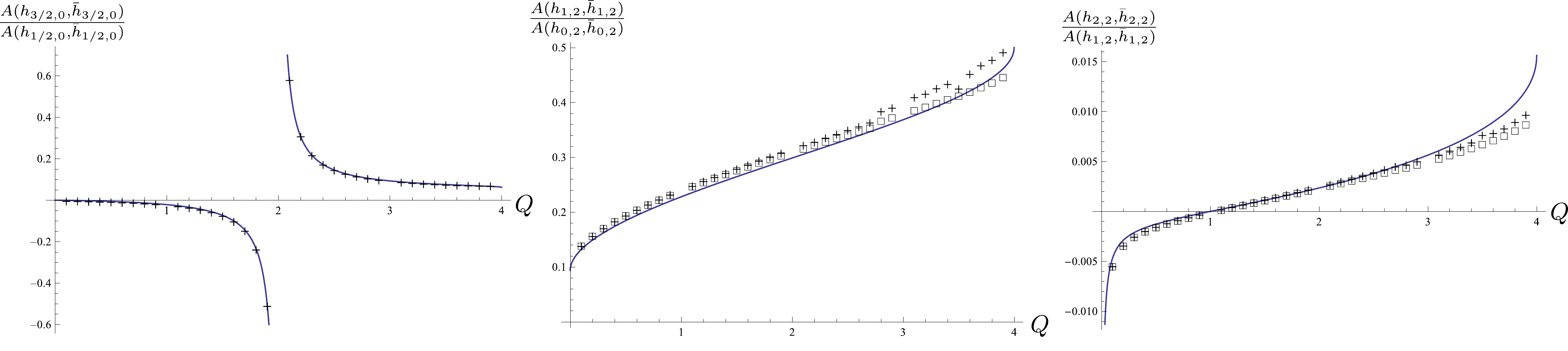}
	\caption{The recursions \eqref{Rshift1N} compared with lattice results. The continuous curves are the analytic expressions. The ``{\tiny $+$}" indicate the lattice results from the amplitudes in $P_{aaaa}$ and ``{\tiny $\square$}" indicate the lattice results from $P_{abab}+P_{abba}$.}
	\label{recursionlattice}
\end{figure}

In figure \ref{recursionlattice}, we plot the analytic recursions \eqref{Rshift1N} compared with the lattice results and find reasonable agreement. In particular, the lattice data for the last two plots are obtained from amplitudes of $\mathcal{W}_{2,1}$ in both $P_{aaaa}$ and $P_{abab}+P_{abba}$. The small discrepancies between these two different determinations appear to be a reasonable measure of the accuracy of the lattice computations, and to within roughly this accuracy the lattice computations are consistent with the analytic results.


\section{$A^L$ and the Liouville recursions from \cite{Migliaccio:2017dch}} \label{AL}
In section \ref{pp}, we have obtained the analytic form of certain Potts amplitudes in terms of the Liouville amplitudes $A^L$, where the latter refers to the non-diagonal generalization of the Liouville theory with the explicit expressions given in \cite{Migliaccio:2017dch}. In addition, we have discussed in section \ref{renormL} how the recursions of the amplitudes in the Potts model with the index $j$ shifted turn out to be a renormalized version of the Liouville recursion and provided the renormalization factors for all the modules up to $j=4$. In this appendix, we give the expression of $A^L$ and the corresponding Liouville recursions as originally obtained in \cite{Migliaccio:2017dch} for the reader's convenience.

The non-diagonal Liouville amplitudes $A^L$ were solved in \cite{Migliaccio:2017dch} for the four-point function \eqref{PRS4pt} and the proportionality in this approximate description of Potts probabilities was fixed to be $\frac{1}{2}$ by comparing with Monte-Carlo simulation in \cite{Picco:2019dkm}. Due to the exact relation \eqref{PRS4ptCFT}, the combination of \eqref{PRSPP} is therefore
\begin{equation}
P_{aaaa}+\tilde{P}_{abab}=2 \left\langle V^D_{\frac{1}{2},0}V^N_{\frac{1}{2},0}V^D_{\frac{1}{2},0}V^N_{\frac{1}{2},0} \right\rangle \,,
\end{equation}
whose conformal block expansion is given by the spectrum $\mathcal{S}_{\mathbb{Z}+\frac{1}{2},2\mathbb{Z}}$:\footnote{The $A^L(h_{r,s},\bar{h}_{r,s})$ here should be identified with $D_{(s,r)}$ in \cite{Picco:2019dkm}.}
\begin{equation}\label{PRSblockexpansion}
\begin{aligned}
\left\langle V^D_{\frac{1}{2},0}V^N_{\frac{1}{2},0}V^D_{\frac{1}{2},0}V^N_{\frac{1}{2},0} \right\rangle
=&\sum_{e \in \mathbb{N}}A^L \left(h_{e+\frac{1}{2},0},\bar{h}_{e+\frac{1}{2},0} \right)
\mathcal{F}^{(\mathsf{c})}_{h_{e+\frac{1}{2},0}}(z)\mathcal{F}^{(\mathsf{c})}_{h_{e+\frac{1}{2},0}}(\bar{z})\\
+&\sum_{e \in \mathbb{N},s\ge 2}A^L \left(h_{e+\frac{1}{2},s},\bar{h}_{e+\frac{1}{2},s} \right)\bigg(\mathcal{F}^{(\mathsf{c})}_{h_{e+\frac{1}{2},s}}(z)\mathcal{F}^{(\mathsf{c})}_{h_{e+\frac{1}{2},-s}}(\bar{z})+\mathcal{F}^{(\mathsf{c})}_{h_{e+\frac{1}{2},-s}}(z)\mathcal{F}^{(\mathsf{c})}_{h_{e+\frac{1}{2},s}}(\bar{z})\bigg).
\end{aligned}
\end{equation}
Comparing \eqref{PRSblockexpansion} with our construction of the interchiral blocks $\mathbb{F}_{j,-1}$ in \eqref{superjm1} and the definition of $A^{L}(\mathcal{W}_{j,-1})$ in \eqref{PRSPPcft}, we can identify:
\begin{equation}
A^L \left(\mathcal{W}_{0,-1} \right)=2A^L \left(h_{\frac{1}{2},0},\bar{h}_{\frac{1}{2},0} \right) \,,
\end{equation}
whose explicit expression is given by \cite{Picco:2019dkm}:
\begin{equation}\label{A0m1}
\begin{aligned}
2A^L \left(h_{\frac{1}{2},0},\bar{h}_{\frac{1}{2},0} \right)=&8\pi^2\upbeta^{\frac{2}{\upbeta^2}-2\upbeta^2}\frac{\Gamma(\upbeta^2)\Gamma(\frac{1}{\upbeta^2})}{\Gamma(2-\upbeta^2)\Gamma(2-\frac{1}{\upbeta^2})}\\
&\times\Gamma_{\upbeta}^3\left(\upbeta+\frac{1}{2\upbeta}\right)\Gamma_{\upbeta}^3\left(\upbeta-\frac{1}{2\upbeta}\right)\Gamma_{\upbeta}^3\left(\frac{1}{2\upbeta}\right)\Gamma_{\upbeta}^3\left(\frac{3}{2\upbeta}\right)\Upsilon_{\upbeta}^2\left(\frac{\upbeta}{2}-\frac{1}{4\upbeta}\right)\Upsilon_{\upbeta}^6\left(\frac{\upbeta}{2}+\frac{1}{4\upbeta}\right),
\end{aligned}
\end{equation}
where $\Gamma_{\upbeta}$ and $\Upsilon_{\upbeta}$ are the double-Gamma and Upsilon functions. Eq.~\eqref{A0m1} is the expression we have used for plotting the analytic curve in figure \ref{aaaa0m1}, where we have found perfect agreement with the bootstrap results with the normalization \eqref{Anorm}. On the other hand, for $s\ge 2$, i.e, $j \ge 2$, the identification is
\begin{equation}
A^L(\mathcal{W}_{j,-1})=4A^L(h_{\frac{1}{2},j},\bar{h}_{\frac{1}{2},j}).
\end{equation} 

The recursion of shifting the index $s$ for $(h_{r,s},\bar{h}_{r,s})$ was given in \cite{Migliaccio:2017dch} as:
\begin{equation}
\begin{aligned}
\frac{A^L(h_{r,s+1},\bar{h}_{r,s+1})}{A^L(h_{r,s-1},\bar{h}_{r,s-1})}=&-\frac{\Gamma(-r-s\upbeta^2)\Gamma(r-s\upbeta^2)\Gamma(-r+(1-s)\upbeta^2)\Gamma(1-r-(1+s)\upbeta^2)}{\Gamma(-r+s\upbeta^2)\Gamma(r+s\upbeta^2)\Gamma(-r+(1+s)\upbeta^2)\Gamma(1-r-(1-s)\upbeta^2)}\\
&\times \frac{\Gamma(-\frac{r}{2}+\frac{s}{2}\upbeta^2)^2\Gamma(\frac{2-r}{2}+\frac{s}{2}\upbeta^2)^2\Gamma(\frac{1-r}{2}+\frac{s}{2}\upbeta^2)^4}{\Gamma(-\frac{r}{2}-\frac{s}{2}\upbeta^2)^2\Gamma(\frac{2-r}{2}-\frac{s}{2}\upbeta^2)^2\Gamma(\frac{1-r}{2}-\frac{s}{2}\upbeta^2)^4}
\end{aligned}
\end{equation}
which gives us the following Liouville recursions for $j\ge 2$:
\begin{subequations}
\begin{eqnarray}
\frac{A^L(\mathcal{W}_{2,-1})}{A^L(\mathcal{W}_{0,-1})}&=&2\frac{A^L(h_{\frac{1}{2},2},\bar{h}_{\frac{1}{2},2})}{A^L(h_{\frac{1}{2},0},\bar{h}_{\frac{1}{2},0})},\label{factor2}\\
\frac{A^L(\mathcal{W}_{4,-1})}{A^L(\mathcal{W}_{2,-1})}&=&\frac{A^L(h_{\frac{1}{2},4},\bar{h}_{\frac{1}{2},4})}{A^L(h_{\frac{1}{2},2},\bar{h}_{\frac{1}{2},2})}.
\end{eqnarray}
\end{subequations}
Notice that the factor of 2 in \eqref{factor2} is due to the special definition of $A^L$ with $j=0$ in \eqref{PRSblockexpansion}. In general, the Liouville recursion is given by
\begin{equation}\label{Lrecursion}
\frac{A^L(\mathcal{W}_{j+1,e^{2i\pi p/M}})}{A^L(\mathcal{W}_{j-1,e^{2i\pi p/M}})}=\frac{A^L(h_{\frac{p}{M},j+1},\bar{h}_{\frac{p}{M},j+1})}{A^L(h_{\frac{p}{M},j-1},\bar{h}_{\frac{p}{M},j-1})}, \quad \mbox{for } j\neq 1,
\end{equation}
which we have used in \eqref{4020ratio} for extracting another renormalized Liouville recursion from the bootstrap. Note that despite the amplitudes in the numerator and denominator on the right-hand side of \eqref{Lrecursion} not being given by the analytic results of \cite{Migliaccio:2017dch}, the recursion exists and coincides with that of \cite{Migliaccio:2017dch}, as a result of the degeneracy of the field $\Phi^D_{1,2}$ there.

\bibliographystyle{hieeetr}
\bibliography{Pottsreferences}

\begin{thebibliography}{10}

\bibitem{Jacobsen:2018pti}
J.~L. Jacobsen and H.~Saleur, ``{Bootstrap approach to geometrical four-point
  functions in the two-dimensional critical $Q$-state Potts model: A study of
  the $s$-channel spectra},'' {\em JHEP}, vol.~01, p.~084, 2019, 1809.02191.

\bibitem{Gainutdinov:2012nq}
A.~Gainutdinov, N.~Read, and H.~Saleur, ``{Associative algebraic approach to
  logarithmic CFT in the bulk: the continuum limit of the
  ${\mathfrak{gl}(1|1)}$ periodic spin chain, Howe duality and the interchiral
  algebra},'' {\em Commun. Math. Phys.}, vol.~341, no.~1, pp.~35--103, 2016,
  1207.6334.

\bibitem{He:2020mhb}
Y.~He, L.~Grans-Samuelsson, J.~L. Jacobsen, and H.~Saleur, ``{Geometrical
  four-point functions in the two-dimensional critical $Q$-state Potts model:
  {C}onnections with the RSOS models},'' 2020, 2002.09071.

\bibitem{Migliaccio:2017dch}
S.~Migliaccio and S.~Ribault, ``{The analytic bootstrap equations of
  non-diagonal two-dimensional CFT},'' {\em JHEP}, vol.~05, p.~169, 2018,
  1711.08916.

\bibitem{Rattazzi:2008pe}
R.~Rattazzi, V.~S. Rychkov, E.~Tonni, and A.~Vichi, ``{Bounding scalar operator
  dimensions in 4D CFT},'' {\em JHEP}, vol.~12, p.~031, 2008, 0807.0004.

\bibitem{ElShowk:2012ht}
S.~El-Showk, M.~F. Paulos, D.~Poland, S.~Rychkov, D.~Simmons-Duffin, and
  A.~Vichi, ``{Solving the 3D Ising model with the conformal bootstrap},'' {\em
  Phys.\ Rev.\ D}, vol.~86, p.~025022, 2012, 1203.6064.

\bibitem{El-Showk:2014dwa}
S.~El-Showk, M.~F. Paulos, D.~Poland, S.~Rychkov, D.~Simmons-Duffin, and
  A.~Vichi, ``{Solving the 3d Ising model with the conformal bootstrap II.
  $c$-minimization and precise critical exponents},'' {\em J.\ Stat.\ Phys.},
  vol.~157, p.~869, 2014, 1403.4545.

\bibitem{Kos:2016ysd}
F.~Kos, D.~Poland, D.~Simmons-Duffin, and A.~Vichi, ``{Precision islands in the
  Ising and $O(N)$ models},'' {\em JHEP}, vol.~08, p.~036, 2016, 1603.04436.

\bibitem{Gliozzi:2013ysa}
F.~Gliozzi, ``{More constraining conformal bootstrap},'' {\em Phys.\ Rev.\
  Lett.}, vol.~111, p.~161602, 2013, 1307.3111.

\bibitem{Gliozzi:2014jsa}
F.~Gliozzi and A.~Rago, ``{Critical exponents of the 3d Ising and related
  models from conformal bootstrap},'' {\em JHEP}, vol.~10, p.~042, 2014,
  1403.6003.

\bibitem{LeClair:2018edq}
A.~Leclair and J.~Squires, ``{Conformal bootstrap for percolation and
  polymers},'' {\em J. Stat. Mech.}, vol.~1812, p.~123105, 2018, 1802.08911.

\bibitem{Hikami:2017sbg}
S.~Hikami, ``{Conformal bootstrap analysis for single and branched polymers},''
  {\em PTEP}, vol.~2018, no.~12, p.~123I01, 2018, 1708.03072.

\bibitem{FORTUIN1972536}
C.~Fortuin and P.~Kasteleyn, ``On the random-cluster model: I. introduction and
  relation to other models,'' {\em Physica}, vol.~57, no.~4, pp.~536 -- 564,
  1972.

\bibitem{Delfino:2010xm}
G.~Delfino and J.~Viti, ``{On three-point connectivity in two-dimensional
  percolation},'' {\em J. Phys.}, vol.~A44, p.~032001, 2011, 1009.1314.

\bibitem{Picco:2013nga}
M.~Picco, R.~Santachiara, J.~Viti, and G.~Delfino, ``{Connectivities of Potts
  Fortuin-Kasteleyn clusters and time-like Liouville correlator},'' {\em Nucl.
  Phys.}, vol.~B875, pp.~719--737, 2013, 1304.6511.

\bibitem{Ikhlef:2015eua}
Y.~Ikhlef, J.~L. Jacobsen, and H.~Saleur, ``Three-point functions in $c \le 1$
  {L}iouville theory and conformal loop ensembles,'' {\em Phys. Rev. Lett.},
  vol.~116, no.~13, p.~130601, 2016, 1509.03538.

\bibitem{Delfino:2011sc}
G.~Delfino and J.~Viti, ``{Potts q-color field theory and scaling random
  cluster model},'' {\em Nucl. Phys.}, vol.~B852, pp.~149--173, 2011,
  1104.4323.

\bibitem{GoriViti:2018}
G.~Gori and J.~Viti, ``{Four-point boundary connectivities in critical
  two-dimensional percolation from conformal invariance},'' {\em J. High Energ.
  Phys.}, vol.~2018, p.~131, 2018, 1806.02330.

\bibitem{Picco:2016ilr}
M.~Picco, S.~Ribault, and R.~Santachiara, ``{A conformal bootstrap approach to
  critical percolation in two dimensions},'' {\em SciPost Phys.}, vol.~1,
  no.~1, p.~009, 2016, 1607.07224.

\bibitem{Picco:2019dkm}
M.~Picco, S.~Ribault, and R.~Santachiara, ``{On four-point connectivities in
  the critical 2d Potts model},'' {\em SciPost Phys.}, vol.~7, no.~4, p.~044,
  2019, 1906.02566.

\bibitem{Javerzat:2019ujh}
N.~Javerzat, M.~Picco, and R.~Santachiara, ``{Two-point connectivity of
  two-dimensional critical $Q-$ Potts random clusters on the torus},'' {\em J.
  Stat. Mech.}, vol.~2002, no.~2, p.~023101, 2020, 1907.11041.

\bibitem{Javerzat:2019ohi}
N.~Javerzat, M.~Picco, and R.~Santachiara, ``{Three- and four-point
  connectivities of two-dimensional critical $Q-$ Potts random clusters on the
  torus},'' 12 2019, 1912.05865.

\bibitem{Dotsenko:2019dcu}
V.~S. Dotsenko, ``{Four spins correlation function of the $q$ states Potts
  model, for general values of $q$. Its percolation model limit $q \to 1$},''
  {\em Nucl. Phys. B}, vol.~953, p.~114973, 2020, 1911.06682.

\bibitem{Zamolodchikov:1995aa}
A.~B. Zamolodchikov and A.~B. Zamolodchikov, ``{Structure constants and
  conformal bootstrap in Liouville field theory},'' {\em Nucl. Phys.},
  vol.~B477, pp.~577--605, 1996, hep-th/9506136.

\bibitem{Teschner:1995yf}
J.~Teschner, ``{On the Liouville three point function},'' {\em Phys. Lett.},
  vol.~B363, pp.~65--70, 1995, hep-th/9507109.

\bibitem{Estienne:2015sua}
B.~Estienne and Y.~Ikhlef, ``{Correlation functions in loop models},'' 2015,
  1505.00585.

\bibitem{potts_1952}
R.~B. Potts, ``Some generalized order-disorder transformations,'' {\em Math.
  Proc. Cambr. Phil. Soc.}, vol.~48, no.~1, pp.~106--109, 1952.

\bibitem{Baxter_1973}
R.~J. Baxter, ``Potts model at the critical temperature,'' {\em J. Phys. C:
  Solid State Phys.}, vol.~6, no.~23, pp.~L445--L448, 1973.

\bibitem{Baxter_1976}
R.~J. Baxter, S.~B. Kelland, and F.~Y. Wu, ``{Equivalence of the Potts model or
  Whitney polynomial with an ice-type model},'' {\em J. Phys. A: Math. Gen.},
  vol.~9, no.~3, pp.~397--406, 1976.

\bibitem{CGreview}
J.~L. Jacobsen, ``Conformal field theory applied to loop models,'' in {\em
  Polygons, polyominoes and polycubes} (A.~J. Guttmann, ed.), vol.~775 of {\em
  {Lecture Notes in Physics}}, pp.~347--424, Heidelberg: Springer, 2009.

\bibitem{denNijs:1983zz}
M.~den Nijs, ``{Extended scaling relations for the magnetic critical exponents
  of the Potts model},'' {\em Phys. Rev.}, vol.~B27, pp.~1674--1679, 1983.

\bibitem{Nienhuis:1984wm}
B.~Nienhuis, ``{Critical behavior of two-dimensional spin models and charge
  asymmetry in the Coulomb gas},'' {\em J. Statist. Phys.}, vol.~34,
  pp.~731--761, 1984.

\bibitem{Zamolodchikov:2005fy}
A.~B. Zamolodchikov, ``{Three-point function in the minimal Liouville
  gravity},'' {\em Theor. Math. Phys.}, vol.~142, pp.~183--196, 2005,
  hep-th/0505063.

\bibitem{Vasseur:2012tz}
R.~Vasseur, J.~L. Jacobsen, and H.~Saleur, ``{Logarithmic observables in
  critical percolation},'' {\em J. Stat. Mech.}, vol.~1207, p.~L07001, 2012,
  1206.2312.

\bibitem{Vasseur:2013baa}
R.~Vasseur and J.~L. Jacobsen, ``{Operator content of the critical Potts model
  in $\mathcal{d}$ dimensions and logarithmic correlations},'' {\em Nucl. Phys.
  B}, vol.~880, pp.~435--475, 2014, 1311.6143.

\bibitem{Couvreur:2017inl}
R.~Couvreur, J.~Lykke~Jacobsen, and R.~Vasseur, ``{Non-scalar operators for the
  Potts model in arbitrary dimension},'' {\em J. Phys. A}, vol.~50, no.~47,
  p.~474001, 2017, 1704.02186.

\bibitem{Belavin:1984vu}
A.~Belavin, A.~M. Polyakov, and A.~Zamolodchikov, ``{Infinite conformal
  symmetry in two-dimensional quantum field theory},'' {\em Nucl. Phys. B},
  vol.~241, pp.~333--380, 1984.

\bibitem{Zamolodchikovrecursion}
A.~B. Zamolodchikov, ``{Conformal symmetry in two-dimensional space: Recursion
  representation of conformal block},'' {\em Theor. Math. Phys.}, vol.~73,
  p.~1088–1093, 1987.

\bibitem{Gainutdinov_2013}
A.~Gainutdinov, D.~Ridout, and I.~Runkel, ``Logarithmic conformal field
  theory,'' {\em Journal of Physics A: Mathematical and Theoretical}, vol.~46,
  p.~490301, nov 2013.

\bibitem{Ribault:2019qrz}
S.~Ribault, ``{The non-rational limit of D-series minimal models},'' 2019,
  1909.10784.

\bibitem{Ribault:2018jdv}
S.~Ribault, ``{On 2d CFTs that interpolate between minimal models},'' {\em
  SciPost Phys.}, vol.~6, p.~075, 2019, 1809.03722.

\bibitem{Gainutdinov:2014foa}
A.~Gainutdinov, N.~Read, H.~Saleur, and R.~Vasseur, ``{The periodic $s
  \ell$(2|1) alternating spin chain and its continuum limit as a bulk
  logarithmic conformal field theory at $c = 0$},'' {\em JHEP}, vol.~05,
  p.~114, 2015, 1409.0167.

\bibitem{Ribault:2015sxa}
S.~Ribault and R.~Santachiara, ``{Liouville theory with a central charge less
  than one},'' {\em JHEP}, vol.~08, p.~109, 2015, 1503.02067.

\end{thebibliography}

\end{document}